\documentclass[a4paper,11pt]{article}

\usepackage{jheppub}
\usepackage{amsfonts, amsmath, amsthm, amssymb, bm, mathrsfs}
\usepackage{capt-of}
\usepackage{float}
\usepackage{subfigure}
\usepackage{graphicx,color,dcolumn,booktabs}
\usepackage{longtable,lscape,comment,braket}
\usepackage{overpic}
\usepackage{tipa}
\usepackage{indentfirst}
\usepackage{feynmf}   
\usepackage{slashed}  
\usepackage{cases}
\usepackage{times} 
\usepackage{epstopdf}
\usepackage{psfrag}
\usepackage{color}
\usepackage{multirow}
\usepackage{setspace}

\allowdisplaybreaks

\newcommand{\beqa}{\begin{eqnarray}}
\newcommand{\eeqa}{\end{eqnarray}}

\newlength{\feynwidth} 

\title{Hidden-charm and hidden-bottom molecular pentaquarks in chiral perturbation theory}
\author[a,b]{Bo Wang,}
\author[a]{Lu Meng,}
\author[a,b]{Shi-Lin Zhu}

\affiliation[a]{School of Physics and State Key Laboratory of
Nuclear Physics and Technology, Peking University, Beijing 100871,
China} \affiliation[b]{Center of High Energy Physics, Peking
University, Beijing 100871, China}

\emailAdd{bo-wang@pku.edu.cn} \emailAdd{lmeng@pku.edu.cn}
\emailAdd{zhusl@pku.edu.cn}

\bigskip

\abstract{The newly observed $P_c(4312)$, $P_c(4440)$ and
$P_c(4457)$ at the LHCb experiment are very close to the
$\Sigma_c\bar{D}$ and $\Sigma_c\bar{D}^\ast$ thresholds. In this
work, we perform a systematic study and give a complete picture on
the interactions between the $\Sigma_c^{(\ast)}$ and
$\bar{D}^{(\ast)}$ systems in the framework of heavy hadron chiral
perturbation theory, where the short-range contact interaction,
long-range one-pion-exchange contribution, and intermediate-range
two-pion-exchange loop diagrams are all considered. We first
investigate the three $P_c$ states without and with considering the
$\Lambda_c$ contribution in the loop diagrams. It is difficult to
simultaneously reproduce the three $P_c$s unless the $\Lambda_c$ is
included. The coupling between the
$\Sigma_c^{(\ast)}\bar{D}^{(\ast)}$ and $\Lambda_c\bar{D}^{(\ast)}$
channels is crucial for the formation of these $P_c$s. Our
calculation supports the $P_c(4312)$, $P_c(4440)$ and $P_c(4457)$ to
be the $S$-wave hidden-charm $[\Sigma_c\bar{D}]_{J=1/2}^{I=1/2}$,
$[\Sigma_c\bar{D}^\ast]_{J=1/2}^{I=1/2}$ and
$[\Sigma_c\bar{D}^\ast]_{J=3/2}^{I=1/2}$ molecular pentaquarks,
respectively. Our calculation disfavors the spin assignment
$J^P=\frac{1}{2}^-$ for $P_c(4457)$ and $J^P=\frac{3}{2}^-$ for
$P_c(4440)$, because the excessively enhanced spin-spin interaction
is unreasonable in the present case. We obtain the complete mass
spectra of the $[\Sigma_c^{(\ast)}\bar{D}^{(\ast)}]_J$ systems with
the fixed low energy constants. Our result indicates the existence
of the
$[\Sigma_c^{\ast}\bar{D}^{\ast}]_J~(J=\frac{1}{2},\frac{3}{2},\frac{5}{2})$
hadronic molecules. The previously reported $P_c(4380)$ might be a
deeper bound one. Additionally, we also study the hidden-bottom
$\Sigma_b^{(\ast)}B^{(\ast)}$ systems, and predict seven bound
molecular states, which could serve as a guidance for future
experiments. Furthermore, we also examine the heavy quark symmetry
breaking effect in the hidden-charm and hidden-bottom systems by
taking into account the mass splittings in the propagators of the
intermediate states. As expected, the heavy quark symmetry in the
bottom cases is better than that in the charmed sectors. We notice
that the heavy quark symmetry in the $\Sigma_c\bar{D}$ and
$\Sigma_c^\ast\bar{D}$ systems is much worse for some fortuitous
reasons. The heavy quark symmetry breaking effect is nonnegligible
in predicting the effective potentials between the charmed hadrons.}

\keywords{Chiral Lagragians, Molecular Pentaquarks, Heavy Quark
Symmetry}

\begin{document}

\maketitle \flushbottom

\section{Introduction}\label{introduction}

The charmonium physics is one of the most charming and interesting
sectors in quantum chromodynamics (QCD). On the one hand, the
charmonium spectra deepen our understanding on the nonperturbative
QCD and serve as a good platform to develop multifarious potential
models. On the other hand, the discoveries of the exotic $XYZ$
states challenge the conventional hadron
spectra~\cite{Chen:2016qju}, since these states cannot be easily
reconciled with the predictions of the conventional quark models.
Furthermore, the heavy quark symmetry in the charm sector is not
good enough, thus the heavy quark symmetry breaking effect would
manifest itself and lead to some novel phenomena sometimes.

In 2015, two pentaquark candidates $P_c(4380)$ and $P_c(4450)$ were
observed by the LHCb Collaboration in the $J/\psi p$ invariant mass
spectrum via the weak decay process of $\Lambda_b^0\to J/\psi
pK^-$~\cite{Aaij:2015tga}. The discovery of these two exotica
triggered many discussions on the their internal structures (for
some related reviews, see
refs.~\cite{Chen:2016qju,Guo:2017jvc,Liu:2019zoy,Lebed:2016hpi,Esposito:2016noz,Brambilla:2019esw}),
among which, the molecular interpretation is the most favored one.
In ref.~\cite{Chen:2015loa}, these two states are interpreted as the
deeply bound $\Sigma_c\bar{D}^\ast$ and $\Sigma_c^\ast\bar{D}^\ast$
molecular states in the framework of one-pion-exchange model.
Whereas in ref.~\cite{He:2015cea}, they are regarded as the
$\Sigma_c^\ast\bar{D}$ and $\Sigma_c^\ast\bar{D}^\ast$ molecules,
respectively. However, the $J^P$ quantum numbers of the $P_c(4380)$
and $P_c(4450)$ remain an open question.

Very recently, the LHCb Collaboration reported the new results with
the updated data~\cite{Aaij:2019vzc}. A new narrow state $P_c(4312)$
is observed in the $J/\psi p$ mass spectrum. In addition, the
previously observed structure $P_c(4450)$ is dissolved into two
narrow peaks $P_c(4440)$ and $P_c(4457)$. Since these three states
lie several to tens MeV below the thresholds of $\Sigma_c\bar{D}$
and $\Sigma_c\bar{D}^\ast$, the molecular explanation is proposed
with the chiral perturbation theory~\cite{Meng:2019ilv},
contact-range effective field theory~\cite{Liu:2019tjn},
one-boson-exchange model~\cite{Chen:2019asm}, local hidden gauge
formalism~\cite{Xiao:2019aya}, and Bethe-Salpeter equation
approach~\cite{He:2019ify}, respectively. The decays and productions
of the $P_c$ states are also studied in
refs.~\cite{Xiao:2019mst,Sakai:2019qph,Voloshin:2019aut,Wang:2019krd}
(one can see
refs.~\cite{Guo:2019fdo,Ali:2019npk,Guo:2019kdc,Weng:2019ynv,Burns:2019iih}
for some other pertinent works).

The interactions between $\Sigma_c^{(\ast)}$ and $\bar{D}^{(\ast)}$
are essential to map out the mass spectra of the
$\Sigma_c^{(\ast)}\bar{D}^{(\ast)}$ molecules. Before the discovery
of these $P_c$ states, the $\Sigma_c^{(\ast)}\bar{D}^{(\ast)}$
interactions have been investigated with the one-boson-exchange
model~\cite{Wu:2010jy,Yang:2011wz} and chiral quark
model~\cite{Wang:2011rga}. In this work, in light of the newly
observed $P_c(4312)$, $P_c(4440)$ and
$P_c(4457)$~\cite{Aaij:2019vzc}, we systematically study the
$\Sigma_c^{(\ast)}$ and $\bar{D}^{(\ast)}$ interactions with chiral
perturbation theory up to the one-loop level.

Nowadays, as the one inheritor of the Yukawa theory, the
one-boson-exchange model is the most popular and economical
formalism for depicting the nucleon-nucleon ($N$-$N$)
systems~\cite{Stoks:1994wp,Machleidt:2000ge} and $XYZ$
states~\cite{Chen:2016qju}. But in this model, one has to include as
many exchanged particles as possible, such as $\pi$, $\sigma$,
$\rho$, $\omega$, or higher states and so on. As the other inheritor
of the Yukawa theory, chiral perturbation theory plays a pivotal
role in the modern theory of nuclear force. Its degrees of freedom
are unambiguous, i.e., the pion and matter field. Another advantage
of chiral perturbation theory is its consistent power counting. The
scattering amplitude can be expanded order by order with a small
parameter $\varepsilon$ (generally, $\varepsilon=m_\pi/\Lambda_\chi$
or $q/\Lambda_\chi$, where $m_\pi$ and $q$ are the mass and momentum
of pion, respectively, and $\Lambda_\chi\simeq1$ GeV is the chiral
breaking scale). Therefore, the error is estimable and controllable.
In the past decades, the chiral perturbation theory has been
extensively exploited to study the $N$-$N$ systems with great
success~\cite{Bernard:1995dp,Epelbaum:2008ga,Machleidt:2011zz,Meissner:2015wva,Hammer:2019poc}.
Moreover, in recently years, this theory is also employed to
investigate the effective potentials of the
$DD^\ast$~\cite{Xu:2017tsr},
$\bar{B}^{(*)}\bar{B}^{(*)}$~\cite{Liu:2012vd,Wang:2018atz}, and
$\Sigma_c\bar{D}^{(*)}$~\cite{Meng:2019ilv} systems.

\begin{figure*}[hptb]
\begin{centering}
    \scalebox{0.6}{\includegraphics[width=\columnwidth]{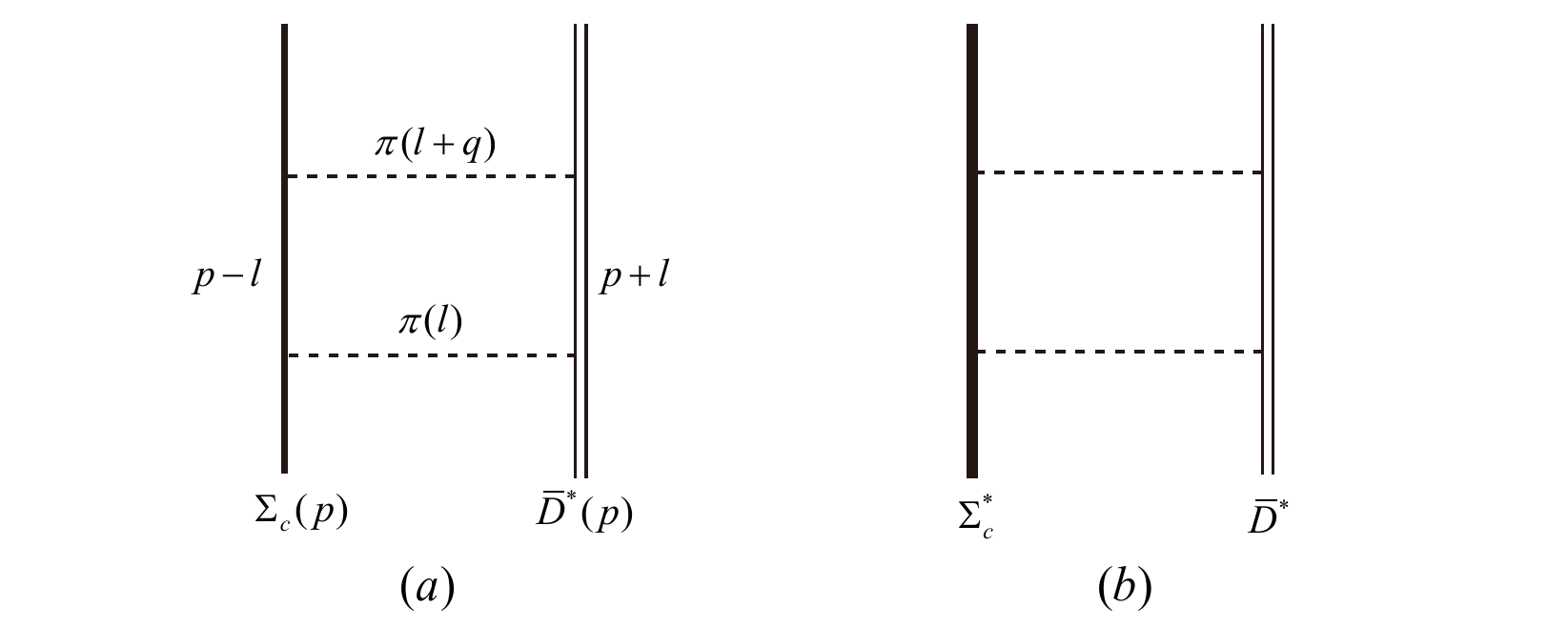}}
    \caption{Two typical Feynman diagrams for the two-pion-exchange process of the $\Sigma_c\bar{D}^\ast$ ($a$) and $\Sigma_c^\ast\bar{D}^\ast$ ($b$) systems. We use the thick line, heavy-thick line, double-thin line and dashed line to denote the $\Sigma_c$, $\Sigma_c^\ast$, $\bar{D}^\ast$ and pion, respectively.\label{Illustration_2PR}}
\end{centering}
\end{figure*}

The interactions between heavy matter fields in the chiral
perturbation theory are clear and straightforward, which generally
include the long-range one-pion-exchange, intermediate-range
two-pion-exchange and short-range contact
interaction~\cite{Machleidt:2011zz,Weinberg:1990rz,Weinberg:1991um}.
The contributions from the heavy degrees of freedom are encoded into
the low energy constants (LECs) of the contact Lagrangians. As we
know, the masses of the heavy matter fields, like
$\Sigma_c^{(\ast)}$ and $\bar{D}^{(\ast)}$, do not vanish in the
chiral limit. The large masses would break the chiral power
counting. Thus, we can adopt the heavy hadron reduction formalism to
integrate out the large mass
scale~\cite{Georgi:1990um,Jenkins:1990jv,Cho:1992cf}. For the loop
diagrams generated by the two-pion-exchange interactions, we will
encounter another trouble, which also destroys the power counting
rule. Considering the one-loop Feynman diagrams illustrated in
figure~\ref{Illustration_2PR}, the scattering amplitude at the
leading order of the nonrelativistic expansion is badly divergent
because of the pinch
singularity~\cite{Machleidt:2011zz,Weinberg:1991um}. Although the
problem of divergence can be solved by including the kinetic
energies of $\Sigma_c^{(*)}$ and $\bar{D}^\ast$ at the leading order
(see some more detailed discussions in
refs.~\cite{Machleidt:2011zz,Weinberg:1991um,Wang:2018atz}), the
amplitude would be finally enhanced by a large factor
$M/|\boldsymbol{p}|$ ($M$ could be the mass of $\Sigma_c^{(*)}$ or
$\bar{D}^\ast$), which will destroy the power counting as well. This
strong enhancement is the manifestation of the nonperturbative
nature of the nuclear force, which is responsible for the existence
of the bound pentaquark states. In other words, a nonperturbative
treatment is required.

In the two seminal works~\cite{Weinberg:1990rz,Weinberg:1991um},
Weinberg pointed out that we shall focus on the {\it effective
potential}, i.e, the contributions from two-particle-irreducible
(2PI) graphs. The two-particle-reducible (2PR) part, that originates
from the on-shell intermediate $\Sigma_c^{(\ast)}$ and
$\bar{D}^\ast$, should be subtracted. On the other hand, the 2PR
part can be automatically recovered when the one-pion-exchange
potential is inserted into the nonperturbative iterative equation,
such as the Schr\"odinger equation or Lippmann-Schwinger equation.
Therefore, the 2PI parts in the diagrams of
figure~\ref{Illustration_2PR} that contribute to the effective
potentials can still be calculated perturbatively. We just need to
solve a nonperturbative iterative equation with the obtained
effective potential eventually.

For the $\Sigma_c^{(*)}\bar{D}^{(*)}$ systems, the mass splittings
in the spin doublets $(\Sigma_c,\Sigma_c^\ast)$ and
$(\bar{D},\bar{D}^\ast)$ do not vanish in the chiral limit, which
only vanish in the strict heavy quark limit. Therefore, except for
the two particular diagrams in figure~\ref{Illustration_2PR}, the
intermediate states in the loops can also be their spin partners. In
this case, the loop integral is well defined, and we do not need to
make the 2PR subtraction, unless the inelastic one-pion-exchange
couple channel is included.

In this work, we try to reproduce the newly observed $P_c(4312)$,
$P_c(4440)$ and $P_c(4457)$ after simultaneously considering the
leading order contact interaction and one-pion-exchange
contribution, as well as the next-to-leading order two-pion-exchange
diagrams. The mass splittings are kept in the loop diagrams. If
these $P_c$ states are shallow bound hadronic molecules, they would
be very sensitive to the subtle changes of the effective potentials.
Furthermore, the nonanalytic structures, such as the terms with the
logarithmic and square root functions, would emerge from the loop
diagrams, which may enhance the two-pion-exchange potential to some
extent. In particular, the mass difference $\delta$ between
$\bar{D}^\ast$ and $\bar{D}$ is larger than the pion mass $m_\pi$.
The heavy quark spin symmetry breaking effect has been noticed for
the charmed sectors in some works~\cite{Meng:2019ilv,Wang:2019mhm}.
Besides, one shall not neglect the role of $\Lambda_c$, since the
$\Lambda_c\pi$ couples strongly with the $\Sigma_c^{(\ast)}$.
Therefore, we also include the contribution of $\Lambda_c$ in the
loop diagrams. We will see the dramatic influences of $\Lambda_c$ on
the $\Sigma_c^{(\ast)}\bar{D}^{(\ast)}$ intermediate-range
potentials.

We use the $P_c(4312)$, $P_c(4440)$ and $P_c(4457)$ as inputs to fix
the unknown LECs. We notice that the three $P_c$ states can be
synchronously reproduced when the $\Lambda_c$ is considered. We then
use the fixed LECs to study the previously reported $P_c(4380)$ and
predict the possible $\Sigma_c^{\ast}\bar{D}^{\ast}$ molecules. We
also investigate the $\Sigma_b^{(\ast)}B^{(\ast)}$ systems, and
predict the possible $P_b$ states.

This paper is organized as follows. In section~\ref{lagrangians}, we
give the effective chiral Lagrangians. In
section~\ref{AnalyticalExpressions}, we present the analytical
expressions for the effective potentials of the
$\Sigma_c^{(\ast)}\bar{D}^{(\ast)}$ systems. In
section~\ref{NumericalResults}, we illustrate the numerical results
and discussions, which contain the results without and with the
$\Lambda_c$, and an investigation on interchanging the spins of
$P_c(4440)$ and $P_c(4457)$. In section~\ref{HiddenBottomSystems},
we study the hidden-bottom systems and predict their mass spectra.
In section~\ref{HQSBreakingEffect}, we give a detailed examination
of the heavy quark symmetry breaking effect in the hidden-charm and
-bottom systems. In section~\ref{SummaryandConclusion}, we conclude
this work with a short summary. In the
appendices~\ref{LoopIntegrals},~\ref{Remove_2PR},~\ref{SpinTransfOperators}
and~\ref{PotentialModel}, we display the definitions and expressions
of the loop integrals, the detailed elucidation on how to remove the
2PR contributions with the mass splittings being kept, the
derivation of the spin-spin terms in the potentials, and a tentative
parameterization of the effective potentials from the quark model,
respectively.

\section{Effective chiral Lagrangians}\label{lagrangians}

In the framework of heavy hadron chiral perturbation theory, the
scattering amplitudes of the $\Sigma_c^{(\ast)}\bar{D}^{(\ast)}$
systems can be expanded order by order in powers of a small
parameter $\varepsilon=q/\Lambda_\chi$, where $q$ is either the
momentum of Goldstone bosons or the residual momentum of heavy
hadrons, and $\Lambda_\chi$ represents either the chiral breaking
scale or the mass of a heavy hadron. The expansion is organized by
the power counting rule~\cite{Weinberg:1990rz,Weinberg:1991um}. One
can get the order $\nu$ of a diagram with
\begin{eqnarray}
\nu=2L-\frac{E_n}{2}+2+\sum_i
V_i\Delta_i,\quad\quad\Delta_i=d_i+\frac{n_i}{2}-2,
\end{eqnarray}
where $L$ and $E_n$ represent the number of loops and external lines
of the matter field. $V_i$ denote the number of the type-$i$ vertex
with the order $\Delta_i$. $d_i$ and $n_i$ stand for the number of
derivatives (or $m_\pi$ factors) and external lines of the matter
field in a type-$i$ vertex.
\subsection{Pion interactions}
In the SU(2) flavor space, the two light quarks in the charmed
baryons can form the antisymmetric isosinglet and symmetric
isotriplet. The corresponding total spins of the light quarks are
$S_l=0$ and $S_l=1$, respectively. We use the notations $\psi_1$,
$\psi_3$ and $\psi_{3^\ast}^{\mu}$ to denote the spin-$\frac{1}{2}$
isosinglet, spin-$\frac{1}{2}$ and spin-$\frac{3}{2}$ isotriplet,
respectively.
\begin{eqnarray}
\psi_1=\left( \begin{array}{cc}
0&\Lambda_c^+\\
-\Lambda_c^+&0
\end{array} \right),\quad\quad
\psi_3=\left( \begin{array}{cc}
\Sigma_c^{++}&\frac{\Sigma_c^+}{\sqrt{2}}\\
\frac{\Sigma_c^+}{\sqrt{2}}&\Sigma_c^0
\end{array} \right),\quad\quad
\psi_{3^\ast}^\mu=\left( \begin{array}{cc}
\Sigma_c^{\ast++}&\frac{\Sigma_c^{\ast+}}{\sqrt{2}}\\
\frac{\Sigma_c^{\ast+}}{\sqrt{2}}&\Sigma_c^{\ast0}
\end{array} \right)^\mu.
\end{eqnarray}

The leading order relativistic chiral Lagrangians for the charmed
baryons have been constructed in
ref.~\cite{Cheng:1992xi,Liu:2012uw}, which are given as
\begin{eqnarray}\label{Baryon_Lag_Rela}
\mathcal{L}_{B\phi}&=&\mathrm{Tr}\left\{\bar{\psi}_{3^\ast}^\mu\left[-g_{\mu\nu}(i\slashed{D}-M_{3^\ast})+i(\gamma_\mu D_\nu+\gamma_\nu D_\mu)-\gamma_\mu(i\slashed{D}+M_{3^\ast})\gamma_\nu\right]\psi_{3^\ast}^\nu\right\}\nonumber\\
&&\mathrm{Tr}\left[\bar{\psi}_3(i\slashed{D}-M_3)\psi_3\right]+g_1\mathrm{Tr}\left(\bar{\psi}_3\slashed{u}\gamma_5\psi_3\right)
+g_3\mathrm{Tr}\left(\bar{\psi}_{3^\ast}^\mu u_\mu\psi_3+\mathrm{H.c.}\right)\nonumber\\
&&+g_5\mathrm{Tr}\left(\bar{\psi}_{3^\ast}^\mu\slashed{u}\gamma_5\psi_{3^\ast\mu}\right)
+\frac{1}{2}\mathrm{Tr}\left[\bar{\psi}_1(i\slashed{D}-M_1)\psi_1\right]+g_2\mathrm{Tr}\left(\bar{\psi}_{3}\slashed{u}\gamma_5\psi_1+\mathrm{H.c.}\right)\nonumber\\
&&+g_4\mathrm{Tr}\left(\bar{\psi}_{3^\ast}^\mu
u_\mu\psi_1+\mathrm{H.c.}\right),
\end{eqnarray}
where $\mathrm{Tr}(X)$ denotes the trace of $X$ in flavor space. The
covariant derivative $D_\mu$ is defined as
$D_\mu\psi=\partial_\mu\psi+\Gamma_\mu\psi+\psi\Gamma_\mu^T$
($\Gamma_\mu^T$ means the transposition of $\Gamma_\mu$). Meanwhile,
the chiral connection $\Gamma_\mu$ and axial current $u_\mu$ are
\begin{eqnarray}
\Gamma_\mu&\equiv&\frac{1}{2}\left[\xi^\dag,\partial_\mu
\xi\right],\quad\quad\quad
u_\mu\equiv\frac{i}{2}\left\{\xi^\dag,\partial_\mu \xi\right\},
\end{eqnarray}
where
\begin{eqnarray}
\xi^2=U=\exp\left(\frac{i\phi}{f_\pi}\right),\quad\quad\quad
\phi=\left( \begin{array}{cc}
\pi^0&\sqrt{2}\pi^+\\
\sqrt{2}\pi^-&-\pi^0
\end{array} \right),
\end{eqnarray}
and $f_\pi=92.4$ MeV is the pion decay constant.

We then adopt the heavy baryon reduction
formalism~\cite{Scherer:2002tk} to get rid of the large baryon
masses in eq.~\eqref{Baryon_Lag_Rela}, where the heavy baryon field
is decomposed into the light and heavy components by the projection
operators $(1\pm\slashed{v})/2$,
\begin{eqnarray}
\mathcal{B}_i=e^{iM_iv\cdot
x}\frac{1+\slashed{v}}{2}\psi_i,\quad\quad\quad
\mathcal{H}_i=e^{iM_iv\cdot x}\frac{1-\slashed{v}}{2}\psi_i,
\end{eqnarray}
where $\psi_i$ denotes the relativistic heavy baryon field $\psi_1$,
$\psi_3$ and $\psi_{3^\ast}$, $M_i$ is their masses, and
$v_\mu=(1,\mathbf{0})$ represents the four-velocity of a slowly
moving heavy baryon. $\mathcal{B}_i$ and $\mathcal{H}_i$ are the
corresponding light and heavy components, respectively.
$\mathcal{H}_i$ disappears at the leading order expansion.

Consequently, the eq.~\eqref{Baryon_Lag_Rela} can then be
reexpressed with the nonrelativistic form as
\begin{eqnarray}\label{Baryon_Lag_nonRela}
\mathcal{L}_{B\phi}&=&\mathrm{Tr}\left[\bar{\mathcal{B}}_3(iv\cdot
D-\delta_c)\mathcal{B}_3\right]-\mathrm{Tr}\left[\bar{\mathcal{B}}_{3^\ast}^\mu\left(iv\cdot
D-\delta_d\right)\mathcal{B}_{3^{\ast}\mu}\right]
+2g_1\mathrm{Tr}\left(\bar{\mathcal{B}}_3\mathcal{S}\cdot u\mathcal{B}_3\right)\nonumber\\
&&+g_3\mathrm{Tr}\left(\bar{\mathcal{B}}_{3^\ast}^\mu
u_\mu\mathcal{B}_3+\mathrm{H.c.}\right)+2g_5\mathrm{Tr}\left(\bar{\mathcal{B}}_{3^\ast}^\mu\mathcal{S}\cdot
u\mathcal{B}_{3^\ast\mu}\right)
+\frac{1}{2}\mathrm{Tr}\left[\bar{\mathcal{B}}_1(iv\cdot D)\mathcal{B}_1\right]\nonumber\\
&&+2g_2\mathrm{Tr}\left(\bar{\mathcal{B}}_{3}\mathcal{S}\cdot
u\mathcal{B}_1+\mathrm{H.c.}\right)+g_4\mathrm{Tr}\left(\bar{\mathcal{B}}_{3^\ast}^\mu
u_\mu\mathcal{B}_1+\mathrm{H.c.}\right),
\end{eqnarray}
where $\mathcal{S}^\mu=\frac{i}{2}\gamma_5\sigma^{\mu\nu}v_\nu$
denotes the spin operator for the spin-$\frac{1}{2}$ particle. We
adopt the mass splittings $\delta_a=M_{3^\ast}-M_3=65$ MeV,
$\delta_c=M_{3}-M_1=168.5$ MeV, and $\delta_d=M_{3^\ast}-M_1=233.5$
MeV~\cite{Tanabashi:2018oca}.

Recall that the $(\psi_3,\psi_{3^\ast})$ form the spin doublet in
the heavy quark limit. Thus eq.~\eqref{Baryon_Lag_nonRela} can be
rewritten as a compact form by introducing the
super-field~\cite{Cheng:1993kp,Cho:1992nt},
\begin{eqnarray}\label{Baryon_Lag_SF}
\mathcal{L}_{B\phi}&=&-\mathrm{Tr}\left(\bar{\psi}^\mu iv\cdot D\psi_\mu\right)+ig_a\epsilon_{\mu\nu\rho\sigma}\mathrm{Tr}\left(\bar{\psi}^\mu u^\rho v^\sigma\psi^\nu\right)+i\frac{\delta_a}{2}\mathrm{Tr}\left(\bar{\psi}^\mu\sigma_{\mu\nu}\psi^\nu\right)\nonumber\\
&&+\frac{1}{2}\mathrm{Tr}\left[\bar{\mathcal{B}}_1(iv\cdot
D)\mathcal{B}_1\right]+g_b\mathrm{Tr}\left(\bar{\psi}^\mu
u_\mu\mathcal{B}_1+\mathrm{H.c.}\right),
\end{eqnarray}
where the super-fields $\psi^\mu$ and $\bar{\psi}^\mu$ are defined
as~\cite{Cho:1992cf,Cho:1992gg}
\begin{eqnarray}
\psi^\mu=\mathcal{B}_{3^\ast}^\mu-\frac{1}{\sqrt{3}}(\gamma^\mu+v^\mu)\gamma^5\mathcal{B}_3,\quad\quad\quad\bar{\psi}^\mu=\bar{\mathcal{B}}_{3^\ast}^\mu+\frac{1}{\sqrt{3}}\bar{\mathcal{B}}_3\gamma^5(\gamma^\mu+v^\mu).
\end{eqnarray}
Expanding eq.~\eqref{Baryon_Lag_SF} and comparing them with the
terms in eq.~\eqref{Baryon_Lag_nonRela}, one can get the relations
among the different coupling constants,
\begin{eqnarray}
g_1=-\frac{2}{3}g_a,\quad g_3=-\frac{1}{\sqrt{3}}g_a,\quad
g_5=g_a;\quad g_2=-\frac{1}{\sqrt{3}}g_b,\quad g_4=g_b.
\end{eqnarray}
The values of $g_2$ and $g_4$ can be calculated with the partial
decay widths of $\Sigma_c\to\Lambda_c\pi$ and
$\Sigma_c^{\ast}\to\Lambda_c\pi$~\cite{Tanabashi:2018oca},
respectively. The other axial couplings $g_1$, $g_3$ and $g_5$ can
be obtained by their relations with $g_{2}$ in the framework of the
quark model~\cite{Meguro:2011nr,Liu:2011xc,Meng:2018gan}, which
yields
\begin{align}\label{BaryonCouplings}
g_2&=-0.60,&\quad\quad\quad g_4&=-\sqrt{3}g_2=1.04; \nonumber\\
g_1&=-\sqrt{\frac{8}{3}}g_2=0.98,&\quad\quad\quad
g_3&=\frac{\sqrt{3}}{2}g_1=0.85,&\quad\quad\quad
g_5=-\frac{3}{2}g_1=-1.47.
\end{align}

The leading order chiral Lagrangians for the interactions between
the anticharmed mesons and light pseudoscalars
read~\cite{Wise:1992hn,Manohar:2000dt}
\begin{eqnarray}\label{Meson_Lag_SF}
\mathcal{L}_{H \phi}=-i\langle\bar{\tilde{\mathcal{H}}} v \cdot
\mathcal{D}
\tilde{\mathcal{H}}\rangle-\frac{1}{8}\delta_b\langle\bar{\tilde{\mathcal{H}}}
\sigma^{\mu \nu} \tilde{\mathcal{H}}\sigma_{\mu \nu}\rangle+
g\langle\bar{\tilde{\mathcal{H}}} \slashed{u}\gamma_{5}
\tilde{\mathcal{H}}\rangle,
\end{eqnarray}
where $\langle X\rangle$ stands for the trace of $X$ in spinor
space. The covariant derivative
$\mathcal{D}_\mu=\partial_\mu+\Gamma_\mu$,
$\delta_b=m_{\bar{D}^\ast}-m_{\bar{D}}$ is the mass splitting
between $\bar{D}^\ast$ and $\bar{D}$. $g=-0.59$ represents the axial
coupling constant, and its value is extracted from the partial decay
width of $D^{\ast+}\to D^0\pi^+$~\cite{Tanabashi:2018oca}, while the
sign is determined by the quark model. We use the
$\tilde{\mathcal{H}}$ to denote the super-field for the anticharmed
mesons, which reads
\begin{eqnarray}
\tilde{\mathcal{H}}&=&\left(\tilde{P}_{\mu}^{*} \gamma^{\mu}+i
\tilde{P} \gamma_{5}\right)
\frac{1-\slashed{v}}{2},\quad\quad\quad\bar{\tilde{\mathcal{H}}}=\gamma^{0}
\tilde{\mathcal{H}}^{\dagger}\gamma^{0}=\frac{1-\slashed{v}}{2}\left(\tilde{P}_{\mu}^{*
\dagger}\gamma^{\mu}+i \tilde{P}^{\dagger} \gamma_{5}\right),
\end{eqnarray}
where $\tilde{P}=(\bar{D}^0,D^-)^T$ and
$\tilde{P}^\ast=(\bar{D}^{\ast0},D^{\ast-})^T$, respectively.

\subsection{Contact interactions}
We then construct the leading order Lagrangians that account for the
interactions between $\Sigma_c^{(\ast)}$ and $\bar{D}^{(\ast)}$ at
the short range. We also use the super-field representations for
$\Sigma_c^{(\ast)}$ and $\bar{D}^{(\ast)}$ to reduce the numbers of
the LECs, which read~\cite{Meng:2019ilv}
\begin{eqnarray}\label{Contact_Lag_BM}
\mathcal{L}_{HB}&=&D_a \langle\bar{\tilde{\mathcal{H}}}\tilde{\mathcal{H}}\rangle\mathrm{Tr}\big(\bar{\psi}^\mu \psi_\mu\big)+iD_b\epsilon_{\sigma\mu\nu\rho}v^\sigma\langle\bar{\tilde{\mathcal{H}}}\gamma^\rho\gamma_5\tilde{\mathcal{H}}\rangle\mathrm{Tr}\left(\bar{\psi}^\mu \psi^\nu\right)\nonumber\\
&&+E_a
\langle\bar{\tilde{\mathcal{H}}}\tau^i\tilde{\mathcal{H}}\rangle\mathrm{Tr}\big(\bar{\psi}^\mu\tau_i
\psi_\mu\big)+iE_b\epsilon_{\sigma\mu\nu\rho}v^\sigma\langle\bar{\tilde{\mathcal{H}}}\gamma^\rho\gamma_5\tau^i\tilde{\mathcal{H}}\rangle\mathrm{Tr}\big(\bar{\psi}^\mu\tau_i
\psi^\nu\big),
\end{eqnarray}
where the $D_a$, $D_b$, $E_a$ and $E_b$ are four independent LECs.
The contact terms contain the residual contributions from the heavy
degrees of freedom, which are integrated out and invisible at the
low energy scale. Their values can be delicately determined from the
experimental data~\cite{Meng:2019ilv} or roughly estimated with the
theoretical models~\cite{Xu:2017tsr,Wang:2018atz}. $D_a$ and $D_b$
contribute to the central potential and spin-spin interaction,
respectively. $E_a$ and $E_b$ are related with the isospin-isospin
interaction and contribute to the central and spin-spin interaction
in spin space, respectively .

At the next-to-leading order, we need the
$\mathcal{O}(\varepsilon^2)$ LECs to absorb the divergences of the
loop diagrams. These $\mathcal{O}(\varepsilon^2)$ contact
Lagrangians shall be proportional to the $m_\pi^2$,
$\boldsymbol{q}^2$, $\delta_a^2$ and $\delta_b^2$. As demonstrated
in ref.~\cite{Liu:2012vd}, there exist a large number of contact
terms at $\mathcal{O}(\varepsilon^2)$. It is very difficult to fix
all these LECs at present. Therefore, in our work, we try to combine
some contributions of the $\mathcal{O}(\varepsilon^2)$ LECs with the
leading ones by fitting the experimental data (At least the ones
that proportional to $m_\pi^2$, $\delta_a^2$ and $\delta_b^2$ can be
absorbed by renormalizing the $\mathcal{O}(\varepsilon^0)$ LECs. The
ones correlated with $\boldsymbol{q}^2$ can be largely compensated
by the cutoff).

\section{Analytical expressions for the effective potentials of the $\Sigma_c^{(\ast)}\bar{D}^{(\ast)}$ systems}\label{AnalyticalExpressions}
The effective potential in momentum space can be obtained from the
scattering amplitude in the following way~\cite{Yang:2011wz},
\begin{eqnarray}
\mathcal{V}(\bm{q})=-\frac{\mathcal{M}(\bm{q})}{\sqrt{2M_1 2M_2 2M_3
2M_4}},
\end{eqnarray}
where the $M_{1,2}$ and $M_{3,4}$ are the masses of the initial and
final particles. The scattering amplitude $\mathcal{M}(\bm{q})$ is
calculated by expanding the Lagrangians in
eqs.~\eqref{Baryon_Lag_nonRela}, \eqref{Meson_Lag_SF} and
\eqref{Contact_Lag_BM}. Recall that at the leading order of the
nonrelativistic expansions, there are the
relations~\cite{Manohar:2000dt}
\begin{eqnarray}
\psi(p)=\sqrt{2m_\psi}\left[\chi(v)+\mathcal{O}(1/m_\psi)\right],\quad\quad\quad\tilde{\mathcal{H}}(p)=\sqrt{m_H}\left[\tilde{\mathcal{H}}(v)+\mathcal{O}(1/m_H)\right],
\end{eqnarray}
where $\psi(p)$ and $\tilde{\mathcal{H}}(p)$ are the relativistic
fields. $\chi(v)$ is the two-component spinor. The
$\tilde{\mathcal{H}}(v)$ is the field in eqs.~\eqref{Meson_Lag_SF} and~\eqref{Contact_Lag_BM}.
We then make the Fourier transformation on $\mathcal{V}(\bm{q})$ to
get the potential $\mathcal{V}(r)$ in the coordinate space,
\begin{eqnarray}\label{Four_Tansf}
\mathcal{V}(r)=\int\frac{d^3\bm{q}}{(2\pi)^3}e^{-i\bm{q\cdot
r}}\mathcal{V}(\bm q)\mathcal{F}(\bm q),
\end{eqnarray}
where the Gauss regulator $\mathcal{F}(\bm
q)=\exp\left(-\bm{q}^{2n}/\Lambda^{2n}\right)$ is introduced to
regularize $\mathcal{V}(\bm q)$
nonperturbatively~\cite{Ordonez:1995rz,Epelbaum:1999dj}. As in
refs.~\cite{Xu:2017tsr,Liu:2012vd,Wang:2018atz,Epelbaum:2014efa}, we
set $n=2$. The cutoff value $\Lambda$ is commonly chosen to be
smaller than $\rho$ meson mass in the $N$-$N$
system~\cite{Epelbaum:2014efa}. We adopt a moderate value
$\Lambda=0.5$ GeV as in ref.~\cite{Meng:2019ilv}.

\subsection{$\Sigma_c\bar{D}$ system}
\begin{figure*}[hptb]
\begin{centering}
    \setlength{\abovecaptionskip}{0.01cm}
    \scalebox{1.0}{\includegraphics[width=\columnwidth]{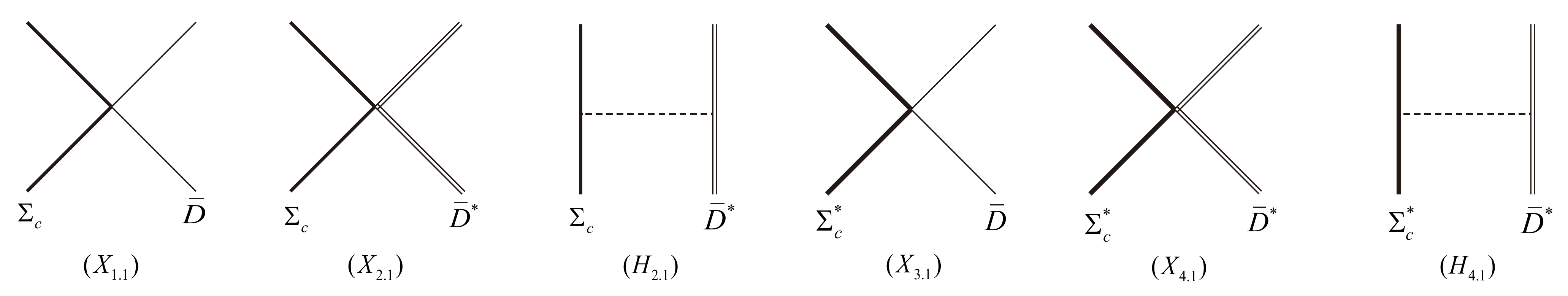}}
    \caption{The leading order Feynman diagrams that account for the $\mathcal{O}(\varepsilon^0)$ effective potentials of the $\Sigma_c\bar{D}$ ($X_{1.1}$), $\Sigma_c\bar{D}^\ast$ ($X_{2.1},H_{2.1}$), $\Sigma_c^\ast\bar{D}$ ($X_{3.1}$) and $\Sigma_c^\ast\bar{D}^\ast$ ($X_{4.1},H_{4,1}$) systems. We use the thin line to denote the $\bar{D}$ meson, and other notations are the same as those in figure~\ref{Illustration_2PR}.\label{Tree_Level_Diagrams}}
\end{centering}
\end{figure*}

Since the $\bar{D}\bar{D}\pi$ vertex is forbidden by the parity
conservation law, the leading order effective potential for the
$\Sigma_c\bar{D}$ system only arises from the contact terms [diagram
$(X_{1.1})$ in the figure~\ref{Tree_Level_Diagrams}]. One can
readily get
\begin{eqnarray}\label{LO_SigmacDb_Con}
\mathcal{V}_{\Sigma_c\bar{D}}^{X_{1.1}}=-D_a-2E_a(\mathbf{I}_1\cdot\mathbf{I}_2),
\end{eqnarray}
where $\mathbf{I}_1$ and $\mathbf{I}_2$ represent the isospin
operators of the $\Sigma_c^{(\ast)}$ and $\bar{D}^{(\ast)}$,
respectively. The matrix element of $\mathbf{I}_1\cdot\mathbf{I}_2$
is
\begin{displaymath}
\langle\mathbf{I}_1\cdot\mathbf{I}_2\rangle = \left\{
\begin{array}{ll}
-1 & \textrm{for $I=\frac{1}{2}$}\\
\frac{1}{2} & \textrm{for $I=\frac{3}{2}$}
\end{array} \right.,
\end{displaymath}
where $I$ is the total isospin of the
$\Sigma_c^{(\ast)}\bar{D}^{(\ast)}$ system. The above values can be
easily obtained with the relation
$\langle\mathbf{I}_1\cdot\mathbf{I}_2\rangle=\frac{1}{2}\left[I(I+1)-I_1(I_1+1)-I_2(I_2+1)\right]$.

\begin{figure*}[hptb]
\begin{centering}
    \setlength{\abovecaptionskip}{0.01cm}
    \scalebox{1.0}{\includegraphics[width=\columnwidth]{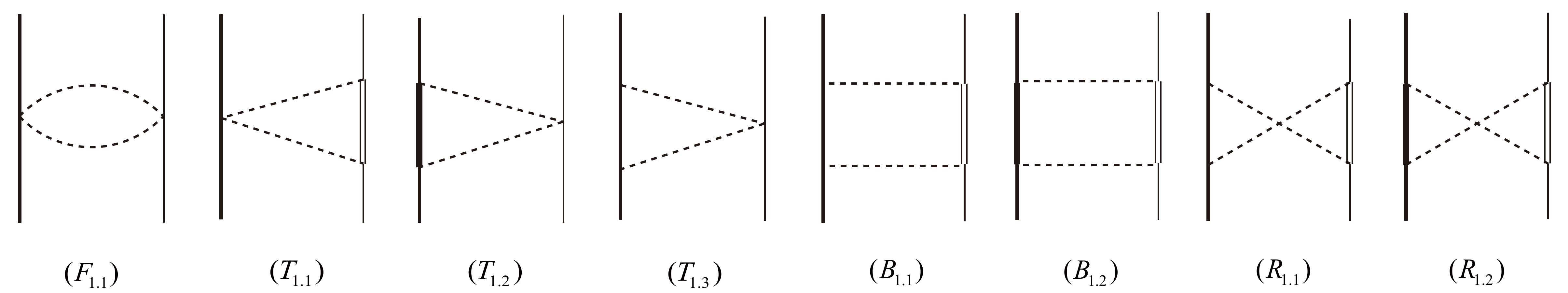}}
    \caption{The two-pion-exchange diagrams of the $\Sigma_c\bar{D}$ system at $\mathcal{O}(\epsilon^2)$. These diagrams are classified as the football diagram ($F_{1.1}$), triangle diagrams ($T_{1.i}$), box diagrams ($B_{1.i}$) and crossed box diagrams ($R_{1.i}$). The internal heavy baryon lines in diagrams $(T_{1.3})$, $(B_{1.1})$ and $(R_{1.1})$ can also be the $\Lambda_c$. The notations are the same as those in figure~\ref{Tree_Level_Diagrams}.\label{TwoPion_Loop1}}
\end{centering}
\end{figure*}

\begin{figure*}[hptb]
\begin{centering}
    \setlength{\abovecaptionskip}{0.01cm}
    \scalebox{1.0}{\includegraphics[width=\columnwidth]{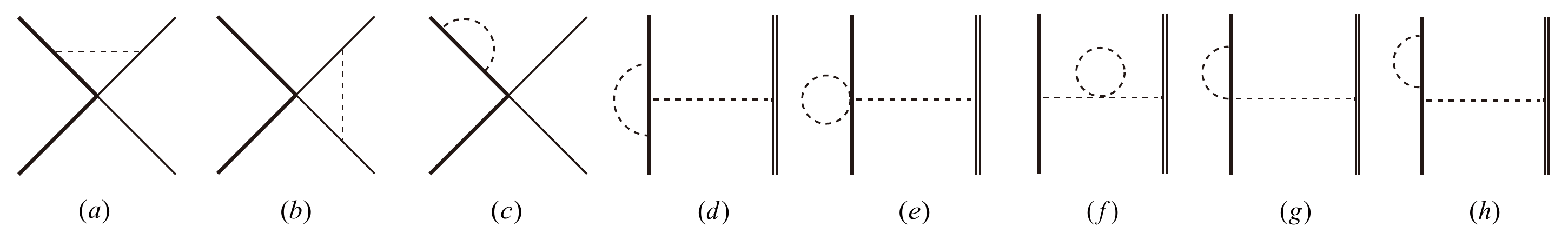}}
    \caption{The next-to-leading order Feynman diagrams that contribute to the vertex corrections and wave function renormalizations. Each graph denotes the one type of diagrams with the same topological structure.\label{Illustration_OneLoop}}
\end{centering}
\end{figure*}

At the next-to-leading order, there are two types of one-loop
diagrams. One is the two-pion-exchange diagrams is
figure~\ref{TwoPion_Loop1}. Another one is the vertex corrections
and wave function renormalizations in
figure~\ref{Illustration_OneLoop}~\cite{Xu:2017tsr,Liu:2012vd,Wang:2018atz}.
The contribution of the diagrams in
figure~\ref{Illustration_OneLoop} could be included by using the
physical values of the parameters in the Lagrangians, such as the
pion mass, decay constant, coupling constants, etc..

The analytical expressions of the two-pion-exchange diagrams in
figure~\ref{TwoPion_Loop1} read
\begin{eqnarray}
\mathcal{V}_{\Sigma_c\bar{D}}^{F_{1.1}}&=&(\mathbf{I}_1\cdot\mathbf{I}_2)\frac{1}{f_\pi^4} J_{22}^F(m_\pi,q),\\[0.3cm]
\mathcal{V}_{\Sigma_c\bar{D}}^{T_{1.1}}&=&(\mathbf{I}_1\cdot\mathbf{I}_2)\frac{g^2}{f_\pi^4}\bigg[(d-1) J_{34}^T-\bm{q}^2\left( J_{24}^T+ J_{33}^T\right)\bigg](m_\pi,\mathcal{E}-\delta_b,q),\\[0.3cm]
\mathcal{V}_{\Sigma_c\bar{D}}^{T_{1.2}}&=&(\mathbf{I}_1\cdot\mathbf{I}_2)\frac{g_3^2}{4f_\pi^4}\bigg[(d-2) J_{34}^T-\bm{q}^2\frac{d-2}{d-1}\left( J_{24}^T+ J_{33}^T\right)\bigg](m_\pi,\mathcal{E}-\delta_a,q),\\[0.3cm]
\mathcal{V}_{\Sigma_c\bar{D}}^{T_{1.3}}&=&(\mathbf{I}_1\cdot\mathbf{I}_2)\frac{g_1^2}{4f_\pi^4}\bigg[(d-1) J_{34}^T-\bm{q}^2\left( J_{24}^T+ J_{33}^T\right)\bigg](m_\pi,\mathcal{E},q),\\[0.3cm]
\mathcal{V}_{\Sigma_c\bar{D}}^{B_{1.1}}&=&(1-\mathbf{I}_1\cdot\mathbf{I}_2)\frac{g^2g_1^2}{8f_\pi^4}\bigg[(d^2-1) J_{41}^B-2\bm{q}^2(d+1)\left( J_{31}^B+ J_{42}^B\right)-\bm{q}^2 J_{21}^B\nonumber\\
&&+\bm{q}^4\left( J_{22}^B+2 J_{32}^B+ J_{43}^B\right)\bigg](m_\pi,\mathcal{E},\mathcal{E}-\delta_b,q),\\[0.3cm]
\mathcal{V}_{\Sigma_c\bar{D}}^{B_{1.2}}&=&(1-\mathbf{I}_1\cdot\mathbf{I}_2)\frac{g^2g_3^2}{8f_\pi^4}\bigg[(d^2-d-2) J_{41}^B-2\bm{q}^2\frac{d^2-d-2}{d-1}\left( J_{31}^B+ J_{42}^B\right)-\bm{q}^2\frac{d-2}{d-1} J_{21}^B\nonumber\\
&&+\bm{q}^4\frac{d-2}{d-1}\left( J_{22}^B+2 J_{32}^B+ J_{43}^B\right)\bigg](m_\pi,\mathcal{E}-\delta_a,\mathcal{E}-\delta_b,q),\\[0.3cm]
\mathcal{V}_{\Sigma_c\bar{D}}^{R_{1.i}}&=&\mathcal{V}_{\Sigma_c\bar{D}}^{B_{1.i}}\Big|_{
J_x^B\to  J_x^R,~\mathbf{I}_1\cdot\mathbf{I}_2\to
-\mathbf{I}_1\cdot\mathbf{I}_2}.
\end{eqnarray}
When the contribution of $\Lambda_c$ is included, it will appear in
the graphs $(T_{1.3})$, $(B_{1.1})$ and $(R_{1.1})$ as the
intermediate state. The expressions read (we use $\bar{T}_{i.j}$,
$\bar{B}_{i.j}$ and $\bar{R}_{i.j}$ to denote the loops with
$\Lambda_c$)
\begin{eqnarray}
\mathcal{V}_{\Sigma_c\bar{D}}^{\bar{T}_{1.3}}&=&(\mathbf{I}_1\cdot\mathbf{I}_2)\frac{g_2^2}{2f_\pi^4}\bigg[(d-1) J_{34}^T-\bm{q}^2\left( J_{24}^T+ J_{33}^T\right)\bigg](m_\pi,\mathcal{E}+\delta_c,q),\\[0.3cm]
\mathcal{V}_{\Sigma_c\bar{D}}^{\bar{B}_{1.1}}&=&(1-2\mathbf{I}_1\cdot\mathbf{I}_2)\frac{g^2g_2^2}{8f_\pi^4}\bigg[(d^2-1) J_{41}^B-2\bm{q}^2(d+1)\left( J_{31}^B+ J_{42}^B\right)-\bm{q}^2 J_{21}^B\nonumber\\
&&+\bm{q}^4\left( J_{22}^B+2 J_{32}^B+ J_{43}^B\right)\bigg](m_\pi,\mathcal{E}+\delta_c,\mathcal{E}-\delta_b,q),\\[0.3cm]
\mathcal{V}_{\Sigma_c\bar{D}}^{\bar{R}_{1.1}}&=&\mathcal{V}_{\Sigma_c\bar{D}}^{\bar{B}_{1.1}}\Big|_{
J_x^B\to  J_x^R,~\mathbf{I}_1\cdot\mathbf{I}_2\to
-\mathbf{I}_1\cdot\mathbf{I}_2}.
\end{eqnarray}
In above equations, the loop functions $ J_x^y$ are defined in
appendix~\ref{LoopIntegrals}. $d$ is the dimension where the loop
integral is performed and approaches four at last. $\mathcal{E}$
represents the residual energies of the $\Sigma_c^{(\ast)}$ and
$\bar{D}^{(\ast)}$, which is defined as
$\mathcal{E}=E_i-M_i~(i=\Sigma_c^{(\ast)},\bar{D}^{(\ast)})$.
$\mathcal{E}$ is set to zero in our calculations.
\subsection{$\Sigma_c\bar{D}^\ast$ system}
The leading order potential for the $\Sigma_c\bar{D}^\ast$ system
stems from the contact interaction and one-pion-exchange diagrams
[graphs $(X_{2.1})$ and $(H_{2.1})$ in
figure~\ref{Tree_Level_Diagrams}], which reads
\begin{eqnarray}\renewcommand{\arraystretch}{2.0}
\mathcal{V}_{\Sigma_c\bar{D}^\ast}^{X_{2.1}}&=&-D_a-2E_a(\mathbf{I}_1\cdot\mathbf{I}_2)+\frac{2}{3}\Big[-D_b-2E_b(\mathbf{I}_1\cdot\mathbf{I}_2)\Big]\boldsymbol{\sigma}\cdot\bm{T},\label{X21_Potential}\\[0.3cm]
\mathcal{V}_{\Sigma_c\bar{D}^\ast}^{H_{2.1}}&=&-(\mathbf{I}_1\cdot\mathbf{I}_2)\frac{gg_1}{2f_\pi^2}\frac{(\bm{q}\cdot\boldsymbol{\sigma})(\bm{q}\cdot\bm{T})}{\bm{q}^2+m_\pi^2},\label{H21_Potential}
\end{eqnarray}
where $\boldsymbol{\sigma}$ is the Pauli matrix. The spin operator
$\bm{S}_1$ of $\Sigma_c$ satisfies
$\bm{S}_1=\frac{1}{2}\boldsymbol{\sigma}$. The operator
$\bm{T}=i\boldsymbol{\varepsilon}^\ast\times\boldsymbol{\varepsilon}$
($\boldsymbol{\varepsilon}$ and $\boldsymbol{\varepsilon}^\ast$ are
the space components of polarization vectors of the initial and
final $\bar{D}^\ast$ meson) is correlated with the spin operator
$\bm{S}_2$ of the $\bar{D}^\ast$ meson by the relation
$\bm{S}_2=-\bm{T}$. Thus the $\boldsymbol{\sigma}\cdot\bm{T}$ term
represents the spin-spin interaction (see
appendix~\ref{SpinTransfOperators}). Since only the $S$-wave
interaction is considered, one can use the following replacement
rules in the potentials,
\begin{eqnarray}
\boldsymbol{\varepsilon}^\ast\cdot\boldsymbol{\varepsilon}\longmapsto
1,\quad\quad\quad q_i
q_j\longmapsto\frac{1}{d-1}\bm{q}^2\delta_{ij}.
\end{eqnarray}
\begin{figure*}[hptb]
\begin{centering}
    \setlength{\abovecaptionskip}{0.01cm}
    \scalebox{1.0}{\includegraphics[width=\columnwidth]{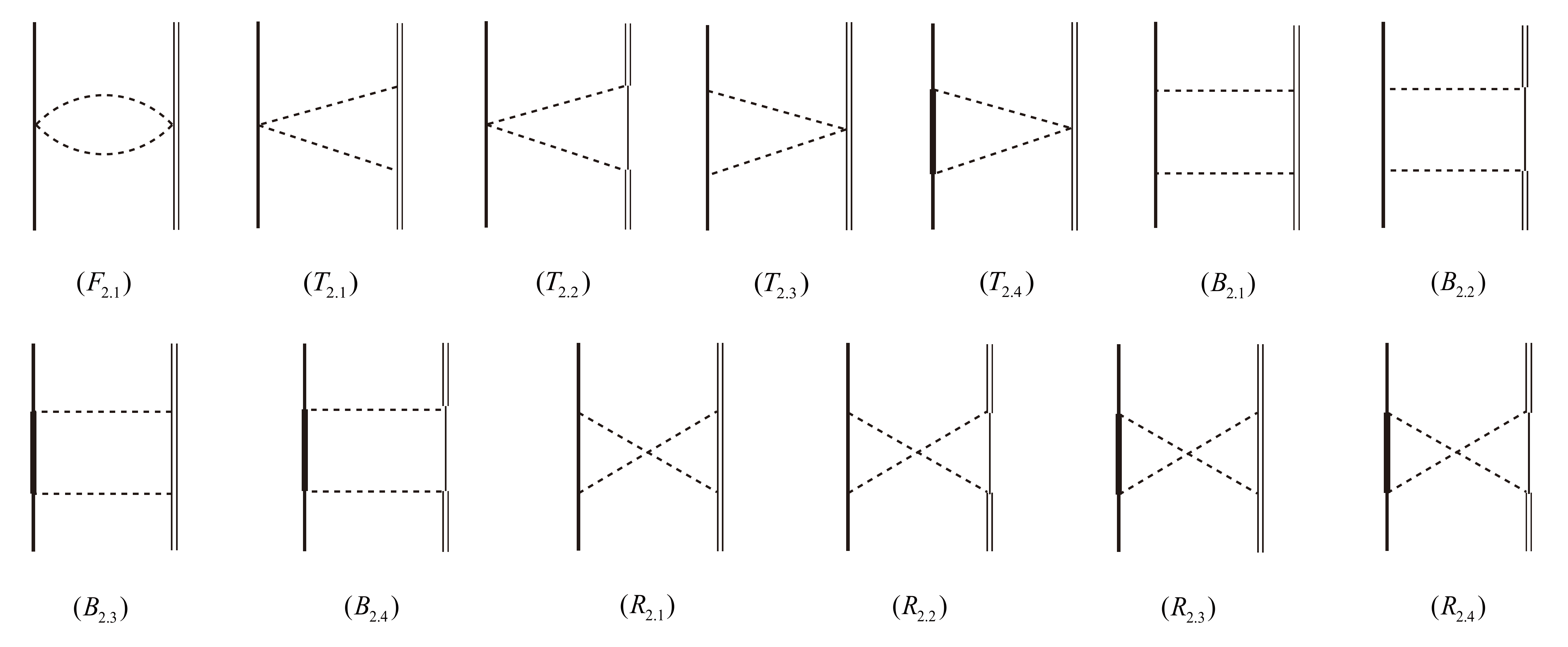}}
    \caption{The two-pion-exchange diagrams of the $\Sigma_c\bar{D}^\ast$ system at $\mathcal{O}(\epsilon^2)$. The internal heavy baryon lines in diagrams $(T_{2.3})$, $(B_{2.1})$, $(B_{2.2})$, $(R_{2.1})$ and $(R_{2.2})$ can also be the $\Lambda_c$. The notations are the same as those in figure~\ref{Tree_Level_Diagrams}.\label{TwoPion_Loop2}}
\end{centering}
\end{figure*}

The two-pion-exchange diagrams for the $\Sigma_c\bar{D}^\ast$ system
are shown in figure~\ref{TwoPion_Loop2}. The potentials from these
graphs read
\begin{eqnarray}
\mathcal{V}_{\Sigma_c\bar{D}^\ast}^{F_{2.1}}&=&(\mathbf{I}_1\cdot\mathbf{I}_2)\frac{1}{f_\pi^4} J_{22}^F(m_\pi,q),\\[0.3cm]
\mathcal{V}_{\Sigma_c\bar{D}^\ast}^{T_{2.1}}&=&(\mathbf{I}_1\cdot\mathbf{I}_2)\frac{g^2}{f_\pi^4}\left[2 J_{34}^T-\bm{q}^2\frac{d-2}{d-1}\left( J_{24}^T+ J_{33}^T\right)\right](m_\pi,\mathcal{E},q),\\[0.3cm]
\mathcal{V}_{\Sigma_c\bar{D}^\ast}^{T_{2.2}}&=&(\mathbf{I}_1\cdot\mathbf{I}_2)\frac{g^2}{f_\pi^4}\left[ J_{34}^T-\frac{\bm{q}^2}{d-1}\left( J_{24}^T+ J_{33}^T\right)\right](m_\pi,\mathcal{E}+\delta_b,q),\\[0.3cm]
\mathcal{V}_{\Sigma_c\bar{D}^\ast}^{T_{2.3}}&=&(\mathbf{I}_1\cdot\mathbf{I}_2)\frac{g_1^2}{4f_\pi^4}\bigg[(d-1) J_{34}^T-\bm{q}^2\left( J_{24}^T+ J_{33}^T\right)\bigg](m_\pi,\mathcal{E},q),\\[0.3cm]
\mathcal{V}_{\Sigma_c\bar{D}^\ast}^{T_{2.4}}&=&(\mathbf{I}_1\cdot\mathbf{I}_2)\frac{g_3^2}{4f_\pi^4}\bigg[(d-2) J_{34}^T-\bm{q}^2\frac{d-2}{d-1}\left( J_{24}^T+ J_{33}^T\right)\bigg](m_\pi,\mathcal{E}-\delta_a,q),\\[0.3cm]
\mathcal{V}_{\Sigma_c\bar{D}^\ast}^{B_{2.1}}&=&(1-\mathbf{I}_1\cdot\mathbf{I}_2)\frac{g^2g_1^2}{8f_\pi^4}\bigg[\frac{4d^2-10d+6}{d-1} J_{41}^B-\bm{q}^2\frac{d^2+3d-8}{d-1}\left( J_{31}^B+ J_{42}^B\right)\nonumber\\
&&-\bm{q}^2\frac{d-2+\boldsymbol{\sigma}\cdot\bm{T}}{d-1} J_{21}^B
+\bm{q}^4\frac{d-2}{d-1}\left( J_{22}^B+2 J_{32}^B+ J_{43}^B\right)\bigg](m_\pi,\mathcal{E},\mathcal{E},q),\\[0.3cm]
\mathcal{V}_{\Sigma_c\bar{D}^\ast}^{B_{2.2}}&=&(1-\mathbf{I}_1\cdot\mathbf{I}_2)\frac{g^2g_1^2}{8f_\pi^4}\bigg[-2\bm{q}^2\frac{d+1}{d-1}\left(
J_{31}^B+ J_{42}^B\right)
-\bm{q}^2\frac{1}{d-1}(1+\boldsymbol{\sigma}\cdot\bm{T}) J_{21}^B\nonumber\\
&&+(d+1) J_{41}^B+\bm{q}^4\frac{1}{d-1}\left( J_{22}^B+2 J_{32}^B+ J_{43}^B\right)\bigg](m_\pi,\mathcal{E},\mathcal{E}+\delta_b,q),\\[0.3cm]
\mathcal{V}_{\Sigma_c\bar{D}^\ast}^{B_{2.3}}&=&(1-\mathbf{I}_1\cdot\mathbf{I}_2)\frac{g^2g_3^2}{8f_\pi^4}\bigg[-\bm{q}^2\frac{(d-2)^2-\boldsymbol{\sigma}\cdot\bm{T}}{(d-1)^2} J_{21}^B-\bm{q}^2\frac{(d-2)(d^2+3d-8)}{(d-1)^2}\left( J_{31}^B+ J_{42}^B\right)\nonumber\\
&&+\frac{2(d^2-2d+2)}{d-1} J_{41}^B+\bm{q}^4\frac{(d-2)^2}{(d-1)^2}\left( J_{22}^B+2 J_{32}^B+ J_{43}^B\right)\bigg](m_\pi,\mathcal{E}-\delta_a,\mathcal{E},q),\\[0.3cm]
\mathcal{V}_{\Sigma_c\bar{D}^\ast}^{B_{2.4}}&=&(1-\mathbf{I}_1\cdot\mathbf{I}_2)\frac{g^2g_3^2}{8f_\pi^4}\frac{1}{d-1}\bigg[-2\bm{q}^2\frac{(d+1)(d-2)}{d-1}\left( J_{31}^B+ J_{42}^B\right)-\bm{q}^2\frac{d-2-\boldsymbol{\sigma}\cdot\bm{T}}{d-1} J_{21}^B\nonumber\\
&&+\bm{q}^4\frac{d-2}{d-1}\left( J_{22}^B+2 J_{32}^B+ J_{43}^B\right)+(d^2-d-2) J_{41}^B\bigg](m_\pi,\mathcal{E}-\delta_a,\mathcal{E}+\delta_b,q),\\[0.3cm]
\mathcal{V}_{\Sigma_c\bar{D}^\ast}^{R_{2.i}}&=&\mathcal{V}_{\Sigma_c\bar{D}^\ast}^{B_{2.i}}\Big|_{
J_x^B\to
J_x^R,~\mathbf{I}_1\cdot\mathbf{I}_2\to-\mathbf{I}_1\cdot\mathbf{I}_2,~\boldsymbol{\sigma}\cdot\bm{T}\to-\boldsymbol{\sigma}\cdot\bm{T}}.\label{Subst_SigmaDast}
\end{eqnarray}
Considering the contribution of $\Lambda_c$:
\begin{eqnarray}
\mathcal{V}_{\Sigma_c\bar{D}^\ast}^{\bar{T}_{2.3}}&=&(\mathbf{I}_1\cdot\mathbf{I}_2)\frac{g_2^2}{2f_\pi^4}\bigg[(d-1) J_{34}^T-\bm{q}^2\left( J_{24}^T+ J_{33}^T\right)\bigg](m_\pi,\mathcal{E}+\delta_c,q),\\[0.3cm]
\mathcal{V}_{\Sigma_c\bar{D}^\ast}^{\bar{B}_{2.1}}&=&(1-2\mathbf{I}_1\cdot\mathbf{I}_2)\frac{g^2g_2^2}{8f_\pi^4}\bigg[\frac{4d^2-10d+6}{d-1} J_{41}^B-\bm{q}^2\frac{d^2+3d-8}{d-1}\left( J_{31}^B+ J_{42}^B\right)\nonumber\\
&&-\bm{q}^2\frac{d-2+\boldsymbol{\sigma}\cdot\bm{T}}{d-1} J_{21}^B+\bm{q}^4\frac{d-2}{d-1}\left( J_{22}^B+2 J_{32}^B+ J_{43}^B\right)\bigg](m_\pi,\mathcal{E}+\delta_c,\mathcal{E},q),\\[0.3cm]
\mathcal{V}_{\Sigma_c\bar{D}^\ast}^{\bar{B}_{2.2}}&=&(1-2\mathbf{I}_1\cdot\mathbf{I}_2)\frac{g^2g_2^2}{8f_\pi^4}\bigg[-2\bm{q}^2\frac{d+1}{d-1}\left( J_{31}^B+ J_{42}^B\right)-\bm{q}^2\frac{1}{d-1}(1+\boldsymbol{\sigma}\cdot\bm{T}) J_{21}^B\nonumber\\
&&+(d+1) J_{41}^B+\bm{q}^4\frac{1}{d-1}\left( J_{22}^B+2 J_{32}^B+ J_{43}^B\right)\bigg](m_\pi,\mathcal{E}+\delta_c,\mathcal{E}+\delta_b,q),\\[0.3cm]
\mathcal{V}_{\Sigma_c\bar{D}^\ast}^{\bar{R}_{2.i}}&=&\mathcal{V}_{\Sigma_c\bar{D}^\ast}^{\bar{B}_{2.i}}\Big|_{
J_x^B\to  J_x^R,~\mathbf{I}_1\cdot\mathbf{I}_2\to
-\mathbf{I}_1\cdot\mathbf{I}_2,~\boldsymbol{\sigma}\cdot\bm{T}\to-\boldsymbol{\sigma}\cdot\bm{T}}.\label{Subst_SigmaDast_Lambda}
\end{eqnarray}
From the above equations we see that, in the $S$-wave interactions,
only the central terms and spin-spin interactions survive in the
effective potentials.
\subsection{$\Sigma_c^\ast\bar{D}$ system}
Like the $\Sigma_c\bar{D}$ system, the leading order potential for
the $\Sigma_c^\ast\bar{D}$ system only stems from the contact terms
[diagram $(X_{3.1})$ in figure~\ref{Tree_Level_Diagrams}]. The
expression reads
\begin{eqnarray}
\mathcal{V}_{\Sigma_c^\ast
\bar{D}}^{X_{3.1}}=-D_a-2E_a(\mathbf{I}_1\cdot\mathbf{I}_2).
\end{eqnarray}
We see that the contact potential of the $\Sigma_c^\ast\bar{D}$
system equals to the one of the $\Sigma_c\bar{D}$ system in
eq.~\eqref{LO_SigmacDb_Con}, because the
$\mathcal{O}(\varepsilon^0)$ contact Lagrangian is constructed in
the heavy quark limit. The heavy quark breaking effect will be
manifested in the loop diagrams when the mass splittings are
considered in the propagators of the heavy matter fields.
\begin{figure*}[hptb]
\begin{centering}
    \setlength{\abovecaptionskip}{0.01cm}
    \scalebox{1.0}{\includegraphics[width=\columnwidth]{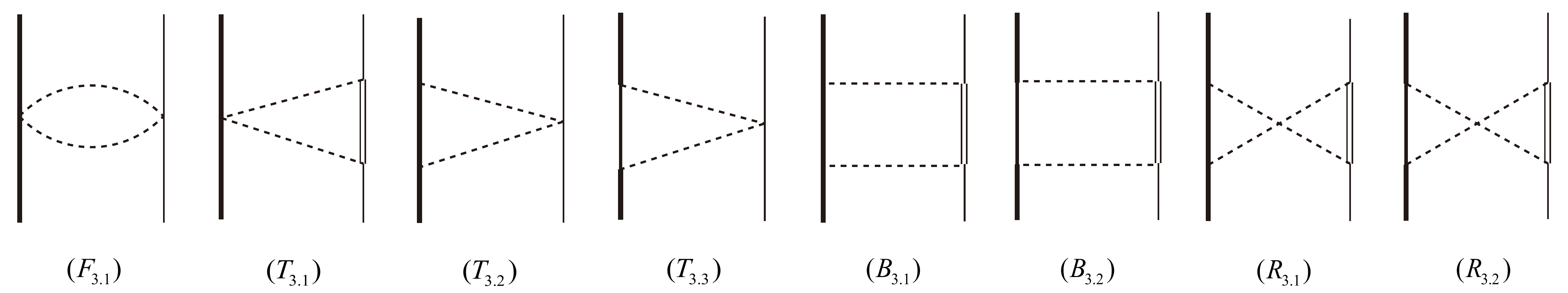}}
    \caption{The two-pion-exchange diagrams of the $\Sigma_c^\ast\bar{D}$ system at $\mathcal{O}(\epsilon^2)$. The internal heavy baryon lines in diagrams $(T_{3.3})$, $(B_{3.2})$ and $(R_{3.2})$ can also be the $\Lambda_c$. The notations are the same as those in figure~\ref{Tree_Level_Diagrams}.\label{TwoPion_Loop3}}
\end{centering}
\end{figure*}

The two-pion-exchange diagrams are illustrated in
figure~\ref{TwoPion_Loop3}. The analytical results for these
diagrams are given as
\begin{eqnarray}
\mathcal{V}_{\Sigma_c^\ast \bar{D}}^{F_{3.1}}&=&(\mathbf{I}_1\cdot\mathbf{I}_2)\frac{1}{f_\pi^4} J_{22}^F(m_\pi,q),\\[0.3cm]
\mathcal{V}_{\Sigma_c^\ast \bar{D}}^{T_{3.1}}&=&(\mathbf{I}_1\cdot\mathbf{I}_2)\frac{g^2}{f_\pi^4}\bigg[(d-1) J_{34}^T-\bm{q}^2\left( J_{24}^T+ J_{33}^T\right)\bigg](m_\pi,\mathcal{E}-\delta_b,q),\\[0.3cm]
\mathcal{V}_{\Sigma_c^\ast \bar{D}}^{T_{3.2}}&=&(\mathbf{I}_1\cdot\mathbf{I}_2)\frac{g_5^2}{4f_\pi^4}\frac{d^2-2d-3}{(d-1)^2}\bigg[(d-1) J_{34}^T-\bm{q}^2\left( J_{24}^T+ J_{33}^T\right)\bigg](m_\pi,\mathcal{E},q),\\[0.3cm]
\mathcal{V}_{\Sigma_c^\ast \bar{D}}^{T_{3.3}}&=&(\mathbf{I}_1\cdot\mathbf{I}_2)\frac{g_3^2}{4f_\pi^4}\bigg[ J_{34}^T-\frac{\bm{q}^2}{d-1}\left( J_{24}^T+ J_{33}^T\right)\bigg](m_\pi,\mathcal{E}+\delta_a,q),\\[0.3cm]
\mathcal{V}_{\Sigma_c^\ast \bar{D}}^{B_{3.1}}&=&(1-\mathbf{I}_1\cdot\mathbf{I}_2)\frac{g^2g_5^2}{8f_\pi^4}\frac{d^2-2d-3}{(d-1)^2}\bigg[(d^2-1) J_{41}^B-2\bm{q}^2(d+1)\left( J_{31}^B+ J_{42}^B\right)\nonumber\\
&&-\bm{q}^2 J_{21}^B+\bm{q}^4\left( J_{22}^B+2 J_{32}^B+ J_{43}^B\right)\bigg](m_\pi,\mathcal{E},\mathcal{E}-\delta_b,q),\\[0.3cm]
\mathcal{V}_{\Sigma_c^\ast \bar{D}}^{B_{3.2}}&=&(1-\mathbf{I}_1\cdot\mathbf{I}_2)\frac{g^2g_3^2}{8f_\pi^4}\bigg[(d+1) J_{41}^B-2\bm{q}^2\frac{d+1}{d-1}\left( J_{31}^B+ J_{42}^B\right)-\bm{q}^2\frac{1}{d-1} J_{21}^B\nonumber\\
&&+\bm{q}^4\frac{1}{d-1}\left( J_{22}^B+2 J_{32}^B+ J_{43}^B\right)\bigg](m_\pi,\mathcal{E}+\delta_a,\mathcal{E}-\delta_b,q),\\[0.3cm]
\mathcal{V}_{\Sigma_c^\ast
\bar{D}}^{R_{3.i}}&=&\mathcal{V}_{\Sigma_c^\ast
\bar{D}}^{B_{3.i}}\Big|_{ J_x^B\to
J_x^R,~\mathbf{I}_1\cdot\mathbf{I}_2\to
-\mathbf{I}_1\cdot\mathbf{I}_2}.
\end{eqnarray}
Including the contribution of $\Lambda_c$:
\begin{eqnarray}
\mathcal{V}_{\Sigma_c^\ast \bar{D}}^{\bar{T}_{3.3}}&=&(\mathbf{I}_1\cdot\mathbf{I}_2)\frac{g_4^2}{2f_\pi^4}\bigg[ J_{34}^T-\frac{\bm{q}^2}{d-1}\left( J_{24}^T+ J_{33}^T\right)\bigg](m_\pi,\mathcal{E}+\delta_d,q),\\[0.3cm]
\mathcal{V}_{\Sigma_c^\ast \bar{D}}^{\bar{B}_{3.2}}&=&(1-2\mathbf{I}_1\cdot\mathbf{I}_2)\frac{g^2g_4^2}{8f_\pi^4}\bigg[(d+1) J_{41}^B-2\bm{q}^2\frac{d+1}{d-1}\left( J_{31}^B+ J_{42}^B\right)-\bm{q}^2\frac{1}{d-1} J_{21}^B\nonumber\\
&&+\bm{q}^4\frac{1}{d-1}\left( J_{22}^B+2 J_{32}^B+ J_{43}^B\right)\bigg](m_\pi,\mathcal{E}+\delta_d,\mathcal{E}-\delta_b,q),\\[0.3cm]
\mathcal{V}_{\Sigma_c^\ast
\bar{D}}^{\bar{R}_{3.2}}&=&V_{\Sigma_c^\ast
\bar{D}}^{\bar{B}_{3.2}}\Big|_{ J_x^B\to
J_x^R,~\mathbf{I}_1\cdot\mathbf{I}_2\to
-\mathbf{I}_1\cdot\mathbf{I}_2}.
\end{eqnarray}
\subsection{$\Sigma_c^\ast\bar{D}^\ast$ system}
The leading order diagrams for $\Sigma_c^\ast\bar{D}^\ast$ system
are the graphs $(X_{4.1})$ and $(H_{4.1})$ in
figure~\ref{Tree_Level_Diagrams}. The potentials from these two
graphs read
\begin{eqnarray}
\mathcal{V}_{\Sigma_c^\ast \bar{D}^\ast}^{X_{4.1}}&=&-D_a-2E_a(\mathbf{I}_1\cdot\mathbf{I}_2)+\Big[-D_b-2E_b(\mathbf{I}_1\cdot\mathbf{I}_2)\Big]\boldsymbol{\sigma}_{rs}\cdot\bm{T},\\[0.3cm]
\mathcal{V}_{\Sigma_c^\ast
\bar{D}^\ast}^{H_{4.1}}&=&(\mathbf{I}_1\cdot\mathbf{I}_2)\frac{gg_5}{2f_\pi^2}\frac{(\bm{q}\cdot\boldsymbol{\sigma}_{rs})(\bm{q}\cdot\bm{T})}{\bm{q}^2+m_\pi^2},
\end{eqnarray}
where the operator $\boldsymbol{\sigma}_{rs}$ is related to the spin
operator $\bm{S}_1$ of the $\Sigma_c^\ast$ with
$\bm{S}_1=\frac{3}{2}\boldsymbol{\sigma}_{rs}$ (see the detailed
derivations in appendix~\ref{SpinTransfOperators}), so the
$\boldsymbol{\sigma}_{rs}\cdot\bm{T}$ term represents the spin-spin
interaction as well.  We see the $\mathcal{O}(\varepsilon^0)$
potentials for $\Sigma_c^\ast\bar{D}^\ast$ resemble the ones for
$\Sigma_c\bar{D}^\ast$ in eqs.~\eqref{X21_Potential} and
\eqref{H21_Potential}.
\begin{figure*}[hptb]
\begin{centering}
    \setlength{\abovecaptionskip}{0.01cm}
    \scalebox{1.0}{\includegraphics[width=\columnwidth]{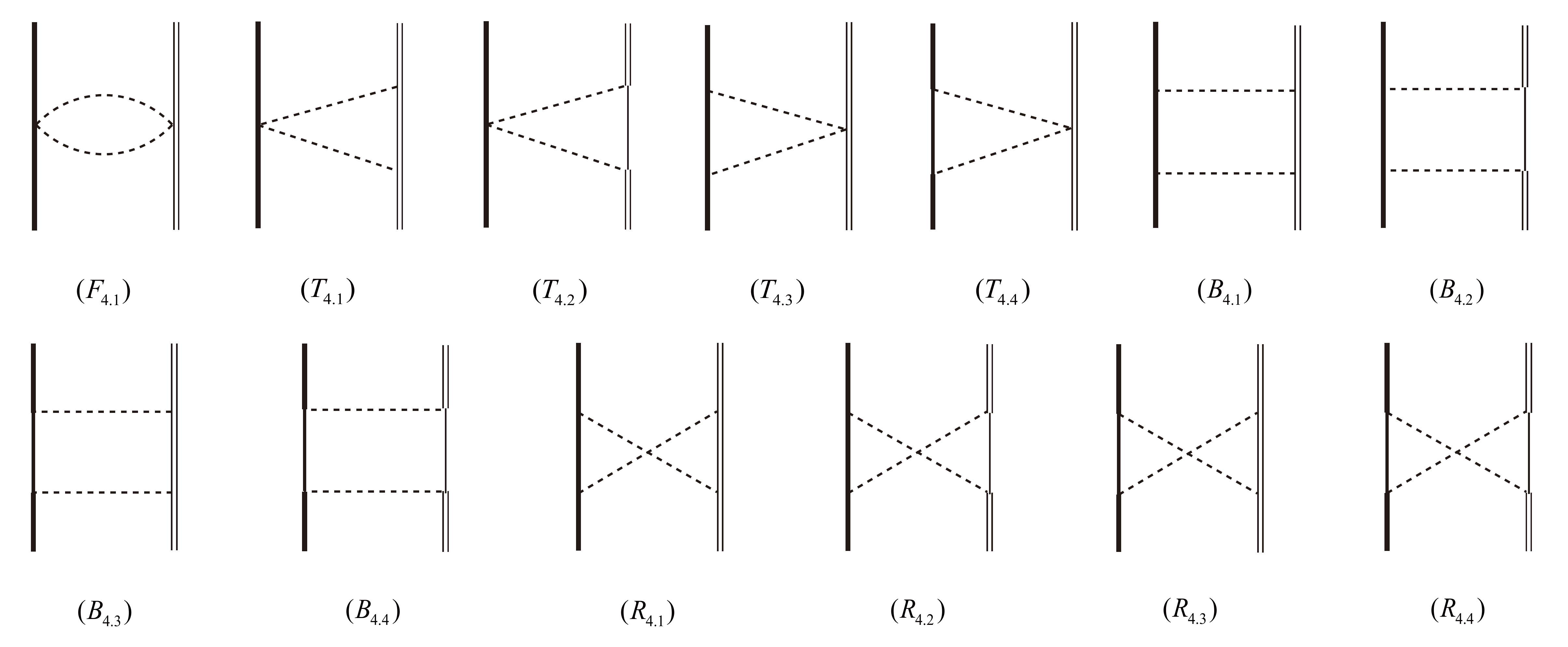}}
    \caption{The two-pion-exchange diagrams of the $\Sigma_c^\ast\bar{D}^\ast$ system at $\mathcal{O}(\epsilon^2)$. The internal heavy baryon lines in diagrams $(T_{4.4})$, $(B_{4.3})$, $(B_{4.4})$, $(R_{4.3})$ and $(R_{4.4})$ can also be the $\Lambda_c$. The notations are the same as those in figure~\ref{Tree_Level_Diagrams}.\label{TwoPion_Loop4}}
\end{centering}
\end{figure*}

The two-pion-exchange diagrams are displayed in
figure~\ref{TwoPion_Loop4}. The potentials originate from these
graphs read
\begin{eqnarray}
\mathcal{V}_{\Sigma_c^\ast \bar{D}^\ast}^{F_{4.1}}&=&(\mathbf{I}_1\cdot\mathbf{I}_2)\frac{1}{f_\pi^4} J_{22}^F(m_\pi,q),\\[0.3cm]
\mathcal{V}_{\Sigma_c^\ast \bar{D}^\ast}^{T_{4.1}}&=&(\mathbf{I}_1\cdot\mathbf{I}_2)\frac{g^2}{f_\pi^4}\bigg[2 J_{34}^T-\bm{q}^2\frac{d-2}{d-1}\left( J_{24}^T+ J_{33}^T\right)\bigg](m_\pi,\mathcal{E},q),\\[0.3cm]
\mathcal{V}_{\Sigma_c^\ast \bar{D}^\ast}^{T_{4.2}}&=&(\mathbf{I}_1\cdot\mathbf{I}_2)\frac{g^2}{f_\pi^4}\bigg[ J_{34}^T-\frac{\bm{q}^2}{d-1}\left( J_{24}^T+ J_{33}^T\right)\bigg](m_\pi,\mathcal{E}+\delta_b,q),\\[0.3cm]
\mathcal{V}_{\Sigma_c^\ast \bar{D}^\ast}^{T_{4.3}}&=&(\mathbf{I}_1\cdot\mathbf{I}_2)\frac{g_5^2}{4f_\pi^4}\frac{d^2-2d-3}{(d-1)^2}\bigg[(d-1) J_{34}^T-\bm{q}^2\left( J_{24}^T+ J_{33}^T\right)\bigg](m_\pi,\mathcal{E},q),\\[0.3cm]
\mathcal{V}_{\Sigma_c^\ast \bar{D}^\ast}^{T_{4.4}}&=&(\mathbf{I}_1\cdot\mathbf{I}_2)\frac{g_3^2}{4f_\pi^4}\bigg[ J_{34}^T-\frac{\bm{q}^2}{d-1}\left( J_{24}^T+ J_{33}^T\right)\bigg](m_\pi,\mathcal{E}+\delta_a,q),\\[0.3cm]
\mathcal{V}_{\Sigma_c^\ast \bar{D}^\ast}^{B_{4.1}}&=&(1-\mathbf{I}_1\cdot\mathbf{I}_2)\frac{g^2g_5^2}{8f_\pi^4}\frac{1}{d-1}\Bigg[2\left(2d^2-5d-7+3(\boldsymbol{\sigma}_{rs}\cdot\bm{T})^2-\boldsymbol{\sigma}_{rs}\cdot\bm{T}\right) J_{41}^B\nonumber\\
&&-\bm{q}^2\frac{d^3+2d^2-15d-16+12(\boldsymbol{\sigma}_{rs}\cdot\bm{T})^2-4\boldsymbol{\sigma}_{rs}\cdot\bm{T}}{d-1}\left( J_{31}^B+ J_{42}^B\right)\nonumber\\
&&-\bm{q}^2\frac{d^2-3d-4+3(\boldsymbol{\sigma}_{rs}\cdot\bm{T})^2+(d-4)\boldsymbol{\sigma}_{rs}\cdot\bm{T}}{d-1} J_{21}^B\nonumber\\
&&+\bm{q}^4\frac{d^3-4d^2+d+6}{(d-1)^2}
\left( J_{22}^B+2 J_{32}^B+ J_{43}^B\right)\Bigg](m_\pi,\mathcal{E},\mathcal{E},q),\\[0.3cm]
\mathcal{V}_{\Sigma_c^\ast
\bar{D}^\ast}^{B_{4.2}}&=&(1-\mathbf{I}_1\cdot\mathbf{I}_2)\frac{g^2g_5^2}{8f_\pi^4}\frac{1}{d-1}\Bigg[\left(d^2-1-6(\boldsymbol{\sigma}_{rs}\cdot\bm{T})^2+2\boldsymbol{\sigma}_{rs}\cdot\bm{T}\right)
J_{41}^B
+\bm{q}^4\frac{d^2-2d-3}{(d-1)^2}\nonumber\\
&&\times\left( J_{22}^B+2 J_{32}^B+ J_{43}^B\right)
-2\bm{q}^2\frac{d^2-1-6(\boldsymbol{\sigma}_{rs}\cdot\bm{T})^2+2\boldsymbol{\sigma}_{rs}\cdot\bm{T}}{d-1}\left( J_{31}^B+ J_{42}^B\right)\nonumber\\
&&-\bm{q}^2\frac{d+1-3(\boldsymbol{\sigma}_{rs}\cdot\bm{T})^2+(d-2)\boldsymbol{\sigma}_{rs}\cdot\bm{T}}{d-1} J_{21}^B\Bigg](m_\pi,\mathcal{E},\mathcal{E}+\delta_b,q),\\[0.3cm]
\mathcal{V}_{\Sigma_c^\ast
\bar{D}^\ast}^{B_{4.3}}&=&(1-\mathbf{I}_1\cdot\mathbf{I}_2)\frac{g^2g_3^2}{32f_\pi^4}\Bigg[\left(20-6(\boldsymbol{\sigma}_{rs}\cdot\bm{T})^2+2\boldsymbol{\sigma}_{rs}\cdot\bm{T}\right)
J_{41}^B
+4\bm{q}^4\frac{d-2}{(d-1)^2}\big( J_{22}^B\nonumber\\
&&+2 J_{32}^B+ J_{43}^B\big)
-4\bm{q}^2\frac{d+6-3(\boldsymbol{\sigma}_{rs}\cdot\bm{T})^2+\boldsymbol{\sigma}_{rs}\cdot\bm{T}}{d-1}\left( J_{31}^B+ J_{42}^B\right)\nonumber\\
&&+3\bm{q}^2\frac{(\boldsymbol{\sigma}_{rs}\cdot\bm{T})^2-\boldsymbol{\sigma}_{rs}\cdot\bm{T}-2}{d-1} J_{21}^B\Bigg](m_\pi,\mathcal{E}+\delta_a,\mathcal{E},q),\\[0.3cm]
\mathcal{V}_{\Sigma_c^\ast
\bar{D}^\ast}^{B_{4.4}}&=&(1-\mathbf{I}_1\cdot\mathbf{I}_2)\frac{g^2g_3^2}{32f_\pi^4}\Bigg[\left(6(\boldsymbol{\sigma}_{rs}\cdot\bm{T})^2-2\boldsymbol{\sigma}_{rs}\cdot\bm{T}\right)
J_{41}^B
+4\bm{q}^2\frac{-3(\boldsymbol{\sigma}_{rs}\cdot\bm{T})^2+\boldsymbol{\sigma}_{rs}\cdot\bm{T}}{d-1}\nonumber\\
&&\times\left( J_{31}^B+ J_{42}^B\right)
+4\bm{q}^4\frac{1}{(d-1)^2}\left( J_{22}^B+2 J_{32}^B+ J_{43}^B\right)\nonumber\\
&&-\bm{q}^2\frac{3(\boldsymbol{\sigma}_{rs}\cdot\bm{T})^2+\boldsymbol{\sigma}_{rs}\cdot\bm{T}-2}{d-1}
J_{21}^B\Bigg](m_\pi,\mathcal{E}+\delta_a,\mathcal{E}+\delta_b,q).
\end{eqnarray}
Unlike the two-pion-exchange potentials of the
$\Sigma_c\bar{D}^\ast$ system, there exists a very simple relation
between $\mathcal{V}_{\Sigma_c\bar{D}^\ast}^{R_{2.i}}$ and
$\mathcal{V}_{\Sigma_c\bar{D}^\ast}^{B_{2.i}}$ [e.g., see
eqs.~\eqref{Subst_SigmaDast} and \eqref{Subst_SigmaDast_Lambda}],
since the $\boldsymbol{\sigma}\cdot\bm{T}$ term only accompanies the
$ J_{21}^B$ and $ J_{21}^R$. For the $\Sigma_c^\ast\bar{D}^\ast$
system, the two-pion-exchange potentials are very complicated, and
we cannot write out the simple relationship as
eqs.~\eqref{Subst_SigmaDast} and \eqref{Subst_SigmaDast_Lambda}. But
there is still a corresponding relation between each
$\mathcal{V}_{\Sigma_c^\ast\bar{D}^\ast}^{R_{4.i}}$ and
$\mathcal{V}_{\Sigma_c^\ast\bar{D}^\ast}^{B_{4.i}}$, which is
\begin{eqnarray}
\mathcal{V}_{\Sigma_c^\ast\bar{D}^\ast}^{R_{4.i}}=\mathcal{V}_{\Sigma_c^\ast\bar{D}^\ast}^{B_{4.i}}\Big|_{
J_x^B\to  J_x^R,~\mathbf{I}_1\cdot\mathbf{I}_2\to
-\mathbf{I}_1\cdot\mathbf{I}_2,\mathcal{C}_{
J_{21}^B}\to\mathcal{C}_{ J_{21}^R}},
\end{eqnarray}
where the substitution rule $\mathcal{C}_{ J_{21}^B}\to\mathcal{C}_{
J_{21}^R}$ represents that only the coefficient of $ J_{21}^B$ in
the square brackets should be replaced with the one of $ J_{21}^R$,
while the other terms remain unchanged. For example, the
$\mathcal{C}_{ J_{21}^B}$s for $\mathcal{V}_{\Sigma_c^\ast
\bar{D}^\ast}^{B_{4.3}}$ and $\mathcal{V}_{\Sigma_c^\ast
\bar{D}^\ast}^{B_{4.4}}$ are
$3\bm{q}^2[(\boldsymbol{\sigma}_{rs}\cdot\bm{T})^2-\boldsymbol{\sigma}_{rs}\cdot\bm{T}-2]/(d-1)$
and
$-\bm{q}^2[3(\boldsymbol{\sigma}_{rs}\cdot\bm{T})^2+\boldsymbol{\sigma}_{rs}\cdot\bm{T}-2]/(d-1)$,
respectively. We write down the $\mathcal{C}_{ J_{21}^R}$s of the
$\mathcal{V}_{\Sigma_c^\ast\bar{D}^\ast}^{R_{4.i}}~(i=1,\dots,4)$ as
follows correspondingly.
\begin{eqnarray}\label{Coe_J21R}
i&=&1:\bm{q}^2\frac{d^2-3d-4+3(\boldsymbol{\sigma}_{rs}\cdot\bm{T})^2-(d-2)\boldsymbol{\sigma}_{rs}\cdot\bm{T}}{1-d},
i=3:3\bm{q}^2\frac{(\boldsymbol{\sigma}_{rs}\cdot\bm{T})^2+\boldsymbol{\sigma}_{rs}\cdot\bm{T}/3-2}{d-1},\nonumber\\
i&=&2:\bm{q}^2\frac{d+1-3(\boldsymbol{\sigma}_{rs}\cdot\bm{T})^2-(d-4)\boldsymbol{\sigma}_{rs}\cdot\bm{T}}{1-d},
i=4:\bm{q}^2\frac{3(\boldsymbol{\sigma}_{rs}\cdot\bm{T})^2-3\boldsymbol{\sigma}_{rs}\cdot\bm{T}-2}{1-d}.
\end{eqnarray}
Including the contribution of $\Lambda_c$:
\begin{eqnarray}
\mathcal{V}_{\Sigma_c^\ast \bar{D}^\ast}^{\bar{T}_{4.4}}&=&(\mathbf{I}_1\cdot\mathbf{I}_2)\frac{g_4^2}{2f_\pi^4}\bigg[ J_{34}^T-\frac{\bm{q}^2}{d-1}\left( J_{24}^T+ J_{33}^T\right)\bigg](m_\pi,\mathcal{E}+\delta_d,q),\\[0.3cm]
\mathcal{V}_{\Sigma_c^\ast
\bar{D}^\ast}^{\bar{B}_{4.3}}&=&(1-2\mathbf{I}_1\cdot\mathbf{I}_2)\frac{g^2g_4^2}{32f_\pi^4}\Bigg[\left(20-6(\boldsymbol{\sigma}_{rs}\cdot\bm{T})^2+2\boldsymbol{\sigma}_{rs}\cdot\bm{T}\right)
J_{41}^B
+4\bm{q}^4\frac{d-2}{(d-1)^2}\big( J_{22}^B\nonumber\\
&&+2 J_{32}^B+ J_{43}^B\big)
-4\bm{q}^2\frac{d+6-3(\boldsymbol{\sigma}_{rs}\cdot\bm{T})^2+\boldsymbol{\sigma}_{rs}\cdot\bm{T}}{d-1}\left( J_{31}^B+ J_{42}^B\right)\nonumber\\
&&+3\bm{q}^2\frac{(\boldsymbol{\sigma}_{rs}\cdot\bm{T})^2-\boldsymbol{\sigma}_{rs}\cdot\bm{T}-2}{d-1} J_{21}^B\Bigg](m_\pi,\mathcal{E}+\delta_d,\mathcal{E},q),\\[0.3cm]
\mathcal{V}_{\Sigma_c^\ast
\bar{D}^\ast}^{\bar{B}_{4.4}}&=&(1-2\mathbf{I}_1\cdot\mathbf{I}_2)\frac{g^2g_4^2}{32f_\pi^4}\Bigg[\left(6(\boldsymbol{\sigma}_{rs}\cdot\bm{T})^2-2\boldsymbol{\sigma}_{rs}\cdot\bm{T}\right)
J_{41}^B
+4\bm{q}^2\frac{-3(\boldsymbol{\sigma}_{rs}\cdot\bm{T})^2+\boldsymbol{\sigma}_{rs}\cdot\bm{T}}{d-1}\nonumber\\
&&\times\left( J_{31}^B+ J_{42}^B\right)
+4\bm{q}^4\frac{1}{(d-1)^2}\left( J_{22}^B+2 J_{32}^B+ J_{43}^B\right)\nonumber\\
&&-\bm{q}^2\frac{3(\boldsymbol{\sigma}_{rs}\cdot\bm{T})^2+\boldsymbol{\sigma}_{rs}\cdot\bm{T}-2}{d-1} J_{21}^B\Bigg](m_\pi,\mathcal{E}+\delta_d,\mathcal{E}+\delta_b,q),\\[0.3cm]
\mathcal{V}_{\Sigma_c^\ast\bar{D}^\ast}^{\bar{R}_{4.i}}&=&\mathcal{V}_{\Sigma_c^\ast\bar{D}^\ast}^{\bar{B}_{4.i}}\Big|_{
J_x^B\to  J_x^R,~\mathbf{I}_1\cdot\mathbf{I}_2\to
-\mathbf{I}_1\cdot\mathbf{I}_2,\mathcal{C}_{
J_{21}^B}\to\mathcal{C}_{ J_{21}^R}},
\end{eqnarray}
where the $\mathcal{C}_{ J_{21}^R}$s are equal to the ones in
eq.~\eqref{Coe_J21R} for $i=3$ and $i=4$, respectively. In the above
equations, we notice that a new spin-spin structure
$(\boldsymbol{\sigma}_{rs}\cdot\bm{T})^2$ arises in the box and
crossed box diagrams, which is the characteristic interaction
structure for the high spin particle systems. Such a structure
cannot appear in the two-body potentials with spin-$1\over2$
particle, such as the $\Sigma_c\bar{D}^\ast$ system. Due to the
constraints of the commutation and anticommutation relations of the
Pauli matrix, the spin operator of a spin-$1\over 2$ particle
appears at most once. On the other hand, the
$(\boldsymbol{\sigma}_{rs}\cdot\bm{T})^2$ terms do not emerge at the
tree level, where the heavy quark symmetry is satisfied. In other
words, this structure is also the reflection of the heavy quark
symmetry breaking effect at the one-loop level, which indeed
disappears if we set the mass splittings in the loops to be zeros
(this structure will persist for the diagrams with $\Lambda_c$ as
the intermediate state, since the mass splittings $\delta_c$ and
$\delta_d$ do not vanish even in the rigorous heavy quark limit).

\section{Numerical results without and with the $\Lambda_c$}\label{NumericalResults}

The newly observed three $P_c$ states, $P_c(4312)$, $P_c(4440)$ and
$P_c(4457)$ have been studied with the same framework in our
previous paper~\cite{Meng:2019ilv}, in which we did not include the
contribution of the $\Lambda_c$. There are three scenarios in
ref.~\cite{Meng:2019ilv}. In scenario I, the LECs are estimated from
the $N$-$N$ data, but the result is not good, because we cannot
reproduce the $P_c(4457)$. In scenario II, the LECs are determined
by fitting the data of the three $P_c$s, yet the result is still
unsatisfactory. In scenario III, the $P_c$s are simultaneously
reproduced in a relatively small parameter region when the couple
channel effect is included. In this part, we revisit these states
without and with the $\Lambda_c$ contribution, and give a comparison
with the result in scenario II of ref.~\cite{Meng:2019ilv}.

\subsection{The three $P_c$ states without the $\Lambda_c$}
Up to now, the four LECs in eq.~\eqref{Contact_Lag_BM} are still
unknown. But we do not have to determine each of them since the
forms of the $\mathcal{O}(\varepsilon^0)$ contact potentials are
homogeneous for definite isospin states. There are only two
independent LECs in nature if the isospin-isospin terms are absorbed
into the relevant LECs with the following redefinitions,
\begin{eqnarray}
\mathbb{D}_1=D_a+2E_a(\mathbf{I}_1\cdot\mathbf{I}_2),\quad\quad\quad
\mathbb{D}_2=D_b+2E_b(\mathbf{I}_1\cdot\mathbf{I}_2).
\end{eqnarray}
Thus the $\mathcal{O}(\varepsilon^0)$ contact potentials of the
$\Sigma_c^{(\ast)}\bar{D}^{(\ast)}$ systems can be rewritten
as\footnote{There is a typo in the eq.~(51) of
ref.~\cite{Meng:2019ilv}. The potential
$\mathcal{V}_{\Sigma_c\bar{D}^\ast}^{X_{2.1}}$ should be revised to
the correct form of this work. But it does not affect the numerical
results in ref.~\cite{Meng:2019ilv}, since the value of
$\mathbb{D}_2$ in the figure 10 of ref.~\cite{Meng:2019ilv} is the
twice of the one used in this work.}
\begin{eqnarray}\label{Contact_Potential}
\mathcal{V}_{\Sigma_c\bar{D}}^{X_{1.1}}&=&-\mathbb{D}_1,\quad\quad\quad \mathcal{V}_{\Sigma_c\bar{D}^\ast}^{X_{2.1}}=-\Big[\mathbb{D}_1+\frac{2}{3}\mathbb{D}_2(\boldsymbol{\sigma}\cdot\bm{T})\Big],\nonumber\\[0.2cm]
\mathcal{V}_{\Sigma_c^\ast\bar{D}}^{X_{3.1}}&=&-\mathbb{D}_1,\quad\quad\quad
\mathcal{V}_{\Sigma_c^\ast\bar{D}^\ast}^{X_{4.1}}=-\Big[\mathbb{D}_1+\mathbb{D}_2(\boldsymbol{\sigma}_{rs}\cdot\bm{T})\Big].
\end{eqnarray}

The masses and widths of the newly observed three $P_c$ states and
the previously reported $P_c(4380)$ are displayed in
table~\ref{Experimental_Data}. The closest thresholds, binding
energies as the $\Sigma_c^{(\ast)}\bar{D}^{(\ast)}$ molecules, and
theoretically favored $I(J^P)$ quantum numbers are also illustrated.
Since the masses of $\Sigma_c^{(\ast)+}$ and $\bar{D}^{(\ast)0}$
have been precisely measured in experiments, their minor errors are
ignored in calculating the uncertainties of binding energies.

\begin{table*}[htbp]
\centering \centering
\renewcommand{\arraystretch}{1.5}
\caption{The experimental and theoretical information of the
$P_c(4312)$, $P_c(4440)$, $P_c(4457)$~\cite{Aaij:2019vzc}, and
$P_c(4380)$~\cite{Aaij:2015tga}. The corresponding binding energies
are obtained with the thresholds of
$\Sigma_c^{(\ast)+}\bar{D}^{(\ast)0}$, such as the binding energy of
$P_c(4312)$ equals to
$m_{P_c(4312)}-(m_{\Sigma_c^+}+m_{\bar{D}^0})$. The masses of
$\Sigma_c^{(\ast)+}$ and $\bar{D}^{(\ast)0}$ are taken from the
Particle Physics Booklet~\cite{Tanabashi:2018oca}. The $I(J^P)$
quantum numbers are the theoretically favored ones, not the
experimental measurements (in units of
MeV).\label{Experimental_Data}} \setlength{\tabcolsep}{2.45mm} {
\begin{tabular}{c|ccccc}
\hline\hline
States& Mass &Width&Threshold&Binding energy&$I(J^P)$\\
\hline
$P_c(4312)$&$4311.9\pm0.7^{+6.8}_{-0.6}$&$9.8\pm2.7^{+3.7}_{-4.5}$&$\Sigma_c^+\bar{D}^0$&$-5.83\pm0.7^{+6.8}_{-0.6}$&$\frac{1}{2}\left(\frac{1}{2}^-\right)$\\
$P_c(4440)$&$4440.3\pm1.3^{+4.1}_{-4.7}$&$20.6\pm2.7^{+8.7}_{-10.1}$&$\Sigma_c^+\bar{D}^{\ast0}$&$-19.45\pm1.3^{+4.1}_{-4.7}$&$\frac{1}{2}\left(\frac{1}{2}^-\right)$\\
$P_c(4457)$&$4457.3\pm0.6^{+4.1}_{-1.7}$&$6.4\pm2.0^{+5.7}_{-1.9}$&$\Sigma_c^+\bar{D}^{\ast0}$&$-2.45\pm0.6^{+4.1}_{-1.7}$&$\frac{1}{2}\left(\frac{3}{2}^-\right)$\\
$P_c(4380)$&$4380\pm8\pm29$&$205\pm18\pm86$&$\Sigma_c^{\ast+}\bar{D}^{0}$&$-2.33\pm8\pm29$&$\frac{1}{2}\left(\frac{3}{2}^-\right)$\\
\hline\hline
\end{tabular}
}
\end{table*}

\begin{figure*}
\begin{minipage}[t]{0.5\linewidth}
\centering
\includegraphics[width=\columnwidth]{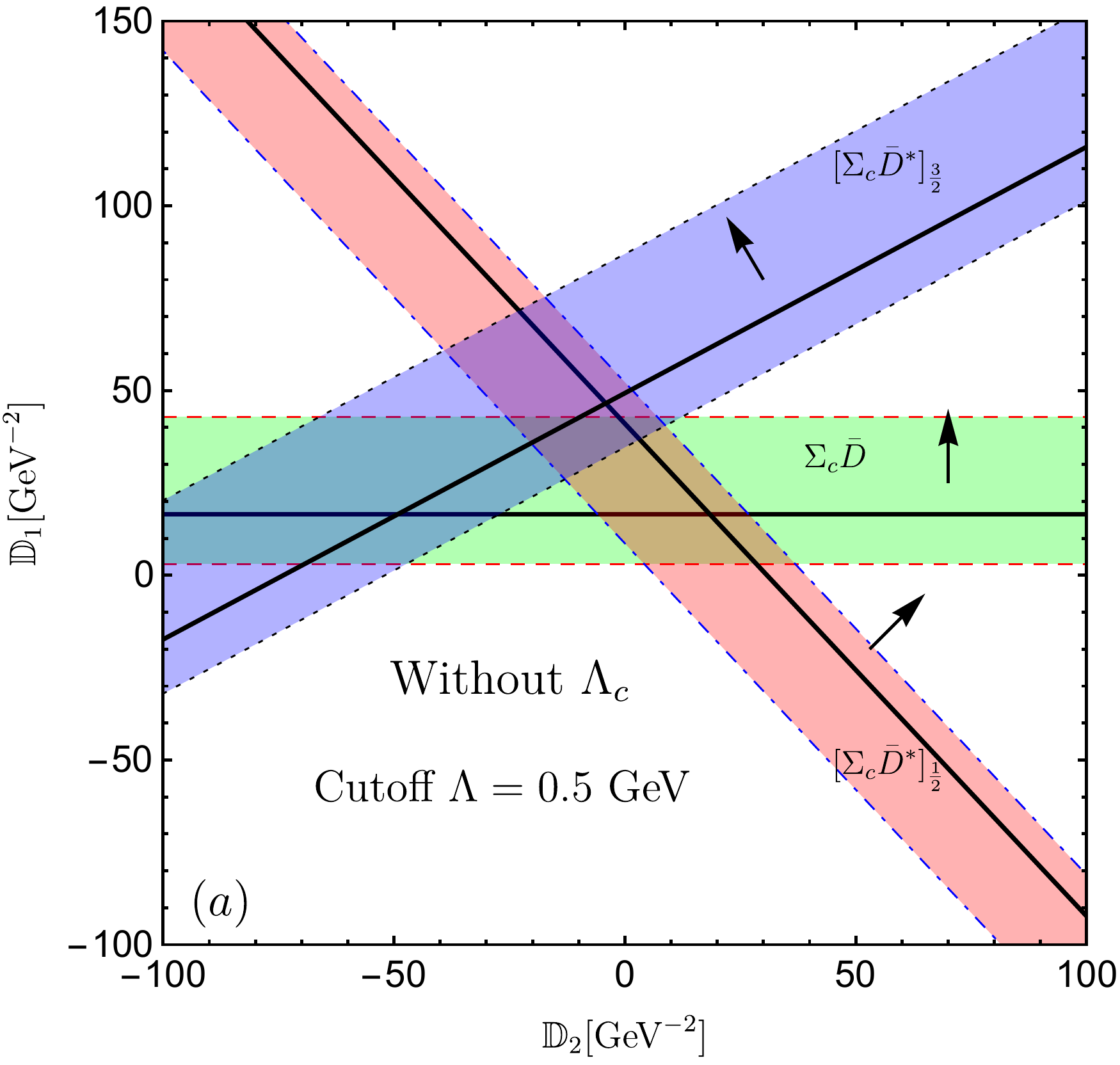}
\end{minipage}%
\begin{minipage}[t]{0.5\linewidth}
\centering
\includegraphics[width=\columnwidth]{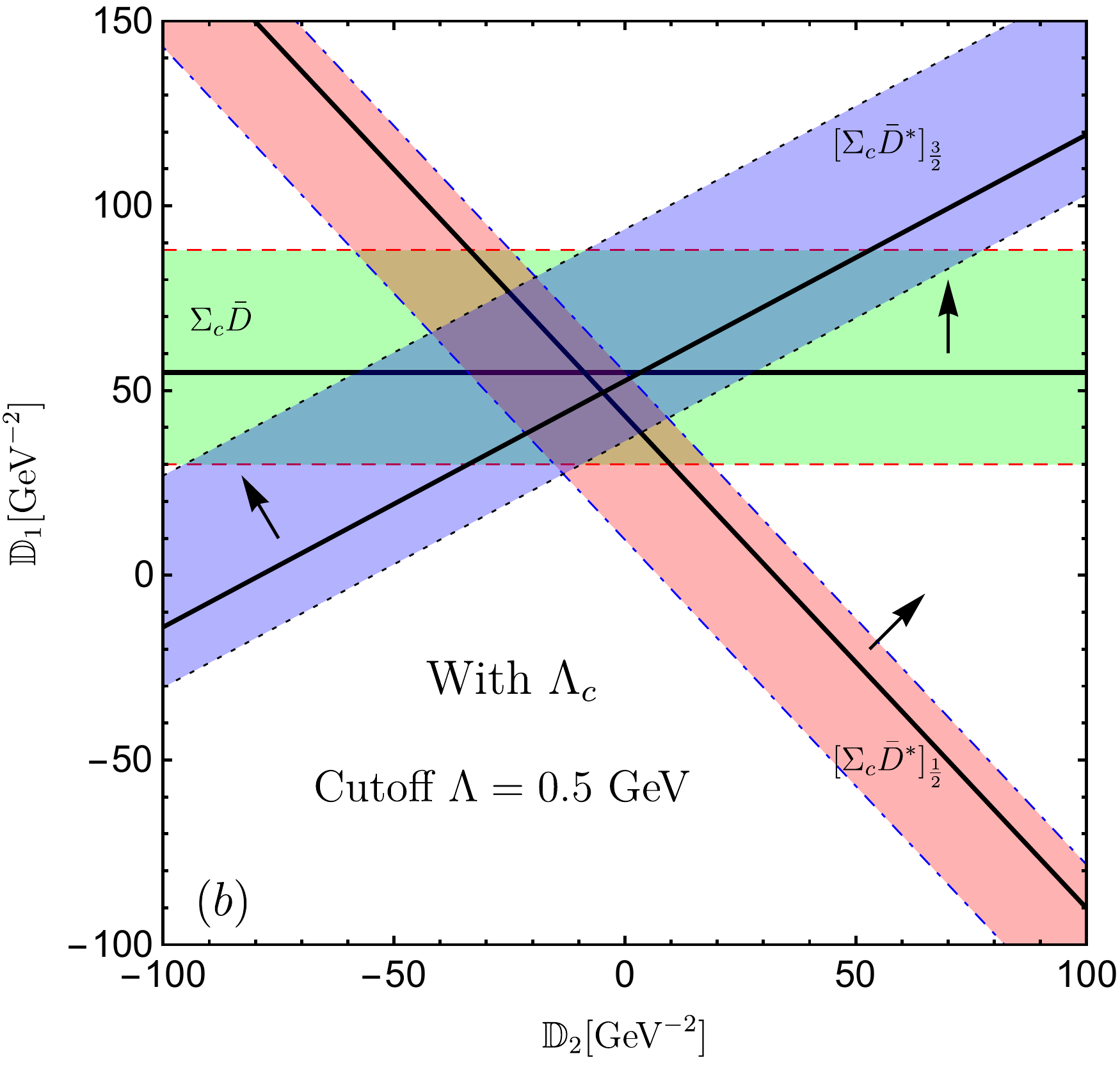}
\end{minipage}
\caption{The dependence of the binding energies of the three $P_c$
states on the redefined LECs $\mathbb{D}_1$ and $\mathbb{D}_2$. The
green, red and blue bands correspond to the
$[\Sigma_c\bar{D}]_{J=\frac{1}{2}}$,
$[\Sigma_c\bar{D}^\ast]_{J=\frac{1}{2}}$ and
$[\Sigma_c\bar{D}^\ast]_{J=\frac{3}{2}}$ systems, respectively. The
three black straight lines represent the central values of the
binding energies obtained from the experimental
data~\cite{Aaij:2019vzc} (the numbers in the fifth column of
table~\ref{Experimental_Data}). The boundaries of the bands that are
parallel to the corresponding straight lines stand for the regions
of parameters with the binding emerges $-30$ MeV and $0$ MeV,
respectively. The accompanied arrow shows the direction that the
each binding becomes deeper. Figures ($a$) and ($b$) illustrate the
results without and with the $\Lambda_c$, respectively. The results
are both calculated with the cutoff $\Lambda=0.5$
GeV.\label{Results_WithoutWith_Lambdac}}
\end{figure*}

With the above preparations, as in ref.~\cite{Meng:2019ilv}, we vary
the $\mathbb{D}_1$ and $\mathbb{D}_2$ in the ranges $[-100,150]$
$\textrm{GeV}^{-2}$ and $[-100,100]$ $\textrm{GeV}^{-2}$
respectively to search for the possible region where the three $P_c$
states can coexist. We mainly focus on the $I=\frac{1}{2}$ states,
because these $P_c$ states are observed in the mass spectra of
$J/\psi p$. To make a comparison, we present both the results
without and with the $\Lambda_c$ in
figures~\ref{Results_WithoutWith_Lambdac}(a) and
\ref{Results_WithoutWith_Lambdac}(b), respectively\footnote{One can
also see the another version in the figure 10(a) of
ref.~\cite{Meng:2019ilv}, where the $x$ and $y$ axes are
interchanged.}. We assume that the $P_c(4312)$, $P_c(4440)$ and
$P_c(4457)$ are the $[\Sigma_c\bar{D}]_{J=\frac{1}{2}}$,
$[\Sigma_c\bar{D}^\ast]_{J=\frac{1}{2}}$ and
$[\Sigma_c\bar{D}^\ast]_{J=\frac{3}{2}}$ molecular states,
respectively. We use three colored bands to denote the region of
parameters with binding energy $[-30,0]$ MeV for each system,
respectively. Considering the hadronic molecules are loosely bound
states, we set $-30$ MeV as the lower limit of the bindings. The
intersection point of two black solid lines designates the
coordinate value $(\mathbb{D}_2,\mathbb{D}_1)$ where the
corresponding two $P_c$s can coexist. Ideally, the three straight
lines should meet at a point if the central value of the mass for
each $P_c$ is exact and these $P_c$s are indeed the molecules of the
corresponding $\Sigma_c^{(\ast)}\bar{D}^{(\ast)}$ systems. However,
the results in figure~\ref{Results_WithoutWith_Lambdac}(a) are not
good. Three intersection points stay far away from each other. It is
hard to reproduce the three $P_c$s in this case, simultaneously.

The line-shape of the effective potentials for the three $P_c$s in
this case have been given in ref.~\cite{Meng:2019ilv}, where a set
of parameters $\mathbb{D}_1=42$ $\textrm{GeV}^{-2}$ and
$\mathbb{D}_2=-12.5$ $\textrm{GeV}^{-2}$ in the overlap region are
adopted. Here, we use these two values to calculate the binding
energies of the $[\Sigma_c^{(\ast)}\bar{D}^{(\ast)}]_{J}$ systems,
the corresponding results are given in the second row of
table~\ref{BindingEnergies}. From table~\ref{BindingEnergies} we see
that only the result for the $[\Sigma_c\bar{D}^\ast]_{\frac{3}{2}}$
system is consistent with the data in table~\ref{Experimental_Data}.
There are large differences for the
$[\Sigma_c\bar{D}]_{\frac{1}{2}}$ and
$[\Sigma_c\bar{D}^\ast]_{\frac{1}{2}}$ systems. In addition, the
$[\Sigma_c^\ast\bar{D}^\ast]_{\frac{1}{2}}$ is very shallowly bound,
and no binding solutions are found for other high spin systems. We
cannot simulate the three $P_c$s simultaneously no matter how we
choose the values of $\mathbb{D}_1$ and $\mathbb{D}_2$ in the
overlapped region of figure~\ref{Results_WithoutWith_Lambdac}(a).
\begin{table*}[htbp]
\centering
\renewcommand{\arraystretch}{1.5}
\caption{The binding energies $\Delta E$ for the $I=\frac{1}{2}$
hidden-charm $[\Sigma_c^{(\ast)}\bar{D}^{(\ast)}]_J$ systems in both
cases with and without the $\Lambda_c$, as well as the case with
$J^P=\frac{1}{2}^-$ for $P_c(4457)$ and $\frac{3}{2}^-$ for
$P_c(4440)$. The values of $(\mathbb{D}_1,\mathbb{D}_2)$ for the
``Without $\Lambda_c$" and ``With $\Lambda_c$" cases are chosen to
be $(42,-12.5)$ $\textrm{GeV}^{-2}$ and $(52,-4)$
$\textrm{GeV}^{-2}$, respectively. ``I.S." stands for the results
when interchanging the spins of $P_c(4440)$ and $P_c(4457)$, where
$(\mathbb{D}_1,\mathbb{D}_2)=(58,-31)$ $\textrm{GeV}^{-2}$ in this
case. ``$\times$" means no binding solution (in units of
MeV).}\label{BindingEnergies} \setlength{\tabcolsep}{2.45mm} {
\begin{tabular}{c|ccccccc}
\hline\hline
$\Delta E $&$[\Sigma_c\bar{D}]_{\frac{1}{2}}$&$[\Sigma_c\bar{D}^\ast]_{\frac{1}{2}}$&$[\Sigma_c\bar{D}^\ast]_{\frac{3}{2}}$&$[\Sigma_c^\ast\bar{D}]_{\frac{3}{2}}$&$[\Sigma_c^\ast\bar{D}^\ast]_{\frac{1}{2}}$&$[\Sigma_c^\ast\bar{D}^\ast]_{\frac{3}{2}}$&$[\Sigma_c^\ast\bar{D}^\ast]_{\frac{5}{2}}$\\
\hline
Without $\Lambda_c$&$-29.05$&$-6.84$&$-2.98$&$-34.30$&$-0.16$&$\times$&$\times$\\
\hline
With $\Lambda_c$&$-4.60$&$-22.48$&$-3.19$&$-34.51$&$-14.34$&$-3.40$&$-0.30$\\
\hline
I.S.&$-7.24$&$-1.47$&$-17.44$&$-40.88$&$\times$&$-0.24$&$-11.20$\\
\hline\hline
\end{tabular}
}
\end{table*}
\subsection{Role of the $\Lambda_c$}
As mentioned above, we cannot give a good description for the $P_c$s
if we only consider the spin partners of $\Sigma_c^{(\ast)}$ and
$\bar{D}^{(\ast)}$ in the two-pion-exchange diagrams. In this part,
we are going to include the contributions of $\Lambda_c$ in the
loops. Since both the $\Sigma_c$ and $\Sigma_c^\ast$ can decay into
$\Lambda_c\pi$, the strong couplings between $\Sigma_c^{(\ast)}$ and
$\Lambda_c\pi$ should not be neglected.

The result with the $\Lambda_c$ being included is illustrated in
figure~\ref{Results_WithoutWith_Lambdac}(b), from which we find that
there exists a very large overlap among the three colored bands. The
small triangle  surrounded by three straight lines just locates in
the overlap. Besides, the intersection points between two of the
three solid lines are very close to each other, and they almost meet
at a point if we consider the experimental errors. In other words,
the three $P_c$ can be synchronously reproduced in this case. The
result in figure~\ref{Results_WithoutWith_Lambdac}(b) is in good
agreement with the experimental data.

We choose the values $\mathbb{D}_1=52$ $\textrm{GeV}^{-2}$ and
$\mathbb{D}_2=-4$ $\textrm{GeV}^{-2}$ in the center of the small
triangle to give the binding energies and effective potentials of
the $[\Sigma_c^{(\ast)}\bar{D}^{(\ast)}]_J$ systems. The binding
energies in this case are shown in the third row of
table~\ref{BindingEnergies}, from which we get the results for the
$[\Sigma_c\bar{D}]_{\frac{1}{2}}$,
$[\Sigma_c\bar{D}^\ast]_{\frac{1}{2}}$ and
$[\Sigma_c\bar{D}^\ast]_{\frac{3}{2}}$ systems that are consistent
with the experimental data. One may note that the
$[\Sigma_c^\ast\bar{D}]_{\frac{3}{2}}$ system is always deeper bound
compared with the other systems regardless of the contribution of
$\Lambda_c$. The $[\Sigma_c^\ast\bar{D}]_{\frac{3}{2}}$ system might
correspond to the previously reported $P_c(4380)$
~\cite{Aaij:2015tga}. Therefore, we urge the experimentalists to
reanalyze the data to see whether $P_c(4380)$ is the most deeply
bound one in the $[\Sigma_c^{(\ast)}\bar{D}^{(\ast)}]_J$ systems.
Moreover, the bound states of the
$[\Sigma_c^\ast\bar{D}^\ast]_{J}~(J=\frac{1}{2}, \frac{3}{2},
\frac{5}{2})$ systems are also predicted. Their binding energies are
determined to be $-14.34$ MeV, $-3.40$ MeV and $-0.30$ MeV,
respectively.

\begin{figure*}
\begin{minipage}[t]{0.33\linewidth}
\centering
\includegraphics[width=\columnwidth]{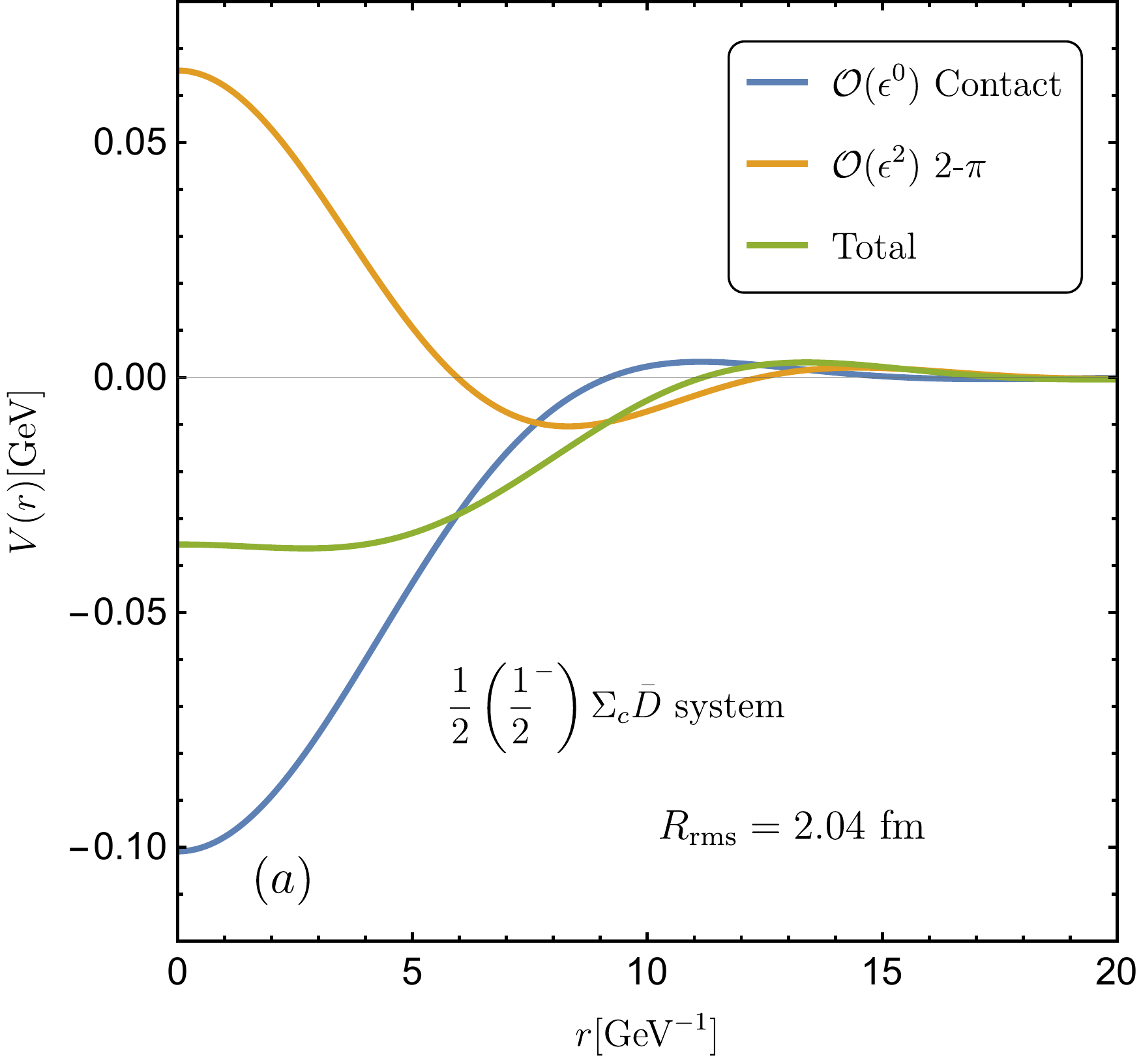}
\end{minipage}%
\begin{minipage}[t]{0.33\linewidth}
\centering
\includegraphics[width=\columnwidth]{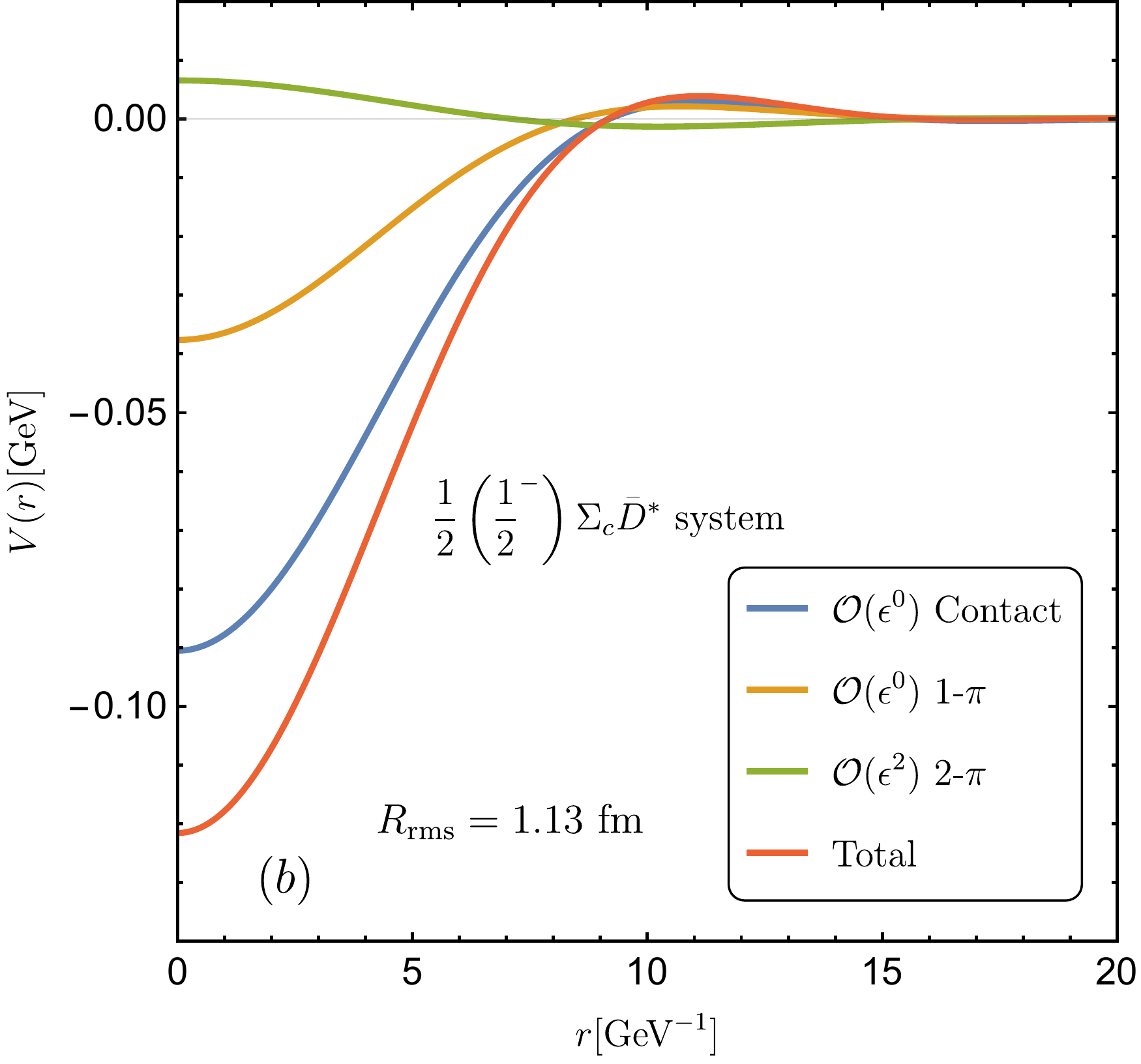}
\end{minipage}
\begin{minipage}[t]{0.33\linewidth}
\centering
\includegraphics[width=\columnwidth]{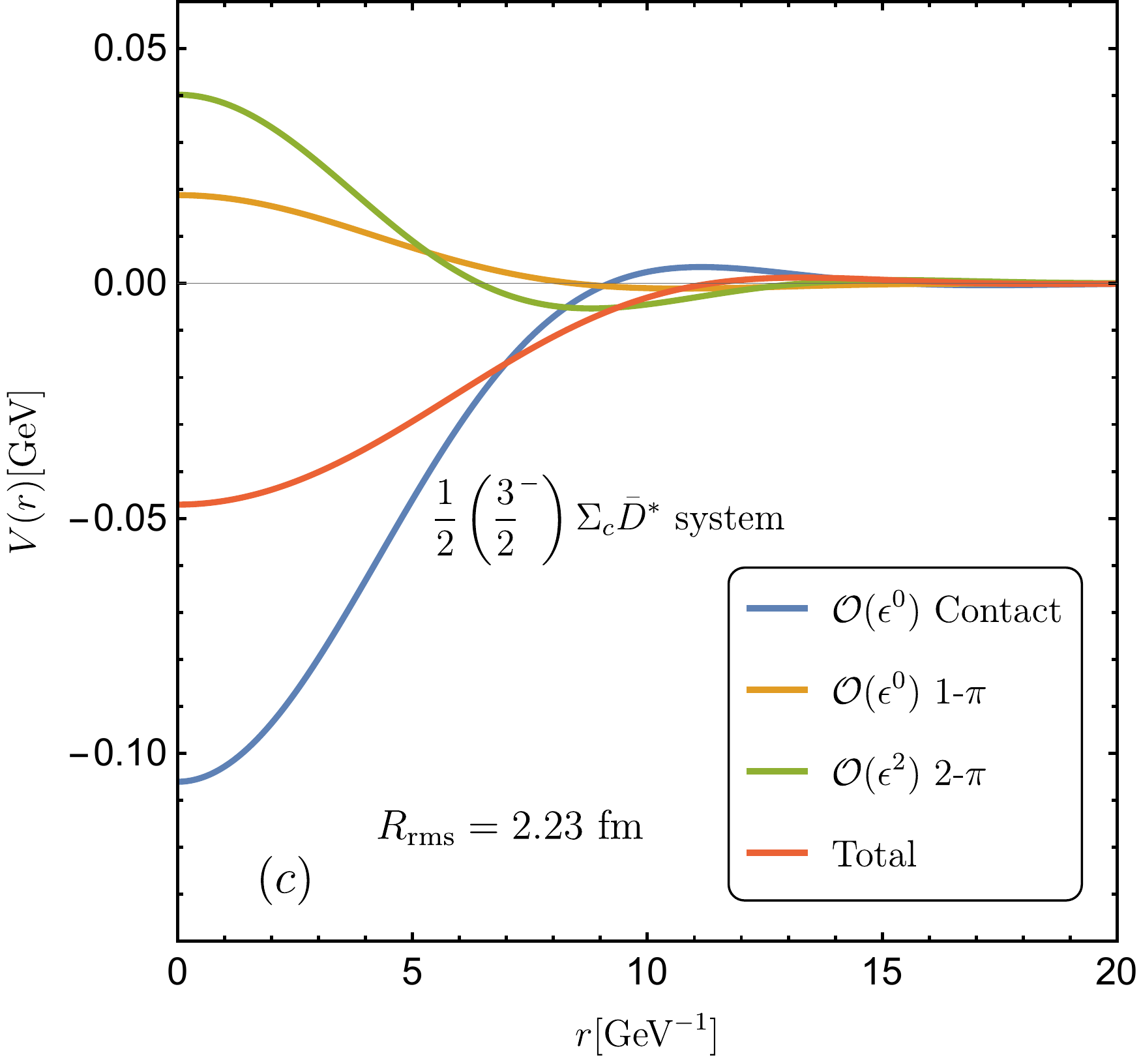}
\end{minipage}
\caption{The effective potentials of the $\Sigma_c\bar{D}^{(\ast)}$
systems. Their $I(J^P)$ are marked in each subfigure. The potentials
are obtained with the cutoff parameter $\Lambda=0.5$ GeV, and the
LECs $\mathbb{D}_1=52$ $\textrm{GeV}^{-2}$, $\mathbb{D}_2=-4$
$\textrm{GeV}^{-2}$. The $R_{\textrm{rms}}$ in each subfigure
denotes the root-mean-square radius of the corresponding
system.\label{Potential1}}
\end{figure*}

\begin{figure*}
\begin{center}
\begin{minipage}[t]{0.4\linewidth}
\centering
\includegraphics[width=\columnwidth]{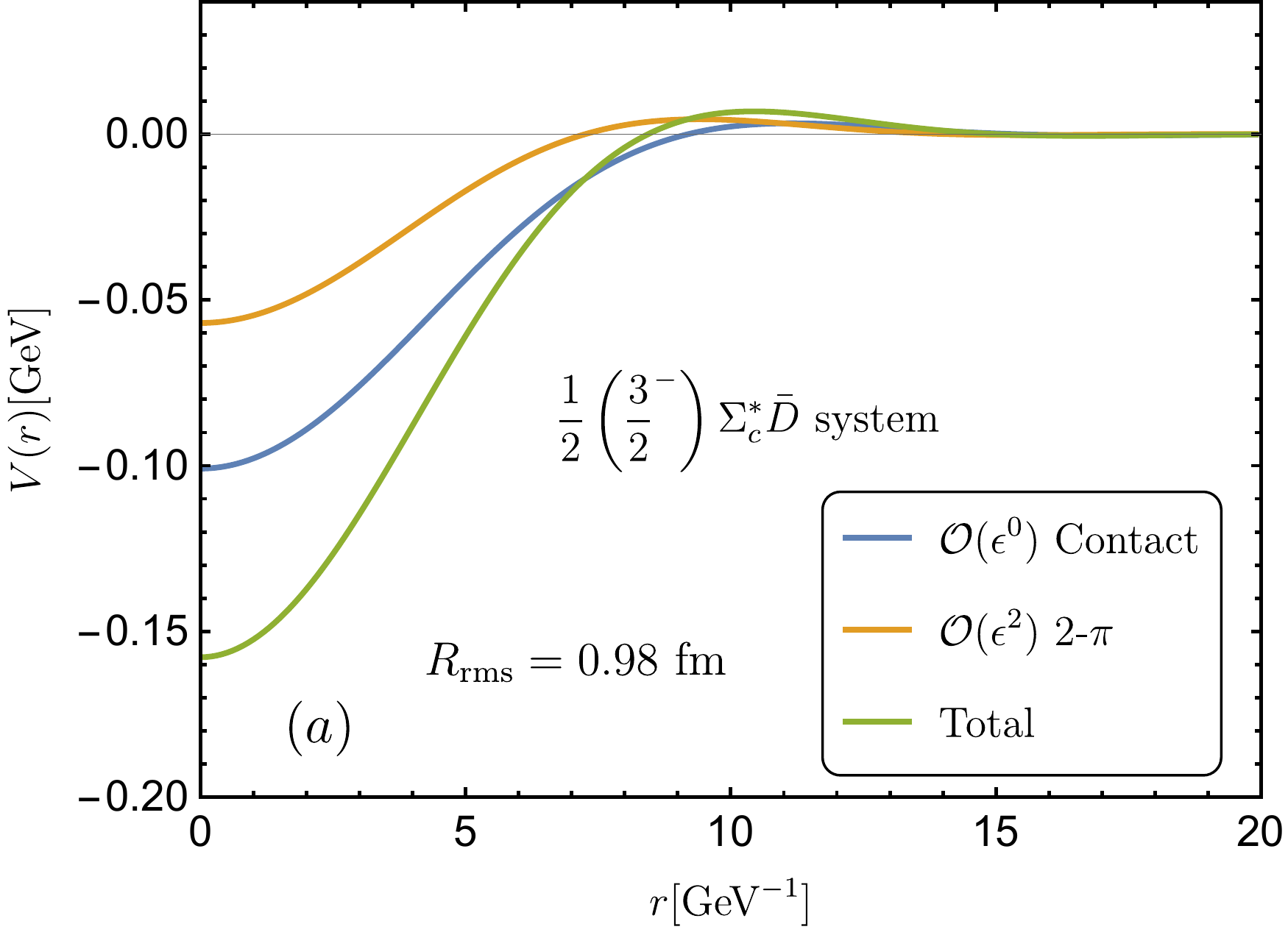}
\end{minipage}%
\hspace{0.5in}
\begin{minipage}[t]{0.4\linewidth}
\centering
\includegraphics[width=\columnwidth]{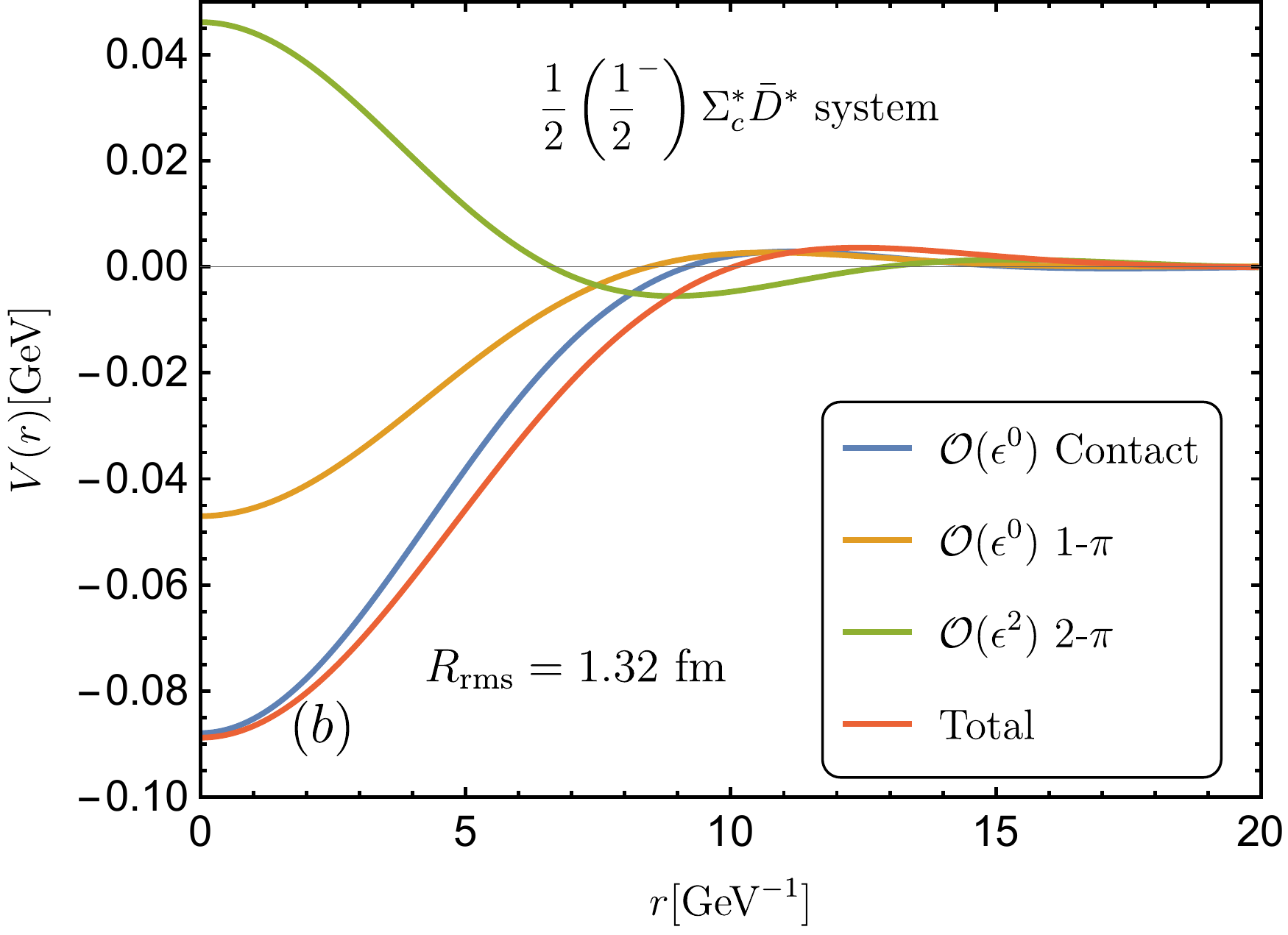}
\end{minipage}
\begin{minipage}[t]{0.4\linewidth}
\centering
\includegraphics[width=\columnwidth]{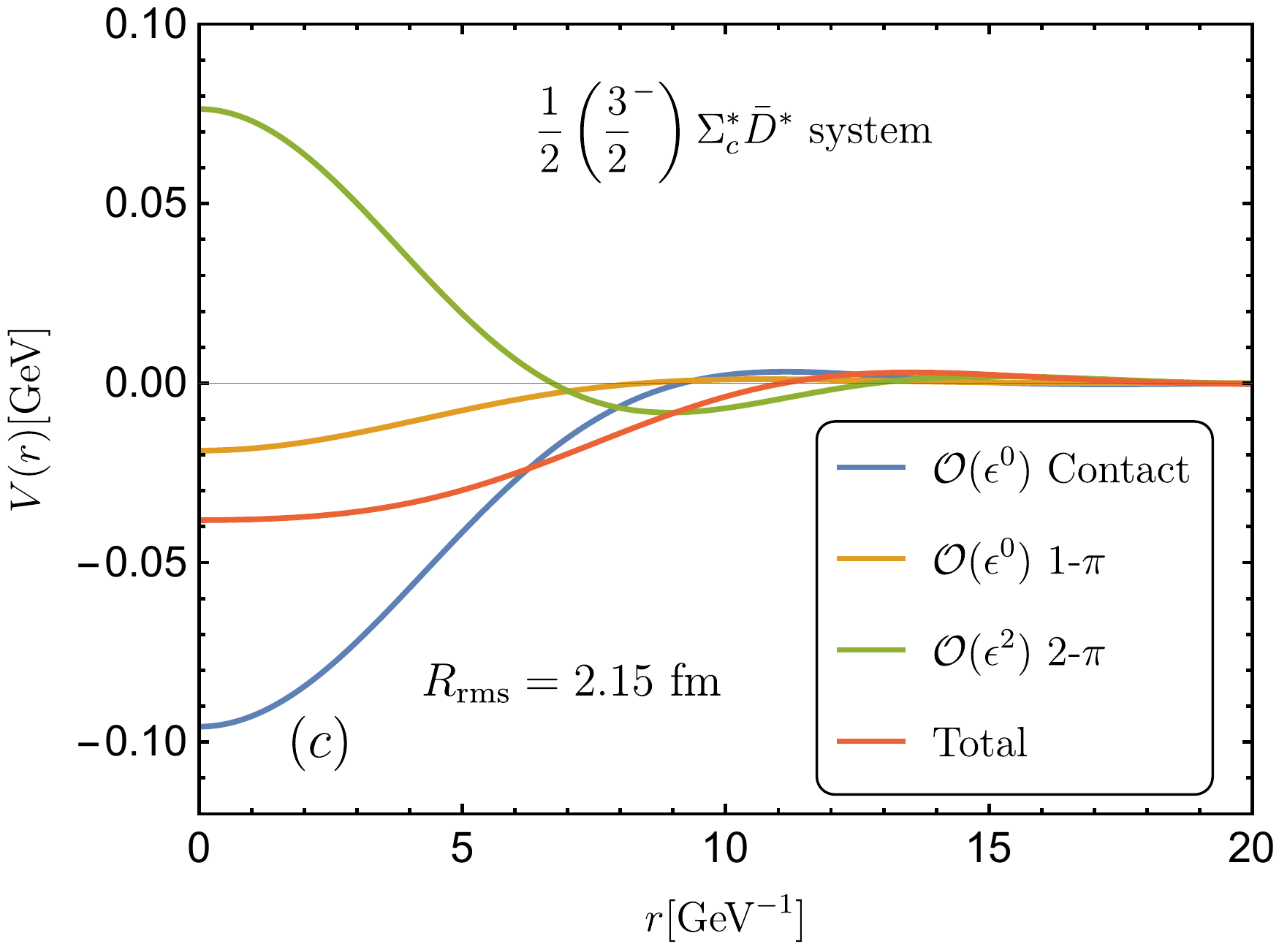}
\end{minipage}%
\hspace{0.5in}
\begin{minipage}[t]{0.4\linewidth}
\centering
\includegraphics[width=\columnwidth]{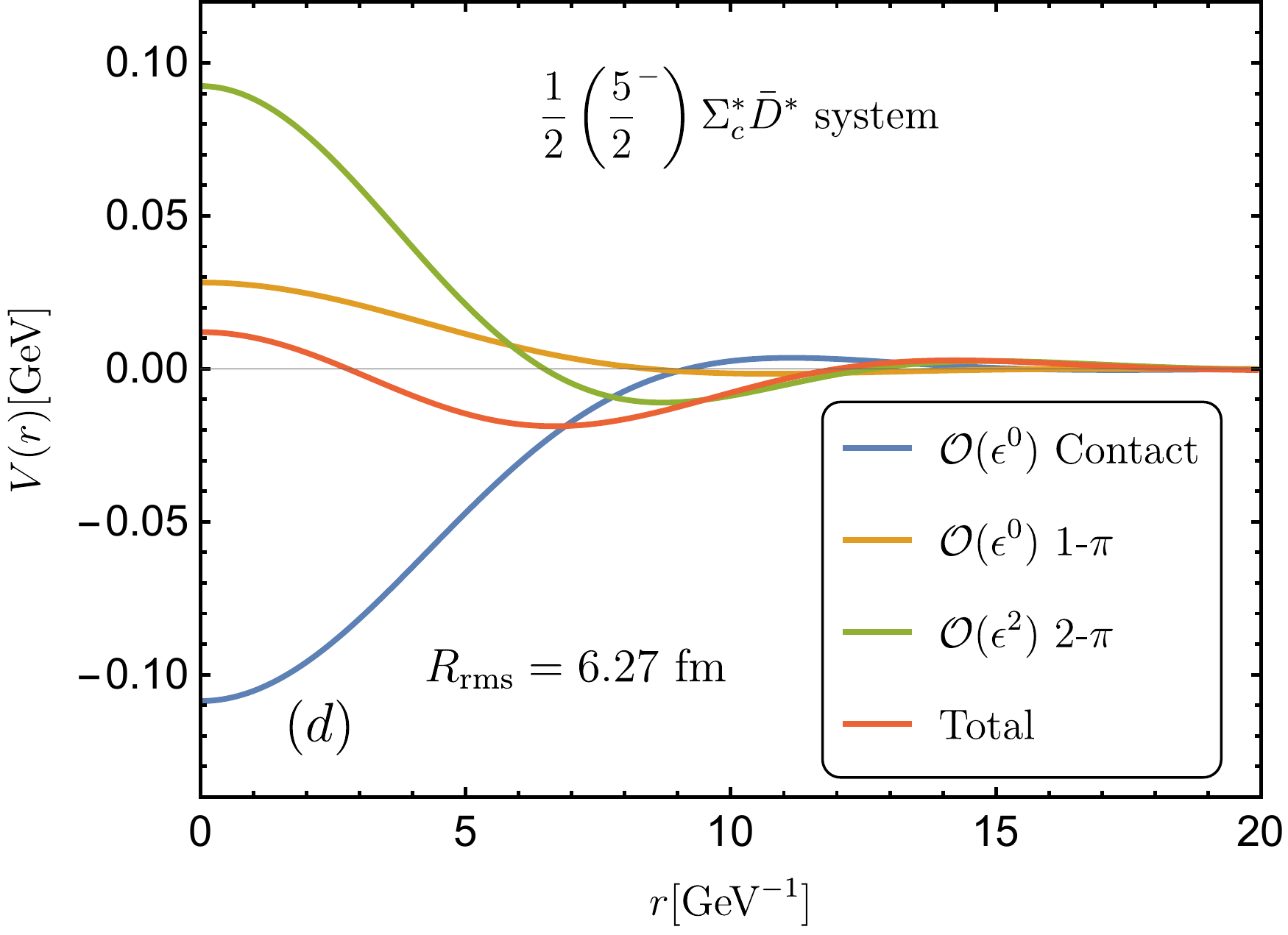}
\end{minipage}
\caption{The effective potentials of the
$\Sigma_c^\ast\bar{D}^{(\ast)}$ systems. Their $I(J^P)$ are marked
in each subfigure. The potentials are obtained with the cutoff
parameter $\Lambda=0.5$ GeV, and the LECs $\mathbb{D}_1=52$
$\textrm{GeV}^{-2}$, $\mathbb{D}_2=-4$ $\textrm{GeV}^{-2}$. The
$R_{\textrm{rms}}$ in each subfigure denotes the root-mean-square
radius of the corresponding system.\label{Potential2}}
\end{center}
\end{figure*}

The effective potentials of the $\Sigma_c\bar{D}^{(\ast)}$ and
$\Sigma_c^{\ast}\bar{D}^{(\ast)}$ systems are shown in
figures~\ref{Potential1} and~\ref{Potential2}, respectively. In the
following, we analyze their behaviors in detail.

{\it$\Sigma_c\bar{D}^{(\ast)}$ systems}: The results in
figures~\ref{Potential1}(a),~\ref{Potential1}(b)
and~\ref{Potential1}(c) all demonstrate that the contact term
supplies the very strong attractive potential. From
eq.~\eqref{Contact_Potential} we know that
$\mathcal{O}(\varepsilon^0)$ contact term for the $\Sigma_c\bar{D}$
system only contains the central potential, while the spin-spin
contact term appears for the $\Sigma_c\bar{D}^\ast$ system. Thus
their difference is mainly caused by the spin-spin interaction.
Meanwhile, the small difference between their
$\mathcal{O}(\varepsilon^0)$ contact potentials indicates that the
spin-spin interaction is rather weak and only serves as the
hyperfine splittings.

There is no one-pion-exchange potential for the $\Sigma_c\bar{D}$
due to the vanishing $\bar{D}\bar{D}\pi$ vertex. The
one-pion-exchange potential for the
$[\Sigma_c\bar{D}^\ast]_{\frac{1}{2}}$ is attractive, while it is
repulsive for the $[\Sigma_c\bar{D}^\ast]_{\frac{3}{2}}$ because of
the different signs of the matrix element of the spin-spin operator
for the spin-$1\over 2$ and spin-$3\over 2$ states.

The contributions of the two-pion-exchange potentials for the
$[\Sigma_c\bar{D}]_{\frac{1}{2}}$ and
$[\Sigma_c\bar{D}^\ast]_{\frac{3}{2}}$ are significant, but it is
marginal for the $[\Sigma_c\bar{D}^\ast]_{\frac{1}{2}}$.
Nevertheless, one can still find the similar behaviours of the
two-pion-exchange potentials, which are repulsive at the short
range, but become weakly attractive at the intermediate range. This
is the typical feature of the nuclear force~\cite{Machleidt:2011zz}.

Finally, the total potentials of the $[\Sigma_c\bar{D}^{(\ast)}]_J$
systems are fully attractive. The subtle interplay among the short-,
intermediate- and long-range interactions yields the experimentally
observed $P_c(4312)$, $P_c(4440)$ and $P_c(4457)$.

{\it$\Sigma_c^{(\ast)}\bar{D}^{\ast}$ systems}: The results in the
figure~\ref{Potential2} are also very interesting, since they are
related with the previously reported $P_c(4380)$ and other
unobserved states. Recalling the binding energies in
table~\ref{BindingEnergies}, the result of $\Sigma_c^\ast\bar{D}$ is
about eight times larger than that of the $\Sigma_c\bar{D}$. These
two systems have the same $\mathcal{O}(\varepsilon^0)$ contact
potentials [e.g., see eq.~\eqref{Contact_Potential}]. The
one-pion-exchange contribution vanishes for both systems. Thus the
difference can only arise from the two-pion-exchange potentials, as
shown in figure~\ref{Potential2}(a). One can notice the behaviors of
the two-pion-exchange potential for the $\Sigma_c^\ast\bar{D}$ is
attractive at the short-range and weakly repulsive at the
intermediate-range, which is in contrast to that of the
$\Sigma_c\bar{D}$ [e.g., see figure~\ref{Potential1}(a)]. Therefore,
if one only considers the $\mathcal{O}(\varepsilon^0)$ contribution,
it is unlikely to obtain the significant difference between the
$\Sigma_c\bar{D}$ and $\Sigma_c^\ast\bar{D}$ systems. So we eagerly
hope the future analysis at LHCb can help us confirm this
observation.

The effective potentials of the $[\Sigma_c^{\ast}\bar{D}^{\ast}]_J$
systems are very similar to those of the
$[\Sigma_c\bar{D}^{\ast}]_J$ systems. For instance, the
$\mathcal{O}(\varepsilon^0)$ contact potentials are attractive. The
one-pion-exchange potentials vary dramatically with the total spins.
The two-pion-exchange potentials have the similar line-shape as the
nuclear force. Although the total potentials are all attractive, the
$[\Sigma_c^{\ast}\bar{D}^{\ast}]_{\frac{5}{2}}$ system is very
shallowly bound with root-mean-square radius $6.27$ fm.

\begin{figure*}
\begin{minipage}[t]{0.33\linewidth}
\centering
\includegraphics[width=\columnwidth]{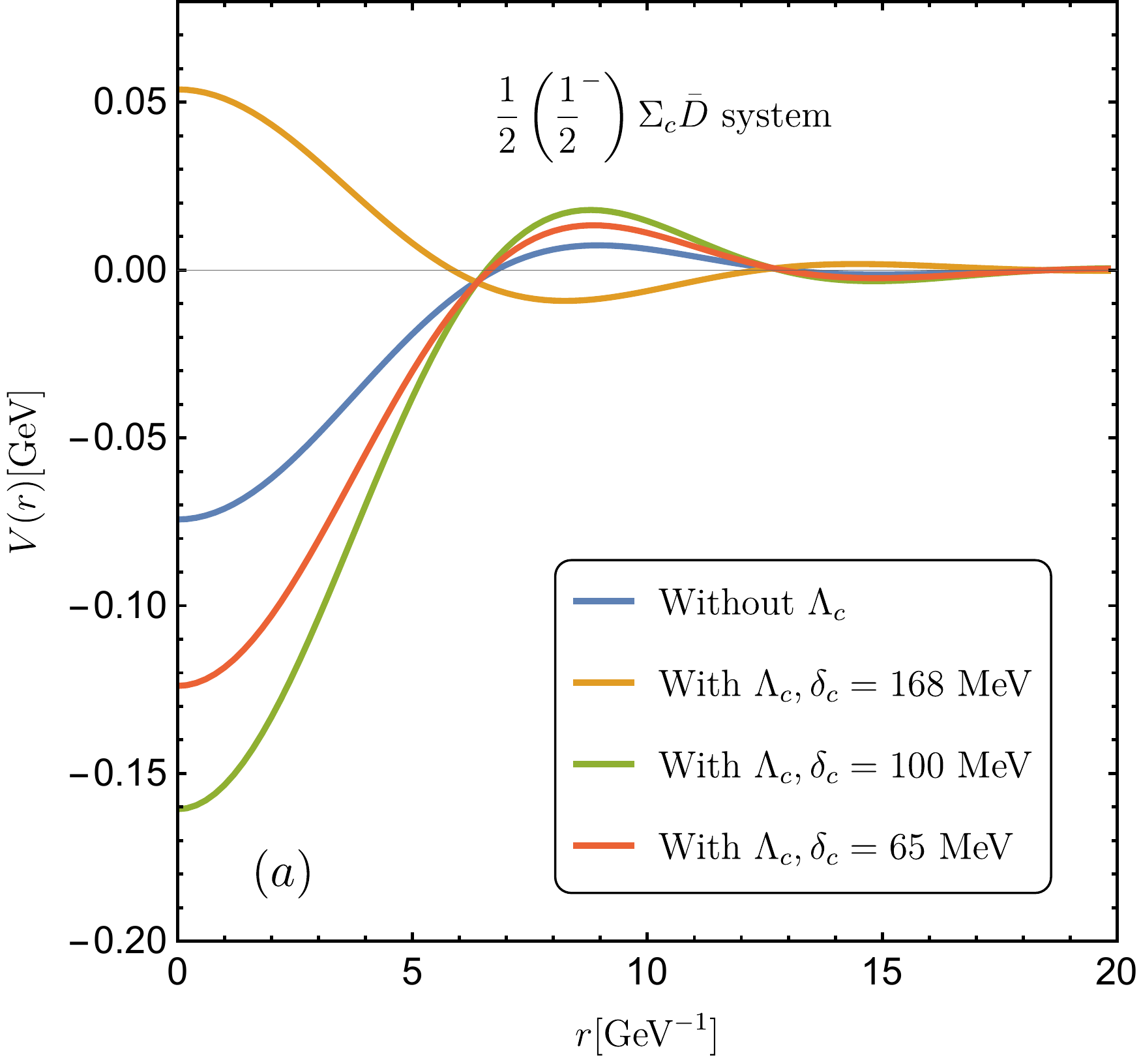}
\end{minipage}%
\begin{minipage}[t]{0.33\linewidth}
\centering
\includegraphics[width=\columnwidth]{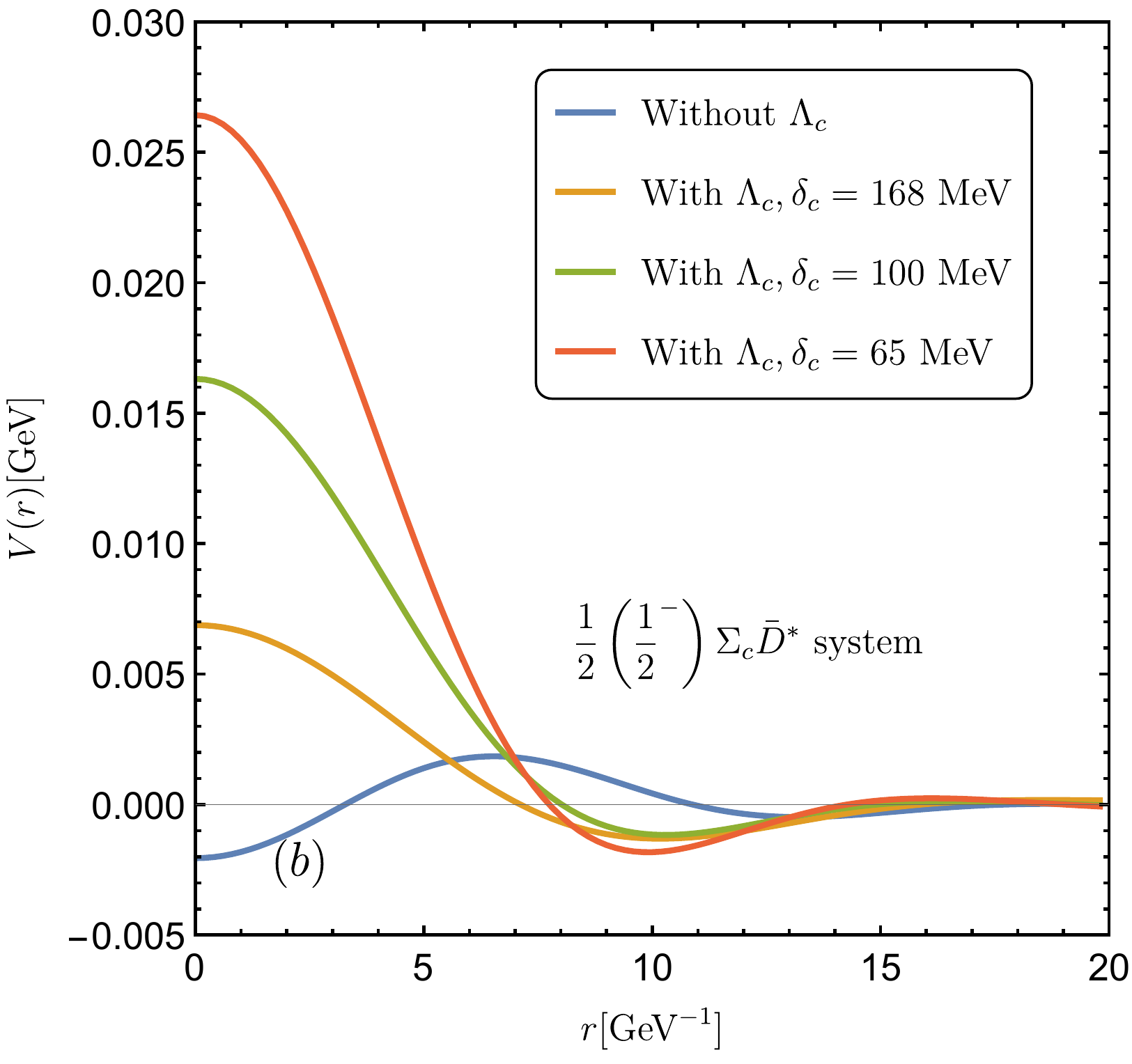}
\end{minipage}
\begin{minipage}[t]{0.33\linewidth}
\centering
\includegraphics[width=\columnwidth]{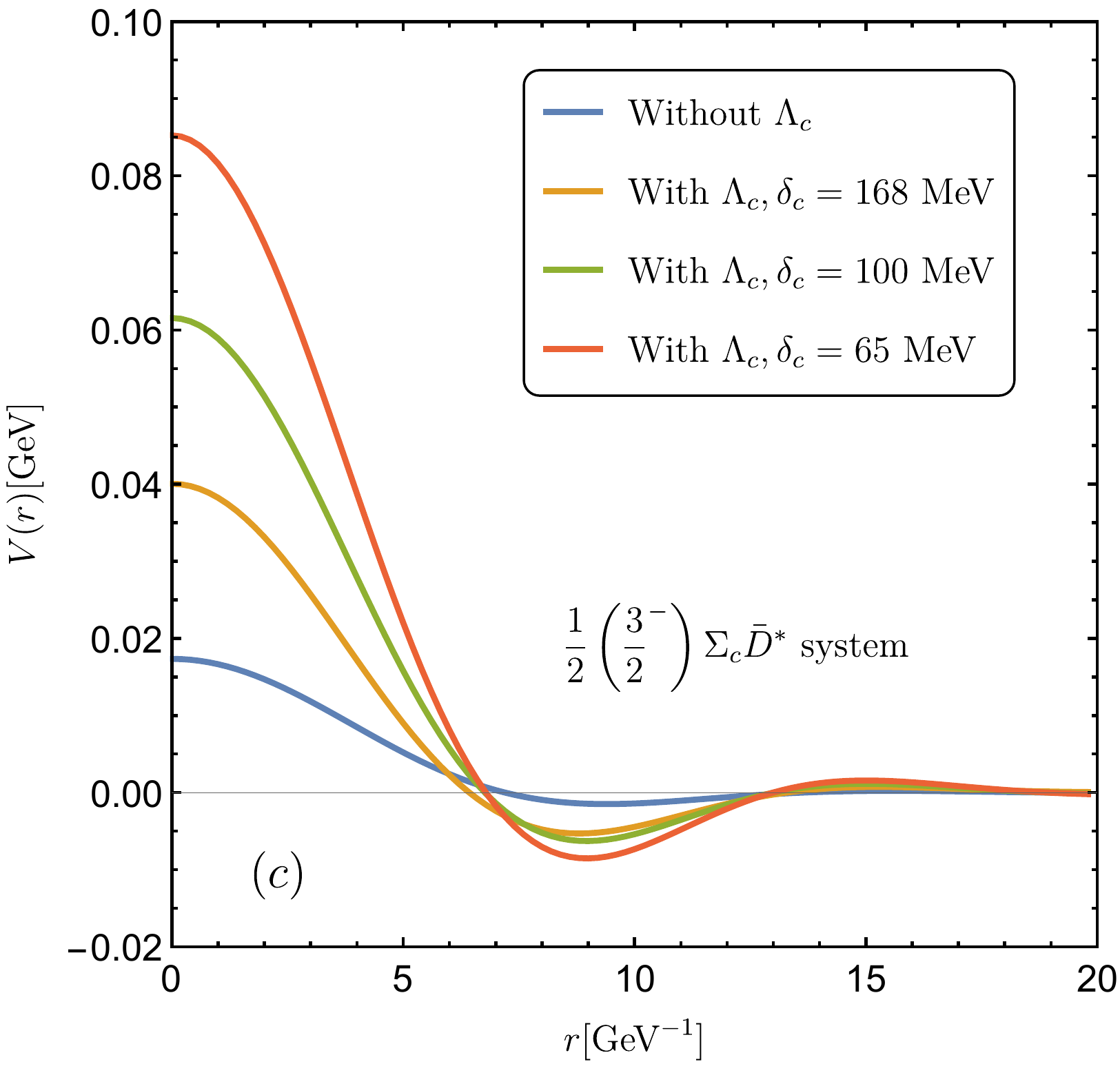}
\end{minipage}
\caption{The variations of the two-pion-exchange potentials for the
$\Sigma_c\bar{D}^{(\ast)}$ systems in the cases of without and with
the $\Lambda_c$. Their $I(J^P)$ are marked in each subfigure. The
dependence on the mass splitting $\delta_c$ is also
illustrated.\label{TPEPotential_Lambdac}}
\end{figure*}

The two-pion-exchange potentials for the $\Sigma_c\bar{D}^{(\ast)}$
systems in different cases are displayed in
figure~\ref{TPEPotential_Lambdac}. We can read the significant
differences when we include the $\Lambda_c$ and not, or vary the
mass splitting $\delta_c$ for the $[\Sigma_c\bar{D}]_{\frac{1}{2}}$
and $[\Sigma_c\bar{D}^\ast]_{\frac{1}{2}}$ systems. We take the
$[\Sigma_c\bar{D}]_{\frac{1}{2}}$ system as an example. The
two-pion-exchange potential is attractive if we do not consider the
$\Lambda_c$, while it becomes repulsive when the $\Lambda_c$ is
involved. This can well explain why the binding of the
$[\Sigma_c\bar{D}]_{\frac{1}{2}}$ state is much deeper without the
$\Lambda_c$ (see table~\ref{BindingEnergies}). The magnitude of the
change from the minimum to the maximum in these two cases is about
$120$ MeV, which is even larger than the minimum of the total
potential [see figure~\ref{Potential1}(a)]. The enhancement is
mainly generated by the accidental degeneration of the
$\Sigma_c\bar{D}$ and $\Lambda_c\bar{D}^\ast$ systems, since the
contribution of the box diagram $(B_{1.1})$ is proportional to
$1/(\delta_c-\delta_b)$, where $\delta_c-\delta_b\simeq28$ MeV is
tiny. Another reason that may cause the enhancement is the
contributing diagrams with the $\Lambda_c$ are only $(T_{1.3})$ and
$(B_{1.1})$. Unlike the $\Sigma_c\bar{D}^\ast$ system, the
accidental cancelations among several diagrams cannot happen. In
other words, the $\Lambda_c$ indeed plays a crucial role in the
formation of the $P_c(4312)$.

For the $[\Sigma_c\bar{D}^\ast]_{\frac{1}{2}}$ system, since the
whole contribution of the two-pion-exchange potential is much weaker
than the $\mathcal{O}(\varepsilon^0)$ contact term [see
figure~\ref{Potential1}(b)], the influence of $\Lambda_c$ on this
state is not so apparent as in $\Sigma_c\bar{D}$. However, it is
still very important to the existence of $P_c(4457)$ and the
possible $[\Sigma_c^\ast\bar{D}^\ast]_J$ bound states (e.g., see the
data in table~\ref{BindingEnergies}).

In figure~\ref{TPEPotential_Lambdac}, we also show the dependence of
the two-pion-exchange potentials on the mass splitting $\delta_c$.
One can see that they are very sensitive to the $\delta_c$. The loop
integrals generally contain two structures. One is the analytic
term, which is the polynomials of the $m_\pi^2$, $\bm{q}^2$,
$\delta^2$, etc.. Another one is the nonanalytic term, which
comprises the typical multivalued functions, such as $\log
\mathcal{X}$ and $\sqrt{\mathcal{X}}$ ($\mathcal{X}$ is the
polynomials of the $m_\pi^2$, $\bm{q}^2$, $\delta^2$.). The physical
value of the $\delta_c$ is about $168$ MeV, which is larger than the
pion mass $m_\pi$. We then decrease its value to $100$ MeV and $65$
MeV. One can anticipate the dependence on $\delta$ is regular if the
terms that make up the potential are only polynomials, but the
variation trend in figure~\ref{TPEPotential_Lambdac} is irregular.
This phenomenon indicates the nonanalytic terms can distort the
$\mathcal{O}(\varepsilon^2)$ potentials, which are vital to the
formations of the $P_c$ states. The contributions of the nonanalytic
terms incorporate the complicated light quark dynamics, which are
almost impossible to estimate from quark models.

\begin{figure*}
\begin{minipage}[t]{0.5\linewidth}
\centering
\includegraphics[width=\columnwidth]{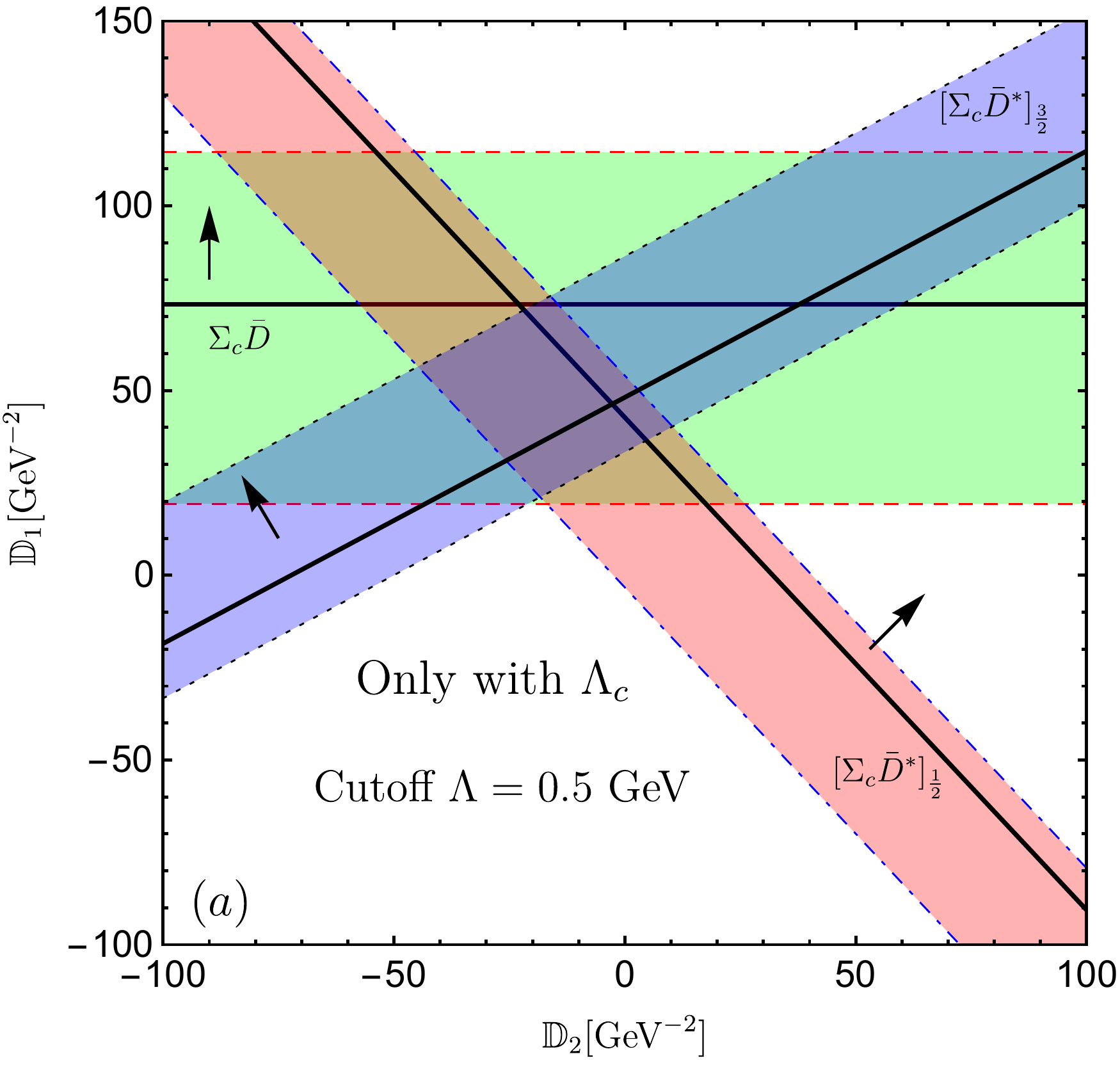}
\end{minipage}%
\begin{minipage}[t]{0.5\linewidth}
\centering
\includegraphics[width=\columnwidth]{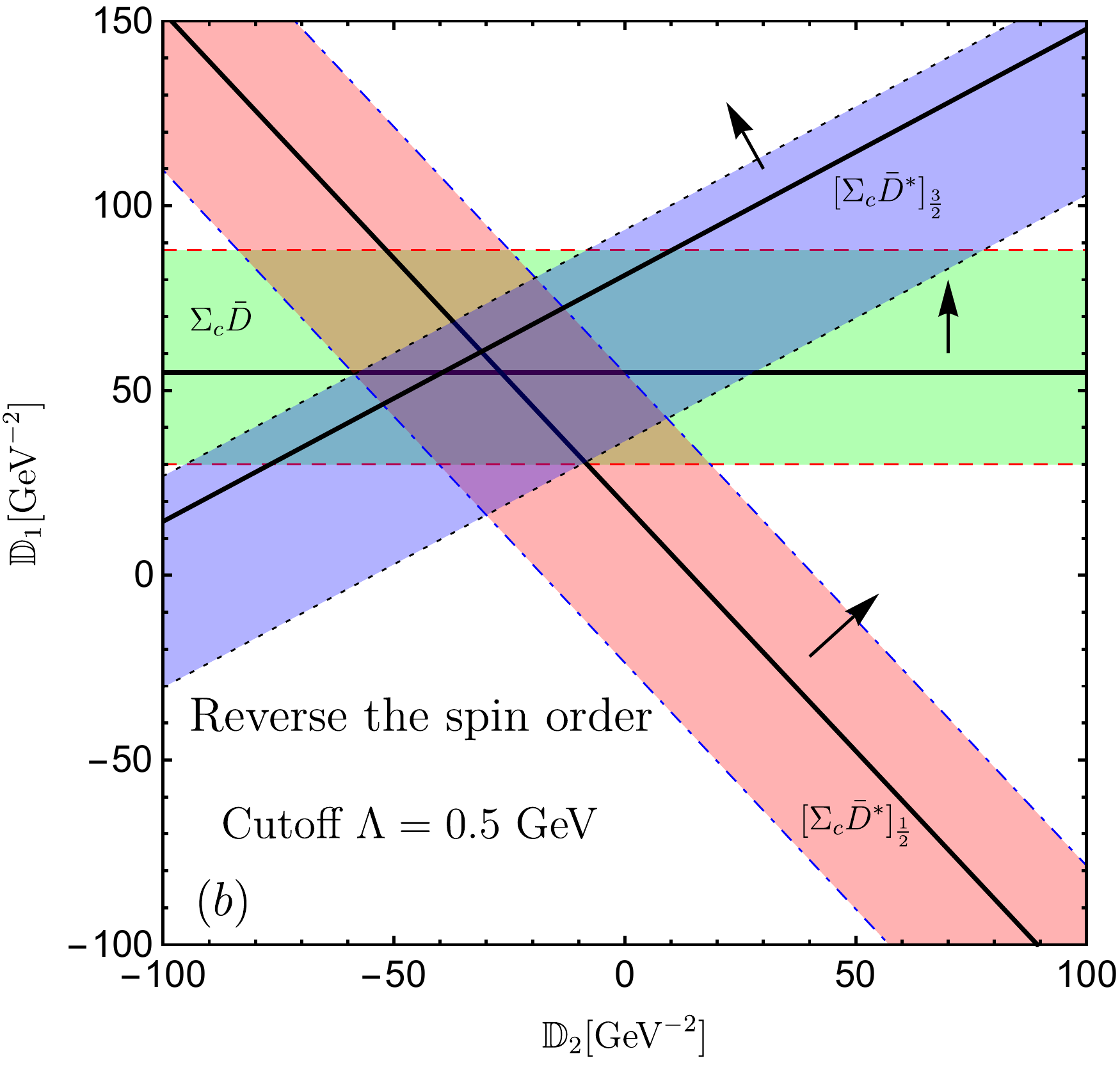}
\end{minipage}
\caption{The dependence of the binding energies of the three $P_c$
states on the redefined LECs $\mathbb{D}_1$ and $\mathbb{D}_2$ in
different cases. Figure ($a$) gives the result that only considering
the contributions of $\Lambda_c$ in the two-pion-exchange diagrams.
Figure ($b$) shows the result when interchanging the spins of
$P_c(4440)$ and $P_c(4457)$. The notations are the same as those in
figure~\ref{Results_WithoutWith_Lambdac}.\label{Results_Sigmac_SpinOrder}}
\end{figure*}

After the above discussions, one may wonder whether it is possible
to reproduce the three $P_c$s simultaneously if we only consider the
contribution of the $\Lambda_c$. The result in this case is given in
figure~\ref{Results_Sigmac_SpinOrder}(a), which is also
unsatisfactory as in the case of
figure~\ref{Results_WithoutWith_Lambdac}(a). Therefore, both the
$\Lambda_c$ and the spin partners of the $\Sigma_c^{(\ast)}$ and
$\bar{D}^{(\ast)}$ are indispensable. Their subtle interaction leads
to the synchronous emergence of the $P_c(4312)$, $P_c(4440)$ and
$P_c(4457)$.

The complete mass spectra of the hidden-charm molecular pentaquarks
are shown in figure~\ref{MassSpectra}(a). We see that the
$P_c(4312)$, $P_c(4440)$ and $P_c(4457)$ can be well interpreted as
the $[\Sigma_c\bar{D}]_{\frac{1}{2}}$,
$[\Sigma_c\bar{D}^{\ast}]_{\frac{1}{2}}$ and
$[\Sigma_c\bar{D}^{\ast}]_{\frac{3}{2}}$ molecules. $P_c(4380)$
might be the deeper bound $[\Sigma_c^\ast\bar{D}]_{\frac{3}{2}}$
molecules. There are also other possible $P_c$s composed of the
$[\Sigma_c^\ast\bar{D}^\ast]_J$. Future search for these states at
LHCb is very important for establishing a complete family of the
hidden-charm pentaquarks.

\subsection{An episode: interchanging the spins of $P_c(4440)$ and $P_c(4457)$}
The $J^P$ quantum numbers of the $P_c(4312)$, $P_c(4440)$ and
$P_c(4457)$ are not determined yet~\cite{Aaij:2019vzc}. The
theoretically favored $J^P$ for $P_c(4440)$ and $P_c(4457)$ in this
paper and some previous
works~\cite{Meng:2019ilv,Liu:2019tjn,Chen:2019asm,Xiao:2019aya,He:2019ify}
are $\frac{1}{2}^-$ and $\frac{3}{2}^-$, respectively. Nevertheless,
in some recent
works~\cite{Yamaguchi:2019seo,Valderrama:2019chc,Liu:2019zvb}, a new
conjecture, that the $J^P=\frac{3}{2}^-$ for $P_c(4440)$ and
$\frac{1}{2}^-$ for $P_c(4457)$, is proposed. In this subsection, we
investigate the possibility of this spin assignment.

The result of interchanging the spin assignment of $P_c(4440)$ and
$P_c(4457)$ is given in figure~\ref{Results_Sigmac_SpinOrder}(b).
Marvelously, the result is comparable with the one in
figure~\ref{Results_WithoutWith_Lambdac}(b), i.e., it seems this
assignment can well describe the experimental data, likewise.
However, one shall note that the values of the LECs
$(\mathbb{D}_1,\mathbb{D}_2)$ in the center of the small triangle
are $(58,-31)$ $\textrm{GeV}^{-2}$, while these values in
figure~\ref{Results_WithoutWith_Lambdac}(b) are $(52,-4)$
$\textrm{GeV}^{-2}$. The shift of $\mathbb{D}_1$ in these two cases
is small, but the $\mathbb{D}_2$ in the first case is about eight
times larger than that of the latter one. One has to largely enhance
the contribution of the $\mathcal{O}(\varepsilon^0)$ spin-spin
interaction to reverse the {\it canonical order}\footnote{An
empirical rule given by the hadron mass spectra is that the higher
spin state always has the larger mass~\cite{Tanabashi:2018oca}.} of
the spins of $P_c(4440)$ and $P_c(4457)$.

The binding energies of the $[\Sigma_c^{(\ast)}\bar{D}^{(\ast)}]_J$
systems with the $(\mathbb{D}_1,\mathbb{D}_2)=(58,-31)$
$\textrm{GeV}^{-2}$ are listed in the fourth row of
table~\ref{BindingEnergies}. Although the binding energies of the
$[\Sigma_c\bar{D}]_{\frac{1}{2}}$,
$[\Sigma_c\bar{D}^\ast]_{\frac{1}{2}}$ and
$[\Sigma_c\bar{D}^\ast]_{\frac{3}{2}}$ can match the ones of
$P_c(4312)$, $P_c(4457)$ and $P_c(4440)$, other predictions are
different from those with the previous spin assignment. The bound
$[\Sigma_c^\ast\bar{D}^\ast]_{\frac{1}{2}}$ state does not exist,
the $[\Sigma_c^\ast\bar{D}^\ast]_{\frac{3}{2}}$ state is very
shallowly bound, and the binding of the
$[\Sigma_c^\ast\bar{D}^\ast]_{\frac{5}{2}}$ state is much deeper.
This result is very theatrical to some extent, since the lowest spin
state of the $\Sigma_c^\ast\bar{D}^\ast$ does not exist. However,
this phenomenon does not occur in the leading order effective field
theory where there does not exist the repulsive core from the
two-pion-exchange diagrams.

The information from figure~\ref{Results_WithoutWith_Lambdac}(b)
indicates the $\mathbb{D}_1$ and $\mathbb{D}_2$ always have the
opposite sign, and the ratio of their absolute values
$\mathcal{R}_{12}=|\mathbb{D}_1|/|\mathbb{D}_2|\simeq13$. We notice
the correspondence and consistence with the $N$-$N$ system. The
leading order contact Lagrangian for the $N$-$N$ system
reads~\cite{Machleidt:2011zz},
\begin{eqnarray}\label{Lag_NN}
\mathcal{L}_{NN}^{(0)}=-\frac{1}{2}C_S(\bar{N}N)(\bar{N}N)-\frac{1}{2}C_T(\bar{N}\boldsymbol{\sigma}N)\cdot(\bar{N}\boldsymbol{\sigma}N),
\end{eqnarray}
where $C_S$ and $C_T$ are two independent LECs. One would see that
they respectively correspond to the $\mathbb{D}_1$ and
$\mathbb{D}_2$ in our work if we write out the
$\mathcal{O}(\varepsilon^0)$ contact potential of the $N$-$N$
system,
\begin{eqnarray}
\mathcal{V}_{NN}=C_S+C_T\boldsymbol{\sigma}\cdot\boldsymbol{\sigma}.
\end{eqnarray}
The values of $C_S$ and $C_T$ have been precisely determined by
fitting the $np$ scattering phase shift at the
next-to-next-to-next-to-leading order of chiral perturbation
theory~\cite{Entem:2003ft}. For the $np$ system, which gives
\footnote{See the data in table F.1 of ref.~\cite{Machleidt:2011zz},
where the similar regulator function as adopted in this work is
used, meanwhile, the cutoff $\Lambda$ is also chosen to be $0.5$
GeV.}
\begin{eqnarray}
C_S=-100.28~\textrm{GeV}^{-2},\qquad C_T=5.61~\textrm{GeV}^{-2}.
\end{eqnarray}
If absorbing the minus sign of eq.~\eqref{Lag_NN} into the $C_S$ and
$C_T$, one would see the redefined $C_S$ and $C_T$ share the same
sign with the $\mathbb{D}_1$ and $\mathbb{D}_2$, correspondingly.
Meanwhile, the ratio of the absolute values for $C_S$ and $C_T$
gives $\mathcal{R}_{ST}=|C_S|/|C_T|\simeq18$, which is compatible
with the $\mathcal{R}_{12}$ for $\mathbb{D}_1$ and $\mathbb{D}_2$.
However, this ratio for the case of interchanging the spins of
$P_c(4440)$ and $P_c(4457)$ is about $1.9$, which is one order of
magnitude smaller than the $\mathcal{R}_{ST}$, because of the
spin-spin term in the contact potential is immoderately enhanced.

On the one hand, from the point of potential model, the spin-spin
term is suppressed by the factor
$1/(m_{\Sigma_c^{(\ast)}}m_{D^{(\ast)}})$ (e.g., see
appendix~\ref{PotentialModel}). On the other hand, one can build a
mandatory connection between the contact terms of chiral effective
field theory and the one-boson-exchange model with the help of
resonance saturation model~\cite{Epelbaum:2001fm,Ecker:1988te}. As
the heavy fields, $\rho$, $\omega$, $f_0$, $a_0$, etc., which are
equally treated in one-boson-exchange model, are integrated out in
chiral effective field theory, and their contributions are packaged
into the LECs. The $(\omega,f_0)$ and $(\rho,a_0)$ mesons account for the
isospin-isospin unrelated $D_a$ and related $E_a$, respectively
[e.g., see eq.~\eqref{Contact_Lag_BM}]~\cite{Xu:2017tsr}. Meanwhile,
the $\omega$ and $\rho$ mesons couple to the matter fields via the
$P$-wave interaction due to the parity conservation. Each vertex
contains one momentum. In other words, the $\omega$ and $\rho$
mesons are responsible for the momentum-dependent spin-spin
interaction, which cannot be matched with the
$\mathcal{O}(\varepsilon^0)$ $D_b$ and $E_b$. Therefore, the
momentum-independent contributions for $D_b$ and $E_b$ can only
come from the axial-vector mesons, such as $(h_1,f_1)$ and
$(b_1,a_1)$. The masses of these states reside around $1.2$ GeV,
which are much heavier than those of $\omega$ and $\rho$, and
suppress the value of $\mathbb{D}_2$.

\section{Hidden-bottom molecular pentaquarks}\label{HiddenBottomSystems}

The above study for the hidden-charm pentaquarks can be extended to
the hidden-bottom case, once the coupling constants and mass
splittings are replaced by the bottomed ones. The coupling constants
$g_2$ and $g_4$ for the bottom baryons can be calculated with the
partial decay widths of $\Sigma_b\to\Lambda_b\pi$ and
$\Sigma_b^\ast\to\Lambda_b\pi$~\cite{Tanabashi:2018oca},
\begin{eqnarray}
\Gamma(\Sigma_b\to\Lambda_b\pi)&=&\frac{g_2^2}{4\pi
f_\pi^2}\frac{m_{\Lambda_b}}{m_{\Sigma_b}}|\bm{q}_\pi|^3,\qquad\qquad
\Gamma(\Sigma_b^\ast\to\Lambda_b\pi)=\frac{g_4^2}{12\pi
f_\pi^2}\frac{m_{\Lambda_b}}{m_{\Sigma_b^\ast}}|\bm{q}_\pi|^3,
\end{eqnarray}
Using the average values of the decay widths of
$\Sigma_b^+\to\Lambda_b^0\pi^+$ and
$\Sigma_b^{\ast+}\to\Lambda_b^0\pi^+$~\cite{Tanabashi:2018oca}, we
get $g_2=-0.51$, $g_4=0.91$. The other couplings can then be
obtained with the relations in eq.~\eqref{BaryonCouplings}, which
yield,
\begin{eqnarray}
g_1=0.83,\qquad g_3=0.72,\qquad g_5=-1.25.
\end{eqnarray}
The axial coupling $g$ of the $B$ mesons cannot be directly derived
from the experiments due to absence of phase space for $B^\ast\to
B\pi$, so we adopt the average value $g=-0.52$ from the lattice
calculations~\cite{Ohki:2008py,Detmold:2012ge}. Similarly, the mass
splittings are correspondingly given by
\begin{align}
\delta_a&=m_{\Sigma_b^\ast}-m_{\Sigma_b}\simeq20~\textrm{MeV},&\delta_b&=m_{B^\ast}-m_B\simeq45~\textrm{MeV},\nonumber\\
\delta_c&=m_{\Sigma_b}-m_{\Lambda_b}\simeq191~\textrm{MeV},&\delta_d&=m_{\Sigma_b^\ast}-m_{\Lambda_b}\simeq211~\textrm{MeV},
\end{align}
where the masses of the $\Sigma_b^{(\ast)+}$ and $B^{(\ast)0}$ are
used \cite{Tanabashi:2018oca}.

The small scale expansion~\cite{Hemmert:1997ye} is used in
eqs.~\eqref{Baryon_Lag_SF} and \eqref{Meson_Lag_SF}, i.e., the mass
splitting $\delta$ is treated as another small scale in the
Lagrangians. This expansion works well for the systems with one
heavy matter field~\cite{Wang:2019mhm}. The loop integrals in these
systems are the polynomials of $\delta$, thus the convergence of the
chiral expansion is not affected as long as the $\delta\sim m_\pi$
or smaller than $m_\pi$. But the situation becomes different for the
systems with two heavy matter fields. The loop integral of the box
diagram is proportional to $1/(\delta_x+\delta_y)$. If
$\delta_x+\delta_y$ is of the order of the pion mass, the
convergence of the expansion could still be good. For example, for
the $\Sigma_c^{(\ast)}\bar{D}^{(\ast)}$ systems~\cite{Meng:2019ilv},
$\delta_b-\delta_a\simeq80$ MeV. However, for the
$\Sigma_b^{(\ast)}B^{(\ast)}$ systems, $\delta_b-\delta_a\simeq25$
MeV, which is much smaller that the pion mass\footnote{The pathosis
does not appear in the diagrams with $\Lambda_b$, because the
differences between $\delta_c(\delta_d)$ and $\delta_b$ are of the
same order as the $m_\pi$.}. Therefore, if we still adopt the same
procedure as used in the $\Sigma_c^{(\ast)}\bar{D}^{(\ast)}$
systems, the amplitudes of some typical box diagrams would be
largely amplified, which results in extremely strong attractive or
repulsive potential. This is unphysical and mainly caused by the
poles of the heavy matter fields. In some previous
works~\cite{Meng:2019ilv,Wang:2018atz}, the mass splittings are
discarded in the box diagrams to subtract the 2PR contributions.
Here, we develop a method to remove the heavy matter field poles in
the box diagrams with the mass splittings being kept (see
appendix~\ref{Remove_2PR} for more details).

In order to predict the possible $P_b$ states, we also need to know
the LECs $\mathbb{D}_1$ and $\mathbb{D}_2$ for the
$\Sigma_b^{(\ast)}B^{(\ast)}$ systems. In principle, they should be
fixed from experimental data or the results from lattice QCD, which
are not available at present. Thus, we estimate the ranges of
$\mathbb{D}_1$ and $\mathbb{D}_2$. Generally, the values of
$\mathbb{D}_1$ and $\mathbb{D}_2$ are different for the
$\Sigma_b^{(\ast)}B^{(\ast)}$ and
$\Sigma_c^{(\ast)}\bar{D}^{(\ast)}$ systems. One explicit example is
that the axial coupling constants for the bottom sectors are about
$17\%$ smaller than those of the charmed sectors. Therefore, we take
the values $(\mathbb{D}_1,\mathbb{D}_2)=(52,-4)$ GeV$^{-2}$ fixed
for the $P_c$s with at most $17\%$ deviation to give the ranges of
$\mathbb{D}_1$ and $\mathbb{D}_2$ in the hidden-bottom case.

\begin{figure*}
\begin{minipage}[t]{0.5\linewidth}
\centering
\includegraphics[width=\columnwidth]{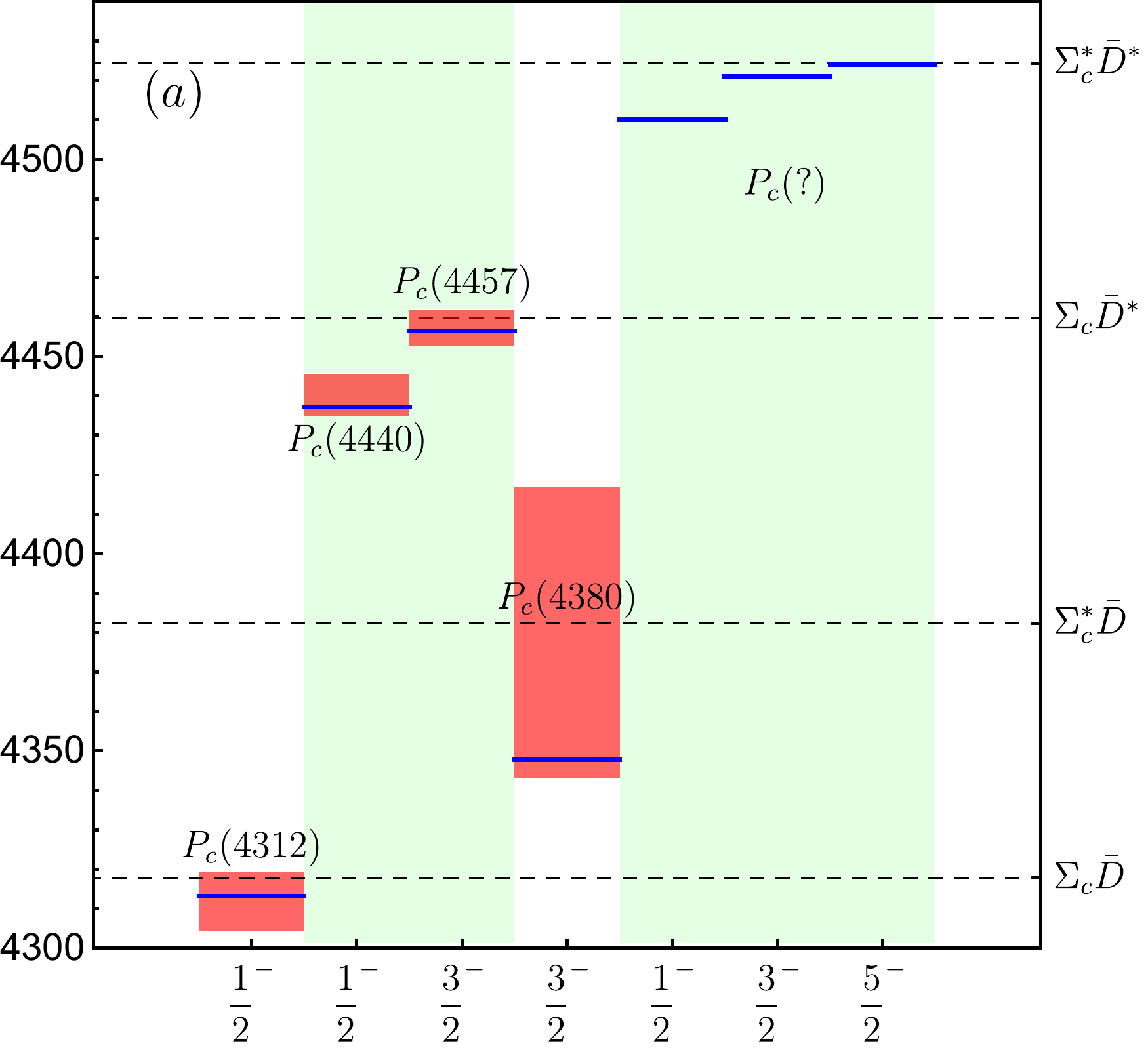}
\end{minipage}%
\begin{minipage}[t]{0.5\linewidth}
\centering
\includegraphics[width=\columnwidth]{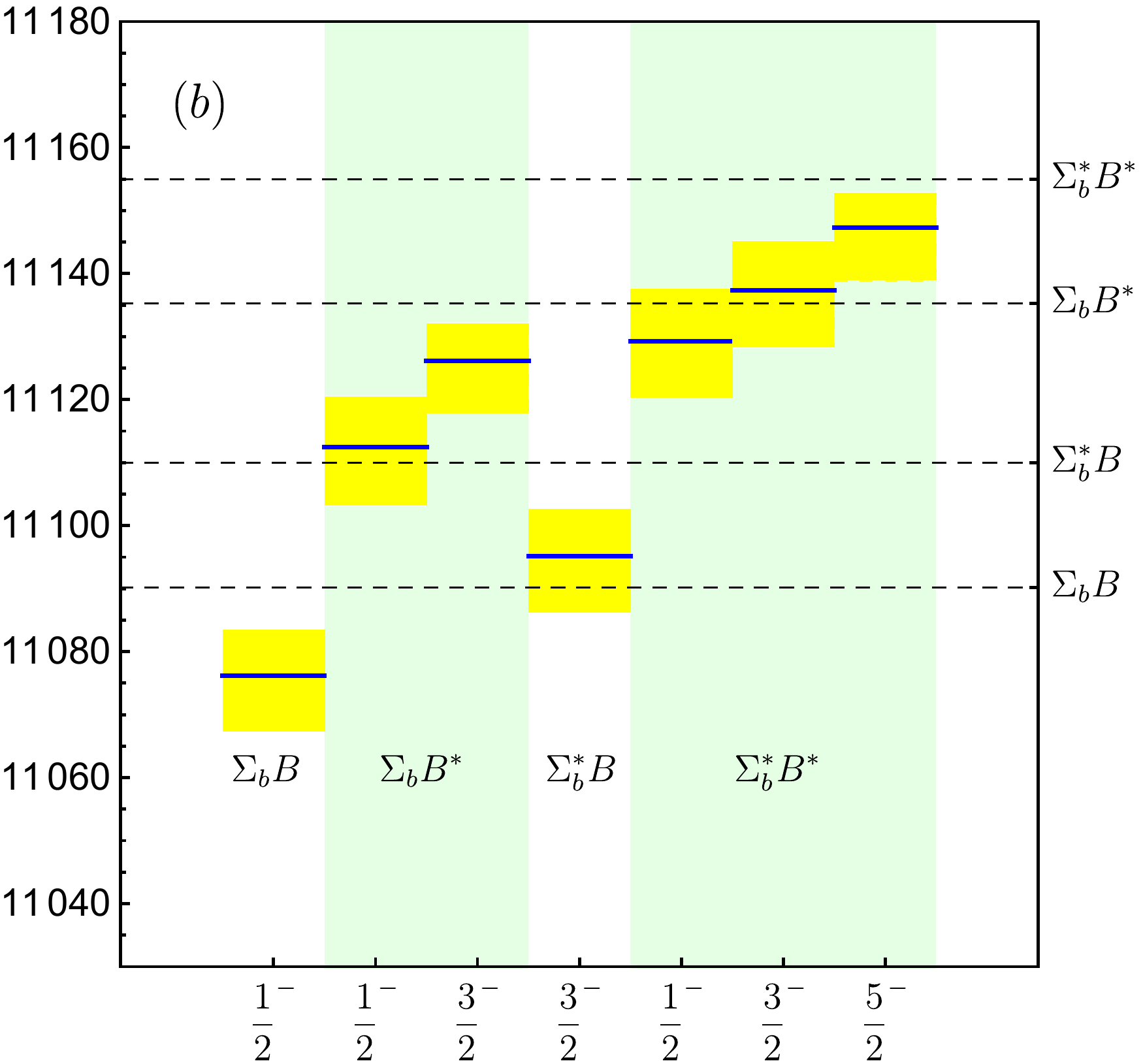}
\end{minipage}
\caption{The mass spectra of the hidden-charm ($a$) and hidden-bottom
($b$) molecular pentaquarks. The red and yellow regions in figures ($a$)
and ($b$) denote the mass ranges obtained from the experimental
measurements and theoretical estimations, respectively. The blue
solid lines represent the central values in our calculations. The
black dashed lines are the corresponding
thresholds.\label{MassSpectra}}
\end{figure*}

We set the $(52,-4)$ GeV$^{-2}$ as the limits of
$(\mathbb{D}_1,\mathbb{D}_2)$ for the bottom case, which deviate
$17\%$ from the central value. Approximately, we have
\begin{eqnarray}
\mathbb{D}_1=43\pm9~\textrm{GeV}^{-2},\qquad
\mathbb{D}_2=-3.3\mp0.7~\textrm{GeV}^{-2}.
\end{eqnarray}
The binding energies and the mass spectra are given in
table~\ref{BindingEnergiesB} and figure~\ref{MassSpectra}(b),
respectively. We notice the hidden-bottom ones are the tightly bound
molecules due to the large masses of their components. Unlike the
$[\Sigma_c^\ast\bar{D}^\ast]_J$ systems, the gaps between the
thresholds of the $[\Sigma_b^\ast B^\ast]_J$ systems are only about
$20$ MeV. Thus the masses of some peculiar states with binding
energies $\Delta E<-20$ MeV may not only lie below its corresponding
threshold but also the lower one. For example, the molecular state
$[\Sigma_b^\ast B^\ast]_{\frac{1}{2}}$ locates below the thresholds
of $\Sigma_b B^\ast$ and $\Sigma_b^\ast B^\ast$ if we only consider
the central value.

The masses of the hidden-bottom molecules are all above $11$ GeV.
Like their $P_c$ partners, they may be observed from the
$\Upsilon(1S) N$ and $\Upsilon(2S) N$ final states. We hope future
experiments to hunt for these $P_b$ states. We conclude this section
by borrowing one of the famous phrases from R. P. Feynman: ``{\it
There is plenty of room at the `bottom'.}"~\cite{Feynman:1959rpf}
\begin{table*}[htbp]
\centering
\renewcommand{\arraystretch}{1.5}
\caption{The binding energies $\Delta E$ for the $I=\frac{1}{2}$
hidden-bottom $[\Sigma_b^{(\ast)}B^{(\ast)}]_J$ systems with the
contribution of the $\Lambda_b$. The values of
$(\mathbb{D}_1,\mathbb{D}_2)$  are chosen to be
$(43\pm9,-3.3\mp0.7)$ $\textrm{GeV}^{-2}$, the cutoff $\Lambda=0.5$
GeV (in units of MeV).}\label{BindingEnergiesB}
\setlength{\tabcolsep}{0.3mm} {
\begin{tabular}{c|ccccccc}
\hline\hline
$\Delta E$&$[\Sigma_b B]_{\frac{1}{2}}$&$[\Sigma_b B^\ast]_{\frac{1}{2}}$&$[\Sigma_b B^\ast]_{\frac{3}{2}}$&$[\Sigma_b^\ast B]_{\frac{3}{2}}$&$[\Sigma_b^\ast B^\ast]_{\frac{1}{2}}$&$[\Sigma_b^\ast B^\ast]_{\frac{3}{2}}$&$[\Sigma_b^\ast B^\ast]_{\frac{5}{2}}$\\
\hline
With $\Lambda_b$&$-14.04^{+7.36}_{-8.92}$&$-22.72^{+8.03}_{-9.34}$&$-9.12^{+6.06}_{-8.34}$&$-14.74^{+7.54}_{-9.05}$&$-25.75^{+8.38}_{-9.06}$&$-17.76^{+7.91}_{-9.07}$&$-7.81^{+5.56}_{-8.41}$\\
\hline\hline
\end{tabular}
}
\end{table*}

\section{Heavy quark symmetry breaking effect}\label{HQSBreakingEffect}

The QCD Lagrangian has heavy quark symmetry (HQS) when the heavy
quark mass $m_Q\to \infty$. For a heavy hadron containing one single
heavy quark, the strong interaction would be independent of the
heavy flavors in this limit. Meanwhile, the heavy quark will
decouple with the light degrees of freedom. The multiplet associated
with the heavy quark spin would be degenerate in the heavy quark
limit. However, the physical masses of the heavy quarks are finite,
such as $m_c\sim1.5$ GeV, $m_b\sim5$ GeV. Therefore, the effects of
the heavy quark flavor symmetry breaking and spin symmetry breaking
are explicit. For example, the axial coupling $g$ for $B^\ast B\pi$
is about $17\%$ smaller that that of the $D^\ast D\pi$. Thus the
value $17\%$ can be roughly regarded as the breaking size of the
heavy quark flavor symmetry. In addition, the heavy quark spin
symmetry (HQSS) breaking is more obvious, such as the mass
splittings of $(B^\ast,B)$ and $(D^\ast,D)$ are about $45$ MeV and
$142$ MeV, respectively.

HQS can be used to relate the coupling constants to one another,
such as the axial coupling constants in the
$\mathcal{O}(\varepsilon^0)$ Lagrangians. Under HQS, the heavy
quarks only serve as the spectators. The interaction between
$\Sigma_c^{(\ast)}$ and $\bar{D}^{(\ast)}$ is mediated by their
inner light degrees of freedom, i.e., the light diquarks in
$\Sigma_c^{(\ast)}$ and the light quark in $\bar{D}^{(\ast)}$.
Therefore, the $S$-wave effective potentials between
$\Sigma_c^{(\ast)}$ and $\bar{D}^{(\ast)}$ at the quark level can be
parameterized as~\cite{Meng:2019ilv}
\begin{eqnarray}
V^{\mathrm{HQS}}_{\mathrm{quark-level}}=V_c+V_s\boldsymbol{l}_1\cdot\boldsymbol{l}_2,
\end{eqnarray}
where $V_c$ and $V_s$ denote the central term and spin-spin term,
respectively. $\boldsymbol{l}_1$ and $\boldsymbol{l}_2$ are the
spins of the light degrees of freedom of the $\Sigma_c^{(\ast)}$ and
$\bar{D}^{(\ast)}$, respectively. With the potentials at the quark
level, one can build the relations between different channels at the
hadron level by parameterizing the hadron level potentials as
\begin{eqnarray}
\mathcal{V}_{\Sigma_c\bar{D}}&=&\mathcal{V}_1,\qquad\mathcal{V}_{\Sigma_c\bar{D}^\ast}=\mathcal{V}_2+\mathcal{V}_2^\prime\bm{S}_1\cdot\bm{S}_2,\nonumber\\
\mathcal{V}_{\Sigma_c^\ast\bar{D}}&=&\mathcal{V}_3,\qquad\mathcal{V}_{\Sigma_c^\ast\bar{D}^\ast}=\mathcal{V}_4+\mathcal{V}_4^\prime\bm{S}_1\cdot\bm{S}_2,
\end{eqnarray}
where $\bm{S}_1$ and $\bm{S}_2$ are the spin operators of the
$\Sigma_c^{(\ast)}$ and $\bar{D}^\ast$, respectively. One can easily
verify\footnote{See more detailed derivations in the appendix A of
ref.~\cite{Meng:2019ilv}}
\begin{eqnarray}\label{HQS_Relations}
\mathcal{V}_1=\mathcal{V}_2=\mathcal{V}_3=\mathcal{V}_4=V_c;\qquad
\mathcal{V}_2^\prime=\frac{2}{3}V_s,\quad
\mathcal{V}_4^\prime=\frac{1}{3}V_s.
\end{eqnarray}
The leading order potentials obviously satisfy the above relations
obtained from HQS. One can also testify the one-loop level
analytical expressions satisfy the above relations as well when
$d\to4$ and $\delta_{a,b}\to0$.

\begin{figure*}[htbp]
\begin{center}
\begin{minipage}[t]{0.4\linewidth}
\centering
\includegraphics[width=\columnwidth]{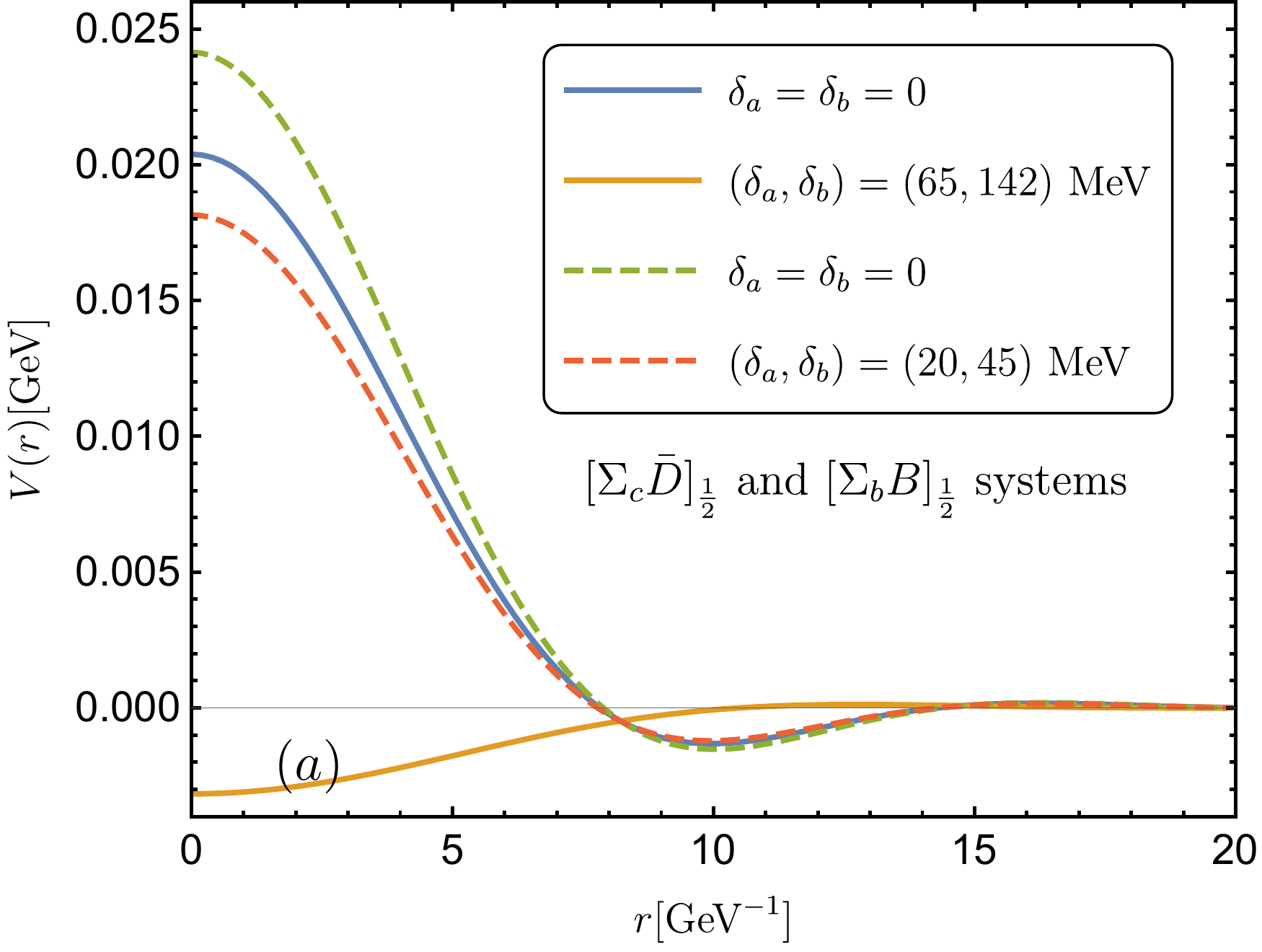}
\end{minipage}%
\hspace{0.5in}
\begin{minipage}[t]{0.4\linewidth}
\centering
\includegraphics[width=\columnwidth]{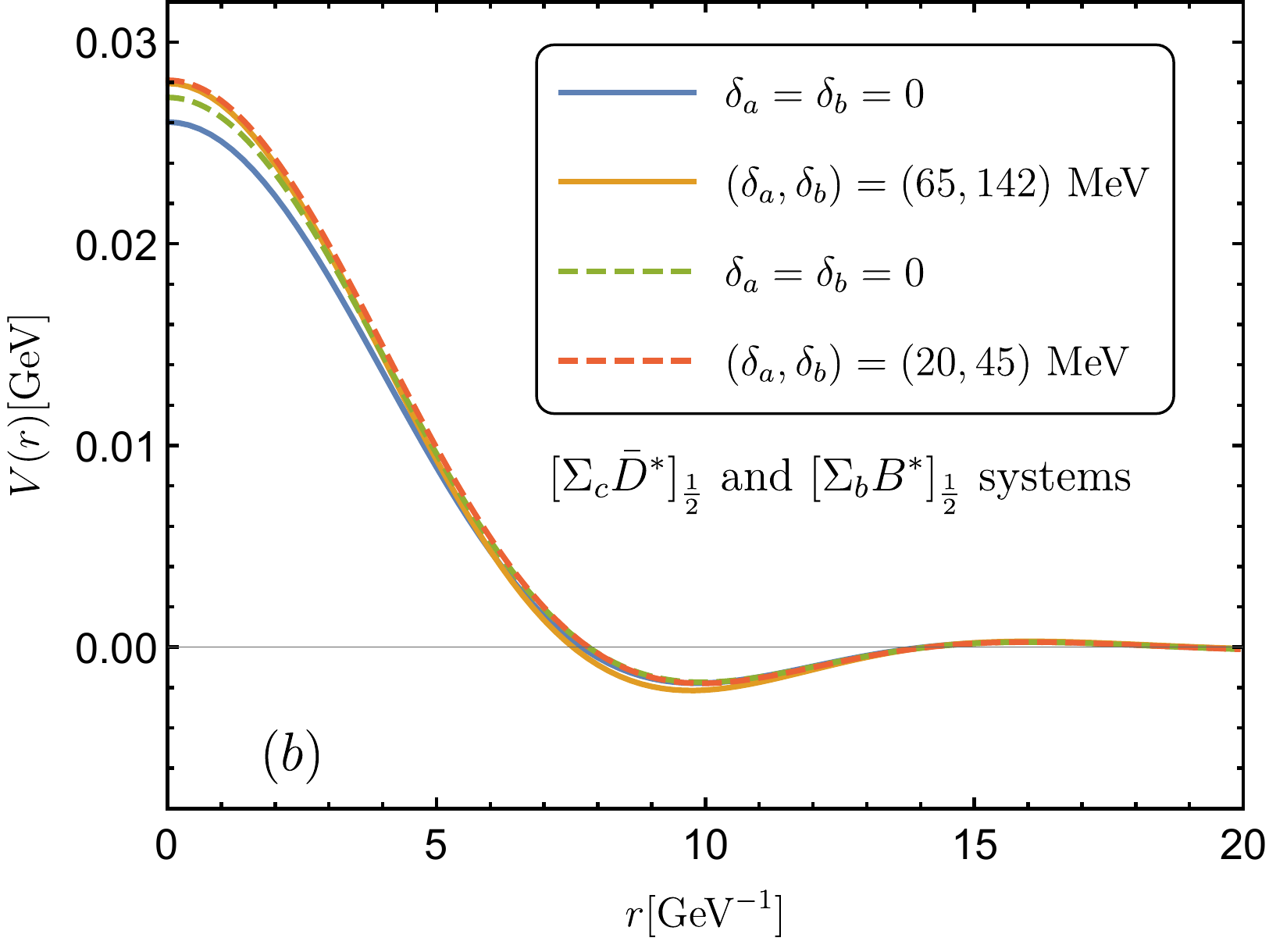}
\end{minipage}
\begin{minipage}[t]{0.4\linewidth}
\centering
\includegraphics[width=\columnwidth]{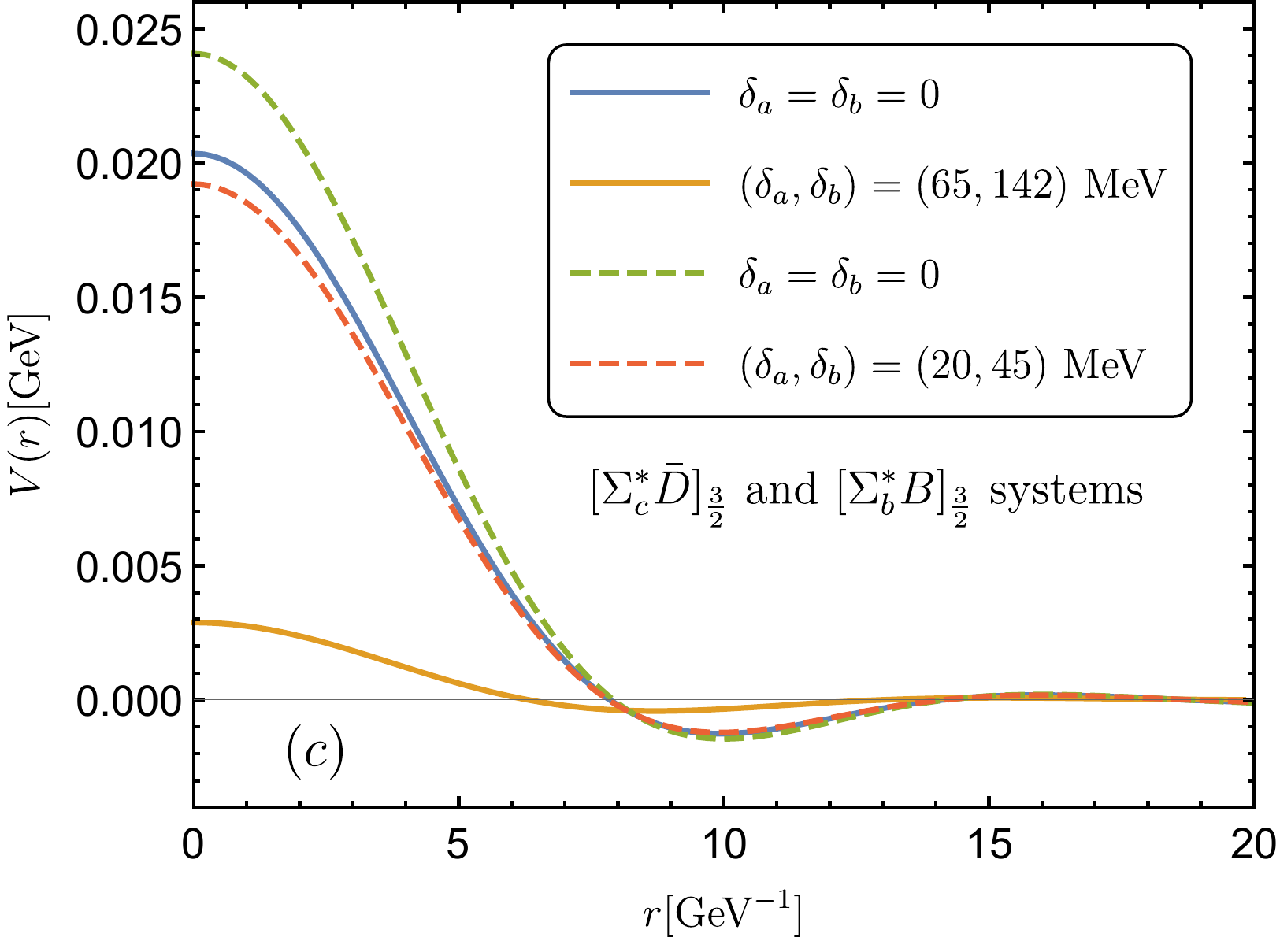}
\end{minipage}%
\hspace{0.5in}
\begin{minipage}[t]{0.4\linewidth}
\centering
\includegraphics[width=\columnwidth]{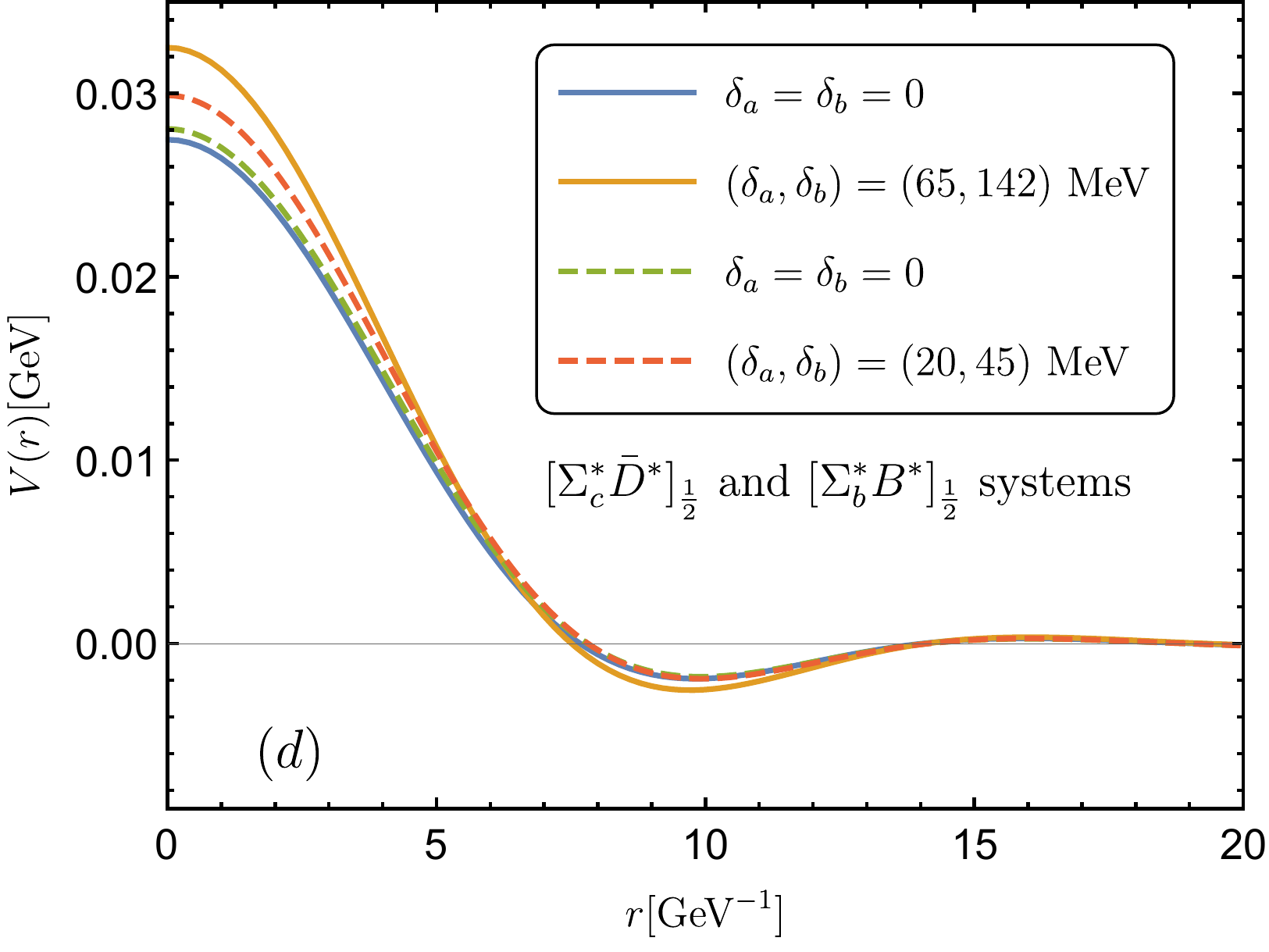}
\end{minipage}
\caption{The heavy quark symmetry breaking phenomena in the
two-pion-exchange diagrams. The solid lines denote the
$\Sigma_c^{(\ast)}\bar{D}^{(\ast)}$ systems with vanishing mass
splittings and physical mass splittings. The dashed lines represent
the same cases but for the $\Sigma_b^{(\ast)}B^{(\ast)}$ systems.
The unlisted systems share the similar behaviors as their spin
partners.\label{HQSBreakingTPE}}
\end{center}
\end{figure*}

The HQS breaking effect would manifest itself in the loop diagrams
if $\delta_{a,b}\neq0$ \footnote{When we are talking about the HQS
in the loop diagrams, the contribution of the $\Lambda_c$ is
ignored, since the mass splittings $\delta_{c,d}$ do not vanish even
in heavy quark limit.}. When $\delta_a=\delta_b=0$, all the box
diagrams would become the 2PR ones, thus we have to remove the 2PR
contributions. In order to compare with the cases of
$\delta_a=\delta_b=0$, we also subtract the 2PR contributions from
the $(\delta_a,\delta_b)\neq0$ cases (see
appendix~\ref{Remove_2PR}). The 2PI two-pion-exchange potentials for
the $\Sigma_c^{(\ast)}\bar{D}^{(\ast)}$ and
$\Sigma_b^{(\ast)}B^{(\ast)}$ systems with and without HQS are
illustrated in figure~\ref{HQSBreakingTPE}. We notice the HQS keeps
relatively good for the $[\Sigma_c\bar{D}^\ast]_J/[\Sigma_b
B^\ast]_J$ and $[\Sigma_c^\ast\bar{D}^\ast]_J/[\Sigma_b^\ast
B^\ast]_J$ systems, while it breaks significantly for the
$[\Sigma_c\bar{D}]_J$ and $[\Sigma_c^\ast\bar{D}]_J$ systems. When
$\delta_a=\delta_b=0$, the two-pion-exchange potential of the
$[\Sigma_c\bar{D}]_J$ system is exactly equal to that of the
$[\Sigma_c^\ast\bar{D}]_J$ system, which satisfies the relations in
eq.~\eqref{HQS_Relations}. However, when we set the physical mas
splitting, $(\delta_a,\delta_b)=(65,142)$ MeV, the line-shapes are
explicitly modified and the relations in eq.~\eqref{HQS_Relations}
are obviously violated. The quantum fluctuation at the loop level
would break the HQS significantly. The predictions inherited from
HQS should be carefully reexamined, at least for the
$[\Sigma_c\bar{D}]_J$ and $[\Sigma_c^\ast\bar{D}]_J$ systems.
Besides, the HQS in the hidden-bottom systems is better than that of
the hidden-charm cases as expected.
\begin{figure*}[tb]
\begin{center}
\begin{minipage}[t]{0.4\linewidth}
\centering
\includegraphics[width=\columnwidth]{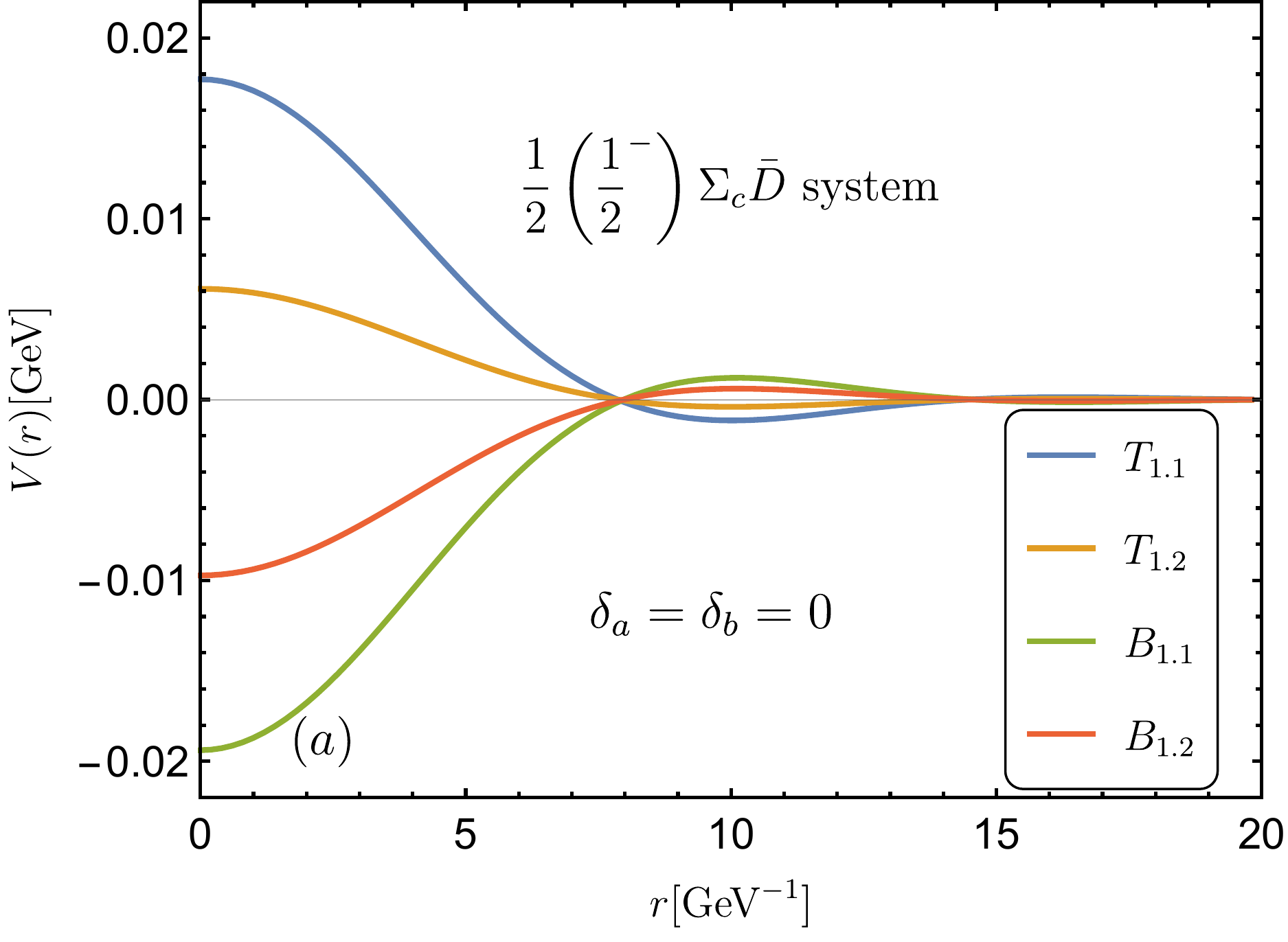}
\end{minipage}%
\hspace{0.5in}
\begin{minipage}[t]{0.4\linewidth}
\centering
\includegraphics[width=\columnwidth]{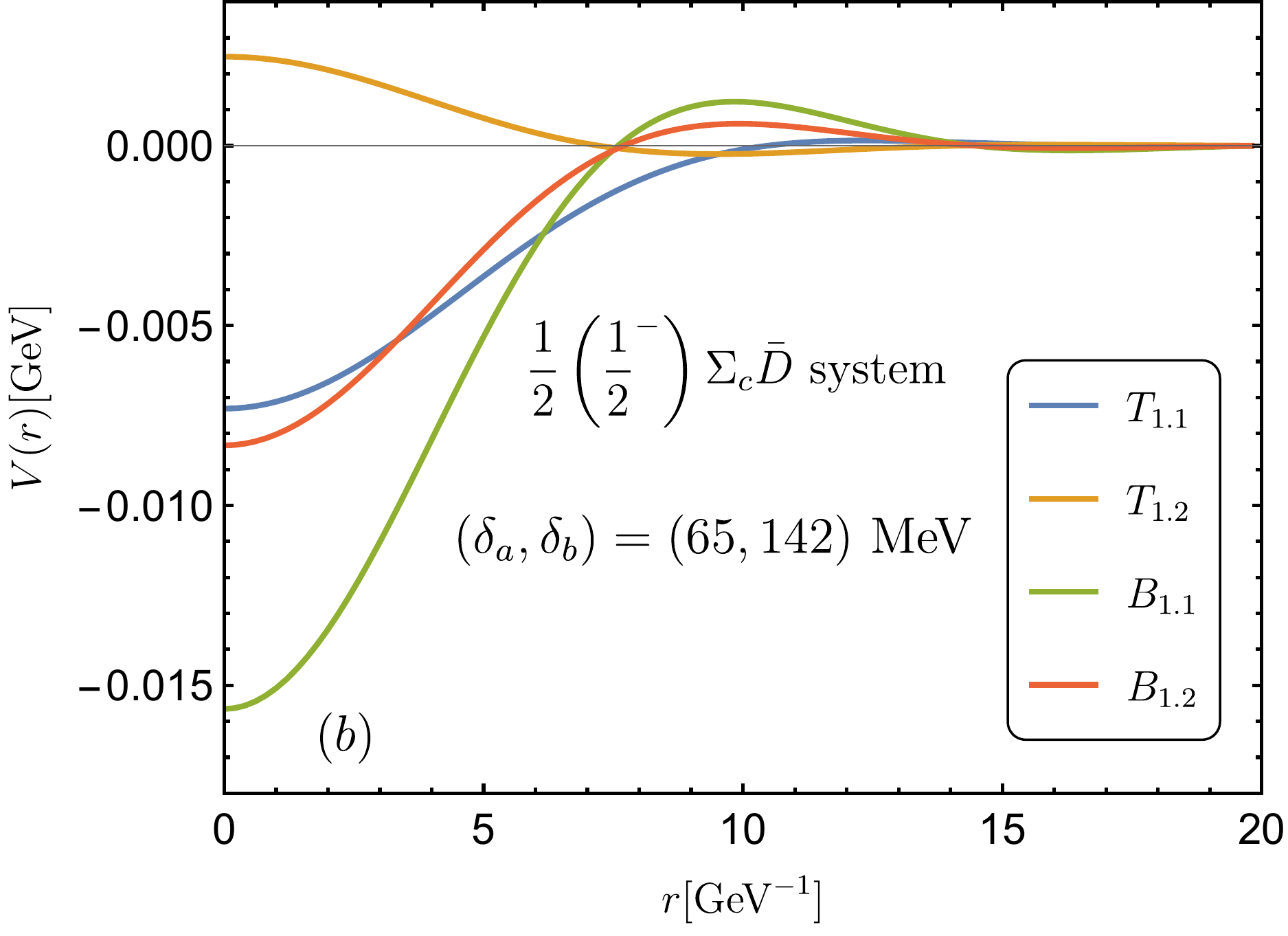}
\end{minipage}
\begin{minipage}[t]{0.4\linewidth}
\centering
\includegraphics[width=\columnwidth]{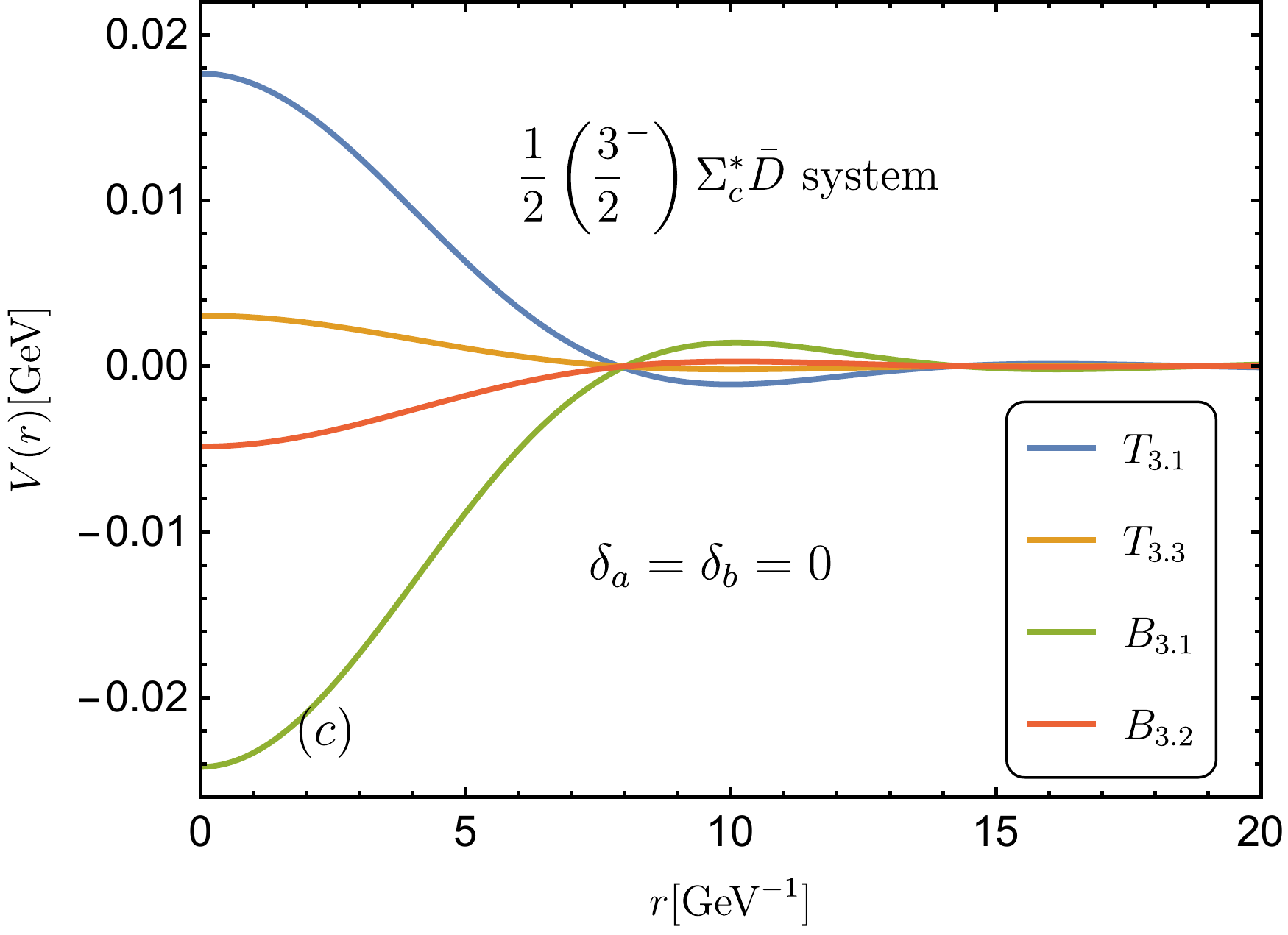}
\end{minipage}%
\hspace{0.5in}
\begin{minipage}[t]{0.4\linewidth}
\centering
\includegraphics[width=\columnwidth]{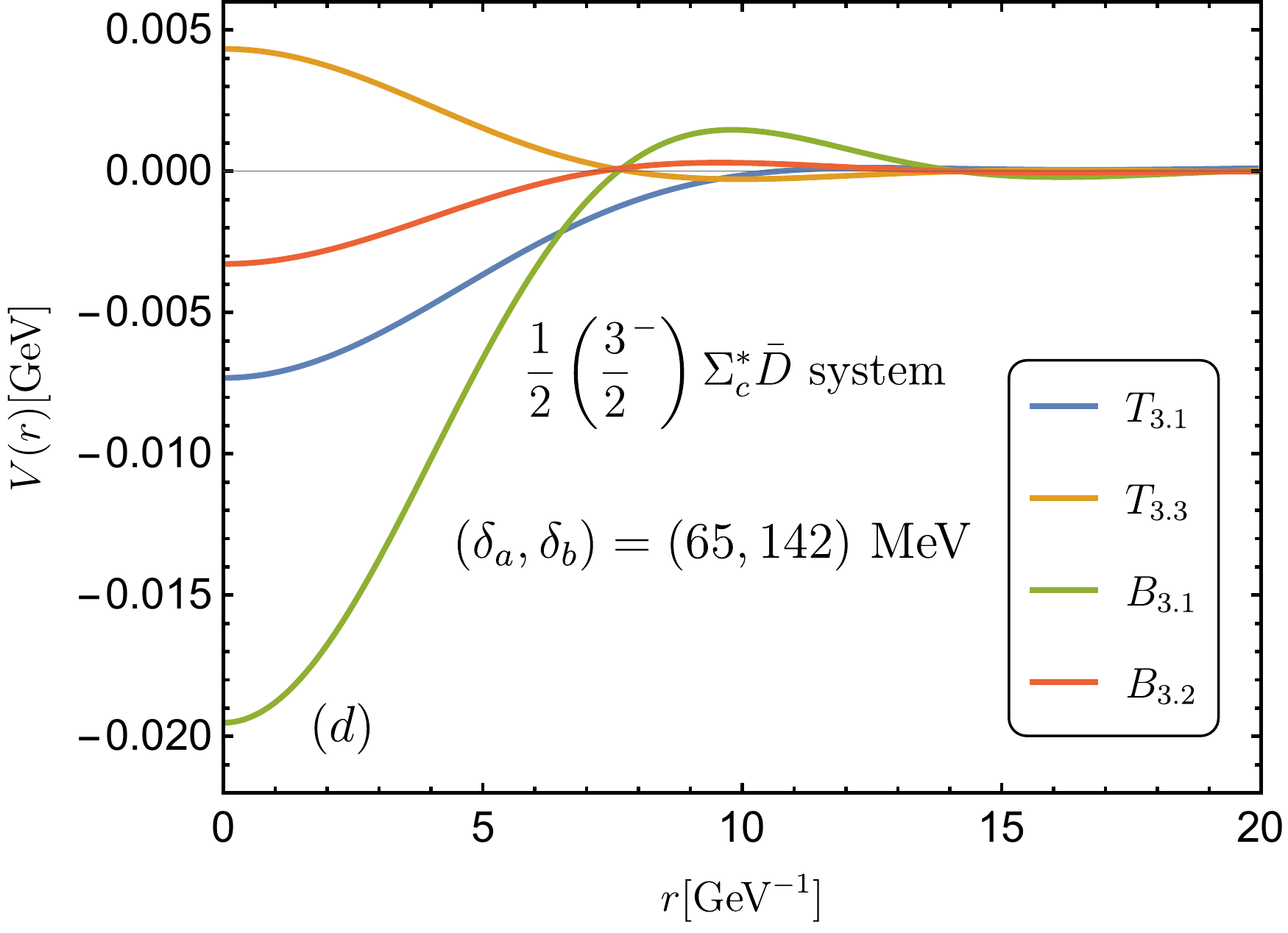}
\end{minipage}
\caption{The behaviors of the two-pion-exchange potentials of each
mass splitting related 2PI diagrams for the
$[\Sigma_c\bar{D}]_{\frac{1}{2}}$ [($a$), ($b$)] and
$[\Sigma_c^\ast\bar{D}]_{\frac{3}{2}}$ [($c$), ($d$)] systems in the
cases of $\delta_a=\delta_b=0$ and $(\delta_a,\delta_b)=(65,142)$
MeV, respectively.\label{HQSBreakingTPE_EachDiagrams}}
\end{center}
\end{figure*}

In the following, we investigate the HQSS violation effect of each
Feynman diagram for the $[\Sigma_c\bar{D}]_{\frac{1}{2}}$ and
$[\Sigma_c^\ast\bar{D}]_{\frac{3}{2}}$ systems. The results are
shown in figure~\ref{HQSBreakingTPE_EachDiagrams}. When
$\delta_a=\delta_b=0$, the contributions from the triangle diagrams
and box diagrams are always repulsive and attractive, respectively.
The differences between the corresponding diagrams, such as
$(B_{1.1})$ and $(B_{3.1})$, are mainly caused by the coupling
constants. However, when the mass splittings are considered, the
magnitudes of most diagrams except for $(T_{1.1})$ and $(T_{3.1})$
would change. The signs of the potentials from $(T_{1.1})$ and
$(T_{3.1})$ are changed. The HQSS breaking mainly originates from
these two diagrams. The repulsive contributions of the two diagrams
in the heavy quark limits become attractive when the mass splittings
are included. Inspecting the analytical expressions of the triangle
diagrams for the $\Sigma_c^{(\ast)}\bar{D}^{(\ast)}$ systems, we
would see that
\begin{eqnarray}
\mathcal{V}_{\Sigma_c^{(\ast)}\bar{D}^{(\ast)}}^{T_{i.j}}\sim(\mathbf{I}_1\cdot\mathbf{I}_2)\Big[A
J_{34}^T-B\bm{q}^2\left( J_{24}^T+
J_{33}^T\right)\Big](m_\pi,\omega,q),
\end{eqnarray}
where $A$ and $B$ are the positive numbers. The corresponding $
J_{ij}^T$ functions generally contain two structures, the odd
function of $\omega$ and the even one (see
appendix~\ref{LoopIntegrals}). The odd part is proportional to
\begin{eqnarray}\label{omegaPart}
\omega\int_0^1dx\mathscr{F}(x,m_\pi,\omega,\bm{q})+\int_{-\omega}^0dy\int_0^1dx\mathscr{G}(x,y,m_\pi,\omega,\bm{q}),
\end{eqnarray}
where $\mathscr{F}(x,\dots)$ and $\mathscr{G}(x,y,\dots)$ denote the
integrands which are the functions of $(x,\dots)$ and $(x,y,\dots)$,
respectively. These two terms vanish in the heavy quark limit, i.e.,
when $\delta_a=\delta_b=0$. The even one is proportional to
\begin{eqnarray}
\int_0^1dx\mathscr{H}(x,m_\pi,\omega,\bm{q}).
\end{eqnarray}
Only this term contributes when $\delta_a=\delta_b=0$. Therefore,
one would see a different scenario when the nonzero mass splittings
are considered, since the two terms in eq.~\eqref{omegaPart} also
contribute. For the diagrams $(T_{1.1})$ and $(T_{3.1})$,
$\omega=-\delta_b$, while for the diagrams $(T_{2.2})$ and
$(T_{4.2})$, $\omega=\delta_b$. Thus the HQS breaking effect is
totally different for the $\Sigma_c^{(\ast)}\bar{D}$ and
$\Sigma_c^{(\ast)}\bar{D}^{\ast}$ systems, because the
eq.~\eqref{omegaPart} is the odd function of $\omega$, which is very
sensitive to the sign of the $\omega$. In addition, the integrands
$\mathscr{F}$, $\mathscr{G}$ and $\mathscr{H}$ always have the
nonanalytic structures, such as the logarithmic and square root
terms. So the variations of the graphs $(T_{1.2})$ and $(T_{3.3})$
are not so dramatic as those of the $(T_{1.1})$ and $(T_{3.1})$,
because $\delta_a<m_\pi$, whereas $\delta_b>m_\pi$. The HQS breaking
effect expounded above issues from the loop diagrams, which is the
quantum physics of the light degrees of freedom at the low energy,
and cannot be modified by any unknown physics that happens at the
high energy.

\section{Summary and conclusion}\label{SummaryandConclusion}

In the April of this year, the LHCb collaboration reported the
observation of the three pentaquark states $P_c(4312)$, $P_c(4440)$
and $P_c(4457)$~\cite{Aaij:2019vzc}. They were subsequently
interpreted as the molecular states by many theoretical
works~\cite{Meng:2019ilv,Liu:2019tjn,Chen:2019asm,Xiao:2019aya,He:2019ify}
due to the proximities to the $\Sigma_c\bar{D}$ and
$\Sigma_c\bar{D}^\ast$ thresholds. In this paper, we have
systematically investigated the interactions between the charmed
baryons $\Sigma_c^{(\ast)}$ and anticharmed mesons
$\bar{D}^{(\ast)}$ in the framework of chiral perturbation theory.
To this end, we have simultaneously considered the short-range
contact interaction, long-range one-pion-exchange contribution,
intermediate-range two-pion-exchange loop diagrams, as well as the
influence of the mass splittings on the effective potentials.

When we fix the total isospin as $I={1\over 2}$, the original four
independent LEC can be reduced to two. These two LECs can be fitted
using the binding energies of the $P_c(4312)$, $P_c(4440)$ and
$P_c(4457)$ as inputs. We first attempt to reproduce the newly
observed three $P_c$s via only considering the spin partners of the
$\Sigma_c\bar{D}^{(\ast)}$ in the loops. But we fall into the same
dilemma as in the scenario II of our previous
work~\cite{Meng:2019ilv}, i.e., it is nearly impossible to reproduce
the three $P_c$s synchronously in this case. Considering the strong
couplings between $\Sigma_c^{(\ast)}$ and $\Lambda_c\pi$, we then
include the contribution of $\Lambda_c$ in the loop diagrams. Three
$P_c$s are simultaneously reproduced at this point. This indicates
the $\Lambda_c$ plays a very important role for the formation of
these $P_c$s. We also notice that only considering the $\Lambda_c$
cannot describe the $P_c$s either. The subtle interplay between the
channels with $\Lambda_c$ and the ones with $\Sigma_c^{(*)}$
determines the existence of these hadronic molecules. Our
calculation supports the $P_c(4312)$, $P_c(4440)$ and $P_c(4457)$ to
be the $S$-wave hidden-charm $[\Sigma_c\bar{D}]_{J=1/2}^{I=1/2}$,
$[\Sigma_c\bar{D}^\ast]_{J=1/2}^{I=1/2}$ and
$[\Sigma_c\bar{D}^\ast]_{J=3/2}^{I=1/2}$ hadronic molecules.

Since the $J^P$ quantum numbers are still unknown in experiment, we
also investigate the possibility of a different spin assignment,
viz, $\frac{1}{2}^-$ for $P_c(4457)$ and $\frac{3}{2}^-$ for
$P_c(4440)$. Although the binding energies can also be well fitted
by changing the LECs in this case, one has to fulfill this
assignment at the cost of largely enhancing spin-spin interaction.
The overwhelming spin-spin term at $\mathcal{O}(\varepsilon^0)$
contradicts the phenomenological considerations of the quark model
and one-boson-exchange model, as well as the empirical conclusions
from the hadron spectra and $N$-$N$ scattering data.

With the fixed LECs, we notice the other four channels
$[\Sigma_c^\ast\bar{D}]_{J=3/2}^{I=1/2}$ and
$[\Sigma_c^\ast\bar{D}^\ast]_{J}^{I=1/2}~(J=\frac{1}{2},\frac{3}{2},\frac{5}{2})$
are also bound ($[\Sigma_c^\ast\bar{D}^\ast]_{5/2}^{1/2}$ is very
shallowly bound). The previously reported
$P_c(4380)$~\cite{Aaij:2015tga}, a candidate of the
$[\Sigma_c^\ast\bar{D}]_{3/2}^{1/2}$ molecular state, is a deeper
bound hadronic molecule in our calculation. This is mainly caused by
the important contribution from the two-pion-exchange diagrams,
which is the essential difference with the predictions from the
quark model and leading order effective field theory. These two
approaches do not contain the nonanalytical terms, such as the
powers of $\log q^2$ and $\sqrt{q^2}$, which are irregular and may
give the enhanced contributions sometimes. These terms cannot be
predicted accurately from the aspects of the quark model.

We also study the hidden-bottom $\Sigma_b^{(\ast)}B^{(\ast)}$
systems. The axial coupling constants for the bottom baryons and
bottom mesons are determined with the partial decay widths of
$\Sigma_b^{(\ast)}\to\Lambda_b\pi$ and the lattice simulations,
respectively. We adopt the fitted LECs in the hidden-charm case as
the limit, and the $17\%$ reduction as the central value for the
hidden-bottom systems. With these fixed parameters, we find the
$[\Sigma_b^{(\ast)}B^{(\ast)}]_J^{1/2}$ systems are more tightly
bound. Because the thresholds of $\Sigma_b^{(\ast)}B^{(\ast)}$ are
very close to each other, so the masses of some states may cross two
thresholds, such as the $[\Sigma_b^{\ast}B^{\ast}]_{1/2}^{1/2}$. The
hidden-bottom ones might be observed from the $\Upsilon N$ final
states. We give a complete picture on the mass spectra of the
hidden-charm and hidden-bottom molecular pentaquarks, and there are
overall fourteen bound states in our calculations. The discovery of
$P_c$s at the LHCb is just the beginning for the community to search
for the exotic multiquark matters.

The heavy quark symmetry is always exploited to predict the mass
spectra of the hidden-charm and hidden-bottom systems. Since $m_b$
is much larger than $m_c$, so the predictions from the HQS in the
bottom sector is more reliable because the correction from the
next-to-leading order heavy quark expansion is very small. But the
reliability of the HQS in the charm sector is still questionable. So
we examine the HQS breaking effect in the loop diagrams by
considering the mass splittings in the propagators of the
intermediate states. As expected, the HQS in the hidden-bottom
systems is much better than that in the hidden-charm cases. Besides,
for some accidental reasons, the HQS as an approximation in the
$\Sigma_c\bar{D}$ and $\Sigma_c^\ast\bar{D}$ systems is not as good
as in the others. The two-pion-exchange potentials become totally
different with the mass splittings or not. One reason is the mass
difference between the initial $\bar{D}$ and intermediate
$\bar{D}^\ast$ is $-\delta_b$ and some triangle diagrams are very
sensitive to the sign of the mass difference. Another reason is
$\delta_b>m_\pi$, so the nonanalytic structures, e.g., logarithmic
and square root terms in the loop functions would be enhanced to
distort the potentials. This enlightens us that the HQS breaking
effect shall not be ignored if we want to give a comprehensive
description of the effective potentials, especially for the
interactions between the charmed hadrons.

We hope the lattice QCD simulations on the hidden-charm and
hidden-bottom pentaquark systems could be carried out in the future,
which can help us to get a deeper insight into the inner structures
of these exotica. The analytical expressions derived in this work
can also be used to perform the chiral extrapolations.

\section*{Acknowledgments}
B. W is very grateful to X. L. Chen and W. Z. Deng for helpful
discussions. This project is supported by the National Natural
Science Foundation of China under Grants 11575008 and 11621131001.

\appendix
\section{Loop integrals}\label{LoopIntegrals}
The various loop functions $ J_{ij}^{F}$, $ J_{ij}^{T}$ and $
J_{ij}^{B}$ in this text are defined in the following. One can find
the complete forms and detailed derivations in
ref.~\cite{Wang:2018atz}.
\begin{eqnarray}\label{LoopIntJF}
&&i\int\frac{d^dl\lambda^{4-d}}{(2\pi)^d}\frac{\{l^\alpha
l^\beta\}}{\left(l^2-m^2+i\epsilon\right)\left[(l+q)^2-m^2+i\epsilon\right]}\equiv\left\{q^\alpha
q^\beta J_{21}^F+g^{\alpha\beta} J_{22}^F\right\}(m,q),
\end{eqnarray}\vspace{-0.5cm}
\begin{eqnarray}
&&i\int\frac{d^dl\lambda^{4-d}}{(2\pi)^d}\frac{\{l^\alpha l^\beta,l^\alpha l^\beta l^\gamma,l^\alpha l^\beta l^\gamma l^\delta\}}{\left(v\cdot l+\omega+i\epsilon\right)\left(l^2-m^2+i\epsilon\right)\left[(l+q)^2-m^2+i\epsilon\right]}\equiv\Big\{g^{\alpha\beta} J_{21}^T+q^\alpha q^\beta  J_{22}^T\nonumber\\
&&\quad+v^\alpha v^\beta  J_{23}^T+(q\vee v) J_{24}^T,(g\vee q) J_{31}^T+q^\alpha q^\beta q^\gamma  J_{32}^T+(q^2\vee v) J_{33}^T+(g\vee v) J_{34}^T\nonumber\\
&&\quad+(q\vee v^2) J_{35}^T+v^\alpha v^\beta v^\gamma  J_{36}^T,(g\vee g) J_{41}^T+(g\vee q^2) J_{42}^T+q^\alpha q^\beta q^\gamma q^\delta  J_{43}^T+(g\vee v^2) J_{44}^T\nonumber\\
&&\quad+v^\alpha v^\beta v^\gamma v^\delta  J_{45}^T+(q^3\vee v) J_{46}^T+(q^2\vee v^2) J_{47}^T+(q\vee v^3) J_{48}^T+(g\vee q\vee v) J_{49}^T\Big\}(m,\omega,q),\nonumber\\
\end{eqnarray}\vspace{-1.2cm}
\begin{eqnarray}\label{LoopIntJRB}
&&i\int\frac{d^dl\lambda^{4-d}}{(2\pi)^d}\frac{\{l^\alpha l^\beta,l^\alpha l^\beta l^\gamma,l^\alpha l^\beta l^\gamma l^\delta\}}{\left(v\cdot l+\omega+i\epsilon\right)\left[(+/-)v\cdot l+\delta+i\epsilon\right]\left(l^2-m^2+i\epsilon\right)\left[(l+q)^2-m^2+i\epsilon\right]}\equiv\nonumber\\
&&\quad\Big\{g^{\alpha\beta} J_{21}^{R/B}+q^\alpha q^\beta  J_{22}^{R/B}+v^\alpha v^\beta  J_{23}^{R/B}+(q\vee v) J_{24}^{R/B},(g\vee q) J_{31}^{R/B}+q^\alpha q^\beta q^\gamma  J_{32}^{R/B}\nonumber\\
&&\quad+(q^2\vee v) J_{33}^{R/B}+(g\vee v) J_{34}^{R/B}+(q\vee v^2) J_{35}^{R/B}+v^\alpha v^\beta v^\gamma  J_{36}^{R/B},(g\vee g) J_{41}^{R/B}\nonumber\\
&&\quad+(g\vee q^2) J_{42}^{R/B}+q^\alpha q^\beta q^\gamma q^\delta  J_{43}^{R/B}+(g\vee v^2) J_{44}^{R/B}+v^\alpha v^\beta v^\gamma v^\delta  J_{45}^{R/B}+(q^3\vee v) J_{46}^{R/B}\nonumber\\
&&\quad+(q^2\vee v^2) J_{47}^{R/B}+(q\vee v^3) J_{48}^{R/B}+(g\vee
q\vee v) J_{49}^{R/B}\Big\}(m,\omega,\delta,q),
\end{eqnarray}
where the notation $X\vee Y\vee Z\vee\cdots$ represents the
symmetrized tensor structure of $X^\alpha Y^\beta
Z^\gamma\cdots+\cdots$, which are given as
\begin{eqnarray}
q \vee v &\equiv& q^\alpha v^\beta+q^\beta v^\alpha,\quad g \vee q
\equiv
g^{\alpha\beta}q^\gamma+g^{\alpha\gamma}q^\beta+g^{\gamma\beta}q^\alpha,\nonumber
\end{eqnarray}\vspace{-1.2cm}
\begin{eqnarray}
g \vee v \equiv
g^{\alpha\beta}v^\gamma+g^{\alpha\gamma}v^\beta+g^{\gamma\beta}v^\alpha,\quad
q^2 \vee v &\equiv& q^{\beta}
q^{\gamma}v^{\alpha}+q^{\alpha}q^{\gamma} v^{\beta}+q^{\alpha}
q^{\beta} v^{\gamma},\nonumber
\end{eqnarray}\vspace{-1.2cm}
\begin{eqnarray}
q \vee v^2 \equiv q^{\gamma} v^{\alpha}v^{\beta}+q^{\beta}
v^{\alpha} v^{\gamma}+q^{\alpha} v^{\beta} v^{\gamma},\quad g \vee g
\equiv g^{\alpha\beta} g^{\gamma\delta}+g^{\alpha\delta}
g^{\beta\gamma }+g^{\alpha\gamma} g^{\beta\delta},\nonumber
\end{eqnarray}\vspace{-1.2cm}
\begin{eqnarray}
g \vee q^2 \equiv q^{\alpha} q^{\beta} g^{\gamma \delta}+q^{\alpha}
q^{\delta} g^{\beta\gamma} +q^{\alpha}
q^{\gamma}g^{\beta\delta}+q^{\gamma}q^{\delta}g^{\alpha\beta}
+q^{\beta} q^{\delta} g^{\alpha\gamma}
+q^{\beta}q^{\gamma}g^{\alpha\delta},\nonumber
\end{eqnarray}\vspace{-1.2cm}
\begin{eqnarray}
g \vee v^2&\equiv& v^{\alpha}
v^{\beta}g^{\gamma\delta}+v^{\alpha}v^{\delta} g^{\beta\gamma
}+v^{\alpha} v^{\gamma} g^{\beta\delta} +v^{\gamma} v^{\delta}
g^{\alpha\beta}+v^{\beta} v^{\delta} g^{\alpha\gamma} +v^{\beta}
v^{\gamma} g^{\alpha\delta},\nonumber
\end{eqnarray}\vspace{-1.2cm}
\begin{eqnarray}
q^3\vee v &\equiv& q^{\beta} q^{\gamma}
q^{\delta}v^{\alpha}+q^{\alpha} q^{\gamma} q^{\delta} v^{\beta}
+q^{\alpha} q^{\beta} q^{\delta} v^{\gamma}+q^{\alpha}q^{\beta}
q^{\gamma} v^{\delta},\nonumber
\end{eqnarray}\vspace{-1.2cm}
\begin{eqnarray}
q\vee v^3 \equiv q^{\delta} v^{\alpha} v^{\beta}
v^{\gamma}+q^{\gamma}v^{\alpha} v^{\beta} v^{\delta} +q^{\beta}
v^{\alpha}v^{\gamma} v^{\delta}+q^{\alpha} v^{\beta}
v^{\gamma}v^{\delta },\nonumber
\end{eqnarray}\vspace{-1.2cm}
\begin{eqnarray}
q^2 \vee v^2 &\equiv& q^{\gamma}q^{\delta} v^{\alpha}
v^{\beta}+q^{\beta} q^{\delta}v^{\alpha} v^{\gamma} +q^{\alpha}
q^{\delta} v^{\beta}v^{\gamma}+q^{\beta} q^{\gamma} v^{\alpha}
v^{\delta} +q^{\alpha} q^{\gamma} v^{\beta}
v^{\delta}+q^{\alpha}q^{\beta} v^{\gamma} v^{\delta},\nonumber
\end{eqnarray}\vspace{-1.2cm}
\begin{eqnarray}
g\vee q \vee v&\equiv& q^{\beta} v^{\alpha}
g^{\gamma\delta}+q^{\alpha}v^{\beta} g^{\gamma\delta} +q^{\delta}
v^{\alpha} g^{\beta\gamma }+q^{\gamma} v^{\alpha} g^{\beta\delta }
+q^{\alpha}v^{\delta} g^{\beta\gamma} +q^{\alpha} v^{\gamma} g^{\beta\delta}+q^{\delta} v^{\gamma} g^{\alpha\beta}\nonumber\\
&&+q^{\delta}v^{\beta} g^{\alpha \gamma}+q^{\gamma} v^{\delta}
g^{\alpha\beta}+q^{\gamma} v^{\beta}g^{\alpha\delta}+q^{\beta}
v^{\delta} g^{\alpha\gamma}+q^{\beta} v^{\gamma}
g^{\alpha\delta}.\nonumber
\end{eqnarray}
These $ J$ functions can be directly calculated with the dimensional
regularization in $d$ dimensions, or by an iterative way as shown in
ref.~\cite{Meng:2019ilv}. Their detailed expressions read
\begin{eqnarray}
 J_{22}^F(m,q)=\left(m^2-\frac{q^2}{6}\right)L+\frac{1}{32\pi^2}\int_0^1\bar{\Delta}\ln\frac{\bar{\Delta}}{\lambda^2}dx,~\textrm{where $\bar{\Delta}=x(x-1)q^2+m^2-i\epsilon$}.
\end{eqnarray}\vspace{-0.7cm}
\begin{eqnarray}
 J_{21}^T(m,\omega,q)=2\omega L+\frac{1}{16\pi^2}\int_0^1dx\int_{-\omega}^0\left(1+\ln\frac{\Delta}{\lambda^2}\right)dy+\frac{1}{16\pi}\int_0^1A^{1/2}dx,
\end{eqnarray}\vspace{-0.7cm}
\begin{eqnarray}
 J_{22}^T(m,\omega,q)=\frac{1}{8\pi^2}\int_0^1dx\int_{-\omega}^0\frac{x^2}{\Delta}dy+\frac{1}{16\pi}\int_0^1x^2A^{-1/2}dx,
\end{eqnarray}\vspace{-0.7cm}
\begin{eqnarray}
 J_{24}^T(m,\omega,q)&=&-L+\frac{1}{8\pi^2}\int_0^1dx\int_{-\omega}^0\frac{x(y+\omega)}{\Delta}dy-\frac{1}{16\pi^2}\int_0^1x\left(1+\ln\frac{A}{\lambda^2}\right)dx\nonumber\\
&&+\frac{\omega}{16\pi}\int_0^1x A^{-1/2}dx,
\end{eqnarray}\vspace{-0.7cm}
\begin{eqnarray}
 J_{31}^T(m,\omega,q)=-\omega L-\frac{1}{16\pi^2}\int_0^1dx\int_{-\omega}^0x\left(1+\ln\frac{\Delta}{\lambda^2}\right)dy-\frac{1}{16\pi}\int_0^1xA^{1/2}dx,
\end{eqnarray}\vspace{-0.7cm}
\begin{eqnarray}
 J_{32}^T(m,\omega,q)=-\frac{1}{8\pi^2}\int_0^1dx\int_{-\omega}^0\frac{x^3}{\Delta}dy-\frac{1}{16\pi}\int_0^1x^3A^{-1/2}dx,
\end{eqnarray}\vspace{-0.7cm}
\begin{eqnarray}
 J_{33}^T(m,\omega,q)&=&\frac{2}{3}L-\frac{1}{8\pi^2}\int_0^1dx\int_{-\omega}^0\frac{x^2(y+\omega)}{\Delta}dy+\frac{1}{16\pi^2}\int_0^1x^2\left(1+\ln\frac{A}{\lambda^2}\right)dx\nonumber\\
&&-\frac{\omega}{16\pi}\int_0^1x^2 A^{-1/2}dx,
\end{eqnarray}\vspace{-0.7cm}
\begin{eqnarray}
 J_{34}^T(m,\omega,q)&=&\left(m^2-\frac{q^2}{6}-2\omega^2\right)L-\frac{1}{16\pi^2}\int_0^1dx\int_{-\omega}^0(y+\omega)\left(1+\ln\frac{\Delta}{\lambda^2}\right)dy\nonumber\\
&&-\frac{\omega}{16\pi}\int_0^1
A^{1/2}dx+\frac{1}{32\pi^2}\int_0^1A\ln\frac{A}{\lambda^2}dx,
\end{eqnarray}\vspace{-0.7cm}
\begin{eqnarray}
 J_{41}^T(m,\omega,q)=\omega\left(m^2-\frac{q^2}{6}-\frac{2}{3}\omega^2\right)L+\frac{1}{32\pi^2}\int_0^1dx\int_{-\omega}^0\Delta\ln\frac{\Delta}{\lambda^2}dy+\frac{1}{48\pi}\int_0^1A^{3/2}dx,\nonumber\\
\end{eqnarray}\vspace{-1.2cm}
\begin{eqnarray}
 J_{42}^T(m,\omega,q)=\frac{2}{3}\omega L+\frac{1}{16\pi^2}\int_0^1dx\int_{-\omega}^0x^2\left(1+\ln\frac{\Delta}{\lambda^2}\right)dy+\frac{1}{16\pi}\int_0^1x^2A^{1/2}dx,
\end{eqnarray}\vspace{-0.7cm}
\begin{eqnarray}
 J_{43}^T(m,\omega,q)=\frac{1}{8\pi^2}\int_0^1dx\int_{-\omega}^0\frac{x^4}{\Delta}dy+\frac{1}{16\pi}\int_0^1x^4A^{-1/2}dx,
\end{eqnarray}\vspace{-0.7cm}
\begin{eqnarray}\label{JB_derivative}
 J_{ij}^B(m,\omega,\delta,q)=\left\{\begin{array}{ll}
\frac{1}{\delta+\omega}\left[ J_{ij}^T(m,\omega,q)+ J_{ij}^T(m,\delta,q)\right] & \textrm{if $\omega\neq-\delta\neq0$}\\
\frac{\partial}{\partial x} J_{ij}^T(m,x,q)\Big|_{x\to0} &
\textrm{if $\omega=\delta=0$}
\end{array} \right.,
\end{eqnarray}\vspace{-0.7cm}
\begin{eqnarray}
 J_{ij}^R(m,\omega,\delta,q)=\left\{\begin{array}{ll}
\frac{1}{\delta-\omega}\left[ J_{ij}^T(m,\omega,q)- J_{ij}^T(m,\delta,q)\right] & \textrm{if $\omega\neq\delta\neq0$}\\
-\frac{\partial}{\partial x} J_{ij}^T(m,x,q)\Big|_{x\to0} &
\textrm{if $\omega=\delta=0$}
\end{array} \right.,
\end{eqnarray}
where $\Delta=y^2+A$, $A=x(x-1)q^2+m^2-\omega^2-i\epsilon$, and
$\lambda=4\pi f_\pi$. The $L$ is defined as
\begin{eqnarray}
L=\frac{1}{16\pi^2}\left[\frac{1}{d-4}+\frac{1}{2}\left(\gamma_E-1-\ln4\pi\right)\right],
\end{eqnarray}
where $\gamma_E$ is the Euler-Mascheroni constant $0.5772157$. We
adopt the $\overline{\mathrm{MS}}$ scheme to renormalize the loop
integrals.

\section{Removing the 2PR contributions}\label{Remove_2PR}

Sometimes, we need to subtract the 2PR contributions from the box
diagrams, which can be recovered by inserting the one-pion-exchange
potentials into the iterative equations. For the case of
$\omega=\delta=0$, the 2PR part must be discarded due to the pinch
singularity. This can be easily done by using the simple derivative
relation given in eq.~\eqref{JB_derivative}. In this part, we
develop a new method to make such a subtraction with the help of the
principal-value integral method. In this way, we can subtract the
2PR part in a diagram with nonvanishing mass splittings, which has
no pinch singularity. Considering the loop integral of a box diagram
with the following form,
\begin{eqnarray}
\mathcal{I}=i\int\frac{d^dl\lambda^{4-d}}{(2\pi)^d}\frac{\mathscr{L}^{\mu\nu\cdots\alpha}(l)}{\left(v\cdot
l+\omega+i\epsilon\right)\left(-v\cdot
l+\delta+i\epsilon\right)\left(l^2-m^2+i\epsilon\right)\left[(l+q)^2-m^2+i\epsilon\right]},
\end{eqnarray}
where the Lorentz structure
$\mathscr{L}^{\mu\nu\cdots\alpha}(l)\equiv l^\mu l^\nu\cdots
l^\alpha$. This integral can be straightforwardly disassembled into
two parts through the following way,
\begin{eqnarray}
\frac{1}{\left(v\cdot l+\omega+i\epsilon\right)\left(-v\cdot
l+\delta+i\epsilon\right)}=\left[\frac{1}{v\cdot
l+\omega+i\epsilon}+\frac{1}{-v\cdot
l+\delta+i\epsilon}\right]\frac{1}{\omega+\delta}.
\end{eqnarray}
The principal-value integral method tells that
\begin{eqnarray}
\lim_{\epsilon\to0^+}\frac{1}{x\pm
i\epsilon}=\mathcal{P}\frac{1}{x}\mp i\pi\delta(x).
\end{eqnarray}
If we replace the $x$ with the $v\cdot l+\omega+i\epsilon$ and
$-v\cdot l+\delta+i\epsilon$, the integral can be divided into two
parts, the principal-vale part and the Dirac delta part. The Dirac
delta part is the pole contribution of the matter fields, which
corresponds to the 2PR part in the time-ordered perturbation
theory.The principal-value part is just the 2PI contribution. In
other words, the 2PI part of the integral $\mathcal{I}$ can be
written as
\begin{eqnarray}\label{2PI_abstract}
\mathcal{I}_{\mathrm{2PI}}=\mathcal{I}+\mathcal{I}_{\mathrm{2PR}}.
\end{eqnarray}
As long as we can derive the form of $\mathcal{I}_{\mathrm{2PR}}$,
we could obtain $\mathcal{I}_{\mathrm{2PI}}$, since the complete
form of $\mathcal{I}$ has been given in
appendix~\ref{LoopIntegrals}. The calculation of the
$\mathcal{I}_{\mathrm{2PR}}$ is simple due to the special property
of the delta function. We take the calculation of the
$\mathcal{I}_{\mathrm{2PR}}$ part of $ J_{21}^B$ as an example. We
first show the concrete form of the $\mathcal{I}_{\mathrm{2PR}}$,
\begin{eqnarray}
\mathcal{I}_{\mathrm{2PR}}&=&i\int_0^1dx\int\frac{d^dl\lambda^{4-d}}{(2\pi)^d}\left\{\frac{\mathscr{L}^{\mu\nu\cdots\alpha}(l-xq)}{l^2-\mathscr{M}^2+i\epsilon}i\pi\Big[\delta(v\cdot
l+\omega)+\delta(v\cdot
l-\delta)\Big]\frac{1}{\omega+\delta}\right\},
\end{eqnarray}
where we have used the Feynman parameterization to combine the
denominators of the propagators of the light pseudoscalars, and
$\mathscr{M}^2=x(x-1)q^2+m^2$. Besides, we have also utilized the
approximation $v\cdot q\simeq0$ in the two delta functions. Choosing
$\mathscr{L}^{\mu\nu\cdots\alpha}(l-xq)$ to be
$\mathscr{L}^{\mu\nu}(l-xq)$ we would be in the position to
calculate the 2PR part of the $ J_{2i}^B$ (denoted by $
J_{2i}^B\big|_{\mathrm{2PR}}$). For the $
J_{21}^B\big|_{\mathrm{2PR}}$, we have
\begin{eqnarray}
(-\pi)\int_0^1dx\int
dl_0\int\frac{d^{d-1}l\lambda^{4-d}}{(2\pi)^d}\frac{l^\alpha
l^\beta}{\left[(l_0+\bar{\mathscr{M}})(l_0-\bar{\mathscr{M}})\right]^2}\left[\delta(l_0+\omega)+\delta(l_0-\delta)\right]\frac{1}{\omega+\delta},
\end{eqnarray}
where $\bar{\mathscr{M}}=\sqrt{\bm{l}^2+x(x-1)q^2+m^2}-i\epsilon$.
This integral can be easily calculated. One finally obtains
\begin{eqnarray}\label{J21B_2PR}
 J_{21}^B\big|_{\mathrm{2PR}}=-\frac{1}{16\pi(\omega+\delta)}\int_0^1dx\left[\sqrt{\mathscr{N}(\omega)}+\sqrt{\mathscr{N}(\delta)}\right],
\end{eqnarray}
where the function
$\mathscr{N}(\omega)=x(x-1)q^2+m^2-\omega^2-i\epsilon$. Following
the same procedure given above, we can get all the 2PR parts of the
$ J_{ij}^B$ functions.

One can avoid the lengthy and tedious calculations by adopting
another trick. The loop integrals of the box diagrams can be
constructed from the ones of the triangle diagrams [e.g., see
eq.~\eqref{JB_derivative}], the finite part of the loop functions $
J_{ij}^T$ that make up the $ J_{ij}^B$ actually contains two types
of functions, one is the odd function of $\omega$, and the other one
is the even function of $\omega$. Therefore, the renormalized $
J_{ij}^T$ can be written as
\begin{eqnarray}
 J_{ij}^T(\omega)=\mathscr{O}_{ij}^T(\omega)+\mathscr{E}_{ij}^T(\omega),
\end{eqnarray}
where $\mathscr{O}_{ij}^T(\omega)$ and $\mathscr{E}_{ij}^T(\omega)$
represent the odd and even parts of the $ J_{ij}^T(\omega)$,
respectively. The other two variables $m$ and $q$ are omitted for
simplicity. It can be proved that $\mathscr{O}_{ij}^T(\omega)$ and
$\mathscr{E}_{ij}^T(\omega)$ account for the 2PI and 2PR parts of
the $ J_{ij}^B$, respectively. For example, we find the
$-\frac{1}{16\pi}\int_0^1dx\sqrt{\mathscr{N}(\omega)}$ in
eq.~\eqref{J21B_2PR} is just the opposite of the
$\mathscr{E}_{12}^T(\omega)$. With the simple properties of the odd
and even functions, we can readily obtain
\begin{eqnarray}\label{JijB_2PI}
 J_{ij}^B(\omega,\delta)\big|_{\mathrm{2PI}}=\frac{1}{\omega-(-\delta)}\left[\mathscr{O}_{ij}^T(\omega)-\mathscr{O}_{ij}^T(-\delta)\right].
\end{eqnarray}
When $\omega$ and $\delta$ approach to zero, this formula evolves
into the derivative relation in eq.~\eqref{JB_derivative}. One can
easily testify the remainder ones indeed satisfy the
eq.~\eqref{JijB_2PI}, likewise.

\section{Spin transition operators}\label{SpinTransfOperators}

In calculating the loop diagrams of the $\Sigma_c^\ast\bar{D}^\ast$
system, we encountered some intractable scalarproducts, such as
$(\bar{u}\cdot\varepsilon^\ast)(u\cdot\varepsilon)$ and
$(\bar{u}\cdot\varepsilon)(u\cdot\varepsilon^\ast)$, where the
$u^\mu$ denotes the spinor-vector of the spin-$3\over2$
Rarita-Schwinger field $\psi^\mu$, and $\varepsilon^\mu$ represents
the polarization vector of the spin-$1$ field $\tilde{P}^{\ast\mu}$.
$\bar{u}^\mu$ and $\varepsilon^{\ast\mu}$ are their conjugations,
respectively. We notice that these structures involving
polarizations can be transformed into the spin-spin interaction
terms by introducing the so-called spin transition operators for the
spin-$3\over2$ and spin-$1$ fields, respectively.

\subsection{Vector field}
In the rest frame of a vector particle, the space components of the
polarization vectors with different helicity $\lambda=0,\pm1$ read,
\begin{eqnarray}
\boldsymbol{\varepsilon}(0)=(0,0,1)^T,\qquad
\boldsymbol{\varepsilon}(\pm1)=\frac{1}{\sqrt{2}}(\mp1,-i,0)^T.
\end{eqnarray}
We define the corresponding eigenfunctions for the $\lambda=0,\pm1$
components, respectively,
\begin{eqnarray}
\phi(+1)=(1,0,0)^T,\quad\phi(0)=(0,1,0)^T,\quad \phi(-1)=(0,0,1)^T.
\end{eqnarray}
The $\boldsymbol{\varepsilon}(\lambda)$ can be obtained with the
following relation,
\begin{eqnarray}
\boldsymbol{\varepsilon}(\lambda)=\boldsymbol{S}_t\phi(\lambda),
\end{eqnarray}
where $\boldsymbol{S}_t$ is the spin transition operator for the
spin-$1$ field. The matrix form of the $\boldsymbol{S}_t$ is
\begin{eqnarray}
S_t^x=\frac{1}{\sqrt{2}}(-1,0,1),\quad
S_t^y=\frac{1}{\sqrt{2}}(-i,0,-i),\quad S_t^z=(0,1,0).
\end{eqnarray}
One can easily verify that
\begin{eqnarray}
\boldsymbol{S}_t^\dagger\cdot\boldsymbol{S}_t=\bm{1}_{3\times3},\qquad
-i\boldsymbol{S}_t^\dagger\times\boldsymbol{S}_t=\bm{S}_{v},
\end{eqnarray}
where $\bm{S}_{v}$ is just the spin operator of the vector field.
One can also testify the following relation,
\begin{eqnarray}\label{StiStj}
S_t^{i\dagger}S_t^j=\frac{i}{2}\epsilon^{ijk}S_v^k-\frac{1}{2}S_v^{\{i}S_v^{j\}}+\delta^{ij}.
\end{eqnarray}
\subsection{Rarita-Schwinger field}
The spin-$3\over2$ Rarita-Schwinger field $\psi^\mu$ can be
constructed by the polarization vector $\varepsilon^\mu$ and
two-component spinor $\chi$ with the following form,
\begin{eqnarray}
\psi^\mu=\sum_{m_\lambda,m_s}\langle
1,m_\lambda;\frac{1}{2},m_s|\frac{3}{2},m_\lambda+m_s\rangle\varepsilon^\mu(m_\lambda)\chi(m_s).
\end{eqnarray}
We can also define the eigenfunctions for helicity
$\lambda=\pm\frac{3}{2},\pm\frac{1}{2}$ components,
\begin{eqnarray}
\varphi(\frac{3}{2})=(1,0,0,0)^T, \varphi(\frac{1}{2})=(0,1,0,0)^T,
\varphi(-\frac{1}{2})=(0,0,1,0)^T,
\varphi(-\frac{3}{2})=(0,0,0,1)^T.
\end{eqnarray}
Then the field $\psi^\mu$ can be reexpressed as follows by
introducing the spin transition operator $\mathscr{S}^\mu_{t}$,
\begin{eqnarray}
\psi^\mu(\lambda)=\mathscr{S}^\mu_{t}\varphi(\lambda).
\end{eqnarray}
We can also get the matrix form of the $\mathscr{S}^\mu_{t}$,
\begin{align}
\mathscr{S}^{0}_{t}&=\bm{0}_{2\times4},&
\mathscr{S}^{x}_{t}&=\frac{1}{\sqrt{2}}\left( \begin{array}{cccc}
-1&0&\frac{1}{\sqrt{3}}&0\\
0&-\frac{1}{\sqrt{3}}&0&1
\end{array} \right),\nonumber\\
\mathscr{S}^{y}_{t}&=\frac{-i}{\sqrt{2}}\left( \begin{array}{cccc}
-1&0&\frac{1}{\sqrt{3}}&0\\
0&\frac{1}{\sqrt{3}}&0&1
\end{array} \right),&
\mathscr{S}^{z}_{t}&=\left( \begin{array}{cccc}
0&\sqrt{\frac{2}{3}}&0&0\\
0&0&\sqrt{\frac{2}{3}}&0
\end{array} \right).
\end{align}
Similarly, one can also obtain
\begin{eqnarray}
\mathscr{S}^\dagger_{t}\cdot\mathscr{S}_{t}=-\bm{1}_{4\times4},\quad
\bm{S}_{rs}=\frac{3}{2}\bm{\sigma}_{rs}=-\frac{3}{2}\mathscr{S}^{\dagger\mu}_{t}\bm{\sigma}\mathscr{S}_{t\mu},
\end{eqnarray}
where $\bm{S}_{rs}$ is the spin operator of the spin-$3\over2$
Rarita-Schwinger field. Analogous to eq.~\eqref{StiStj}, there also
exists a similar relation for
$\mathscr{S}^{i\dagger}_{t}\mathscr{S}_{t}^j$,
\begin{eqnarray}
\mathscr{S}^{i\dagger}_{t}\mathscr{S}_{t}^j=\frac{i}{3}\epsilon^{ijk}S_{rs}^k-\frac{1}{6}S_{rs}^{\{i}S_{rs}^{j\}}+\frac{3}{4}\delta^{ij}.
\end{eqnarray}

With the above preparations, the scalarproducts
$(\bar{u}\cdot\varepsilon^\ast)(u\cdot\varepsilon)$ and
$(\bar{u}\cdot\varepsilon)(u\cdot\varepsilon^\ast)$ can be breezily
worked out,
\begin{eqnarray}
(\bar{u}\cdot\varepsilon^\ast)(u\cdot\varepsilon)&=&-\frac{1}{6}\bm{S}_{rs}\cdot\bm{S}_v+\frac{1}{3}(\bm{S}_{rs}\cdot\bm{S}_v)^2-\frac{1}{2},\nonumber\\
(\bar{u}\cdot\varepsilon)(u\cdot\varepsilon^\ast)&=&\frac{1}{2}\bm{S}_{rs}\cdot\bm{S}_v+\frac{1}{3}(\bm{S}_{rs}\cdot\bm{S}_v)^2-\frac{1}{2}.
\end{eqnarray}
The emergence of the $(\bm{S}_{rs}\cdot\bm{S}_v)^2$ term is the
unique feature of the interactions between the high spin states.

\section{A tentative parameterization of the effective potential from the quark model}\label{PotentialModel}

Assuming a pair of $c$ and $\bar{c}$ quarks are produced in the high
energy colliding process, and they are surrounded by the largely
separated light quarks $u$ and $d$. At the very short $c$ and
$\bar{c}$ separation $r$, the $c$ and $\bar{c}$ quarks interact with
the perturbative one-gluon-exchange Coulomb potential. There is
essentially no screening of the $c\bar{c}$ interaction due to the
much farther separated $u$ and $d$ quarks. Before the hadronization
occurs, the effective potential at this size scale can be written
as~\cite{DeRujula:1975qlm,Godfrey:1985xj}
\begin{eqnarray}
V_{ij}(\bm{r}_i,\bm{s}_i,\bm{r}_j,\bm{s}_j)=-C\frac{\alpha_s}{4}\left(\frac{1}{|\bm{r}_i-\bm{r}_j|}-\delta^3(\bm{r})\frac{8\pi}{3m_im_j}\bm{s}_i\cdot\bm{s}_j+\cdots\right)+\cdots,
\end{eqnarray}
where we only show the central term and spin-spin interaction. Other
terms such as the tensor force and spin-orbit interaction are
omitted for the $S$-wave case. The $C$ denotes the color factor.
$\alpha_s$ is the strong coupling constant. $\bm{r}_i$, $\bm{s}_i$
and $m_i$ represent the position, spin, and mass of the $i$-th
quark, respectively. We need the $c\bar{c}$ color singlet to supply
an attractive core, thus $C=\frac{16}{3}$.

In order to avoid the $c$ and $\bar{c}$ pair to rapidly move far
away from each other with large velocity, we assume that the
$c\bar{c}$ pair is produced near the threshold. When the distance
between the slowly moving $c$ and $\bar{c}$ increases, the light
quarks $u$ and $d$ start to screen the color interaction at this
point. Then the five quarks form two weakly interacting color
singlet clusters $\Sigma_c^{(\ast)}$ and $\bar{D}^{(\ast)}$. The
force between them is nothing but just the residual color
interaction similar to the van der Waals force between neutral
molecules. At this size scale, the attractive core from $c$ and
$\bar{c}$ still works, but attenuates rapidly with the increase of
the separation $r$. At the same time, the heavy quark spin
decouples, and the spin-spin interaction is transferred to their
inner light degrees of freedom. If ignoring other higher order
contributions, one could roughly parameterize the potential as
follows,
\begin{eqnarray}
V=-\frac{e^{-\left(\frac{r}{d}\right)^x}}{r}\left[\frac{A}{\Lambda_\chi^2}+\frac{B}{m_{\Sigma_c^{(\ast)}}m_{D^{(\ast)}}}\boldsymbol{l}_1\cdot\boldsymbol{l}_2\right],
\end{eqnarray}
where $A$ and $B$ are two independent constants with the same
dimension, which can be determined by fitting the data.
$\boldsymbol{l}_1$ and $\boldsymbol{l}_2$ denote the spins of the
inner light degrees of freedom of $\Sigma_c^{(\ast)}$ and
$\bar{D}^{(\ast)}$, respectively\footnote{The matrix element of
$\boldsymbol{l}_1\cdot\boldsymbol{l}_2$ can be found in
ref.~\cite{Meng:2019ilv}.}. $d\in[1,2]$ fm stands for the
characteristic size of a hadronic molecule, we choose the upper
limit $d=2$ fm. $x$ is always chosen to be $1.5$ or $2$ for some
phenomenological considerations. Here we use $x=2$ as in
ref.~\cite{Bicudo:2015vta}. Obviously, the strength of the spin-spin
term is suppressed by the factor
$1/(m_{\Sigma_c^{(\ast)}}m_{D^{(\ast)}})$.
\begin{table*}[htbp]
\centering
\renewcommand{\arraystretch}{1.5}
\caption{The binding energies and masses of the $I=\frac{1}{2}$
hidden-charm $[\Sigma_c^{(\ast)}\bar{D}^{(\ast)}]_J$ systems in the
quark model (in units of MeV).}\label{MassSpectra_QuarkModel}
\setlength{\tabcolsep}{2.8mm} {
\begin{tabular}{c|ccccccc}
\hline\hline
System&$[\Sigma_c\bar{D}]_{\frac{1}{2}}$&$[\Sigma_c\bar{D}^\ast]_{\frac{1}{2}}$&$[\Sigma_c\bar{D}^\ast]_{\frac{3}{2}}$&$[\Sigma_c^\ast\bar{D}]_{\frac{3}{2}}$&$[\Sigma_c^\ast\bar{D}^\ast]_{\frac{1}{2}}$&$[\Sigma_c^\ast\bar{D}^\ast]_{\frac{3}{2}}$&$[\Sigma_c^\ast\bar{D}^\ast]_{\frac{5}{2}}$\\
\hline
$\Delta E$&$-5.32$&$-19.81$&$-4.38$&$-6.07$&$-24.54$&$-14.61$&$-3.55$\\
\hline
$M$&$4312.41$&$4439.94$&$4455.37$&$4376.25$&$4499.81$&$4509.74$&$4520.80$\\
\hline\hline
\end{tabular}
}
\end{table*}

By fitting the binding energies of the $P_c(4312)$, $P_c(4440)$ and
$P_c(4457)$, we obtain $A\simeq2.45$ GeV$^{2}$, $B\simeq-1.83$
GeV$^{2}$, i.e., their absolute values have the similar size. The
predictions for the masses of the $I=\frac{1}{2}$ hidden-charm
$[\Sigma_c^{(\ast)}\bar{D}^{(\ast)}]_J$ systems in the quark model
are listed in table~\ref{MassSpectra_QuarkModel}. We see the newly
observed three $P_c$s can be simultaneously reproduced, and other
four systems all have the binding solutions. The $\Delta E$ for
$[\Sigma_c^\ast\bar{D}]_{\frac{3}{2}}$ in the quark model is smaller
than that of the chiral perturbation theory. In addition, the
bindings of the $[\Sigma_c^\ast\bar{D}^\ast]_{J}$ systems are larger
than the predictions of the chiral perturbation theory. These
deviations mainly arise from the quantum fluctuations in the loop
diagrams, which can hardly be accommodated in the quark models.

\end{document}